\documentclass{aastex63}

\accepted{\today}
\submitjournal{AJ}

\shorttitle{BL Galactic Center survey}
\shortauthors{Gajjar et al.}
\graphicspath{{./}{figures/}}
\usepackage{graphics,graphicx,url,verbatim,xspace}
\usepackage[caption=false]{subfig}
\usepackage{glossaries}
\usepackage{comment}
\usepackage{makecell}
\usepackage{multirow}
\usepackage{todonotes}
\usepackage{amsmath}

\glsdisablehyper

\begin{document}

\title{The Breakthrough Listen Search For Intelligent Life Near the Galactic Center I}

\newacronym{alfa}{ALFA}{Arecibo L-Band Feed Array}
\newacronym{dm}{DM}{dispersion measure}
\newacronym{frb}{FRB}{fast radio burst}
\newacronym{fwhm}{FWHM}{full-width at half-maximum}
\newacronym{gbt}{GBT}{Green Bank Telescope}
\newacronym{gc}{GC}{Galactic Center}
\newacronym{hpbw}{HPBW}{half-power beam width}
\newacronym{if}{IF}{intermediate frequency}
\newacronym{igm}{IGM}{intergalactic medium}
\newacronym{ism}{ISM}{interstellar medium}
\newacronym{iism}{IISM}{ionized interstellar medium}
\newacronym{lst}{LST}{local sidereal time}
\newacronym{msp}{MSP}{millisecond pulsar}
\newacronym{nip}{NIP}{non-image Processing}
\newacronym{rfi}{RFI}{radio-frequency interference}
\newacronym{rrat}{RRAT}{rotating radio transients}
\newacronym{rm}{RM}{rotation measure}
\newacronym{ska}{SKA}{Square Kilometre Array}
\newacronym{sefd}{SEFD}{system equivalent flux density}
\newacronym{seti}{SETI}{search for extraterrestrial intelligence}
\newacronym{snr}{S/N}{signal-to-noise ratio}
\newacronym{sps}{SPS}{single pulse search}
\newacronym{tab}{TAB}{tied-array beam}
\newacronym{trapum}{TRAPUM}{TRansients And PUlsars with MeerKAT}
\newacronym{uwl}{UWL}{Ultra Wideband Low}
\newacronym{vlbi}{VLBI}{very long baseline interferometry}
\newacronym{bl}{BL}{Breakthrough Listen}
\newacronym{eti}{ETI}{extra-terrestrial intelligence}

\newcommand{\BI}{Breakthrough Initiatives\xspace}
\newcommand{\BLI}{Breakthrough Listen Initiative\xspace}
\newcommand{\BL}{Breakthrough Listen\xspace}
\newcommand{\SKA}{\textit{Square Kilometre Array}\xspace}
\newcommand{\MK}{\textit{MeerKAT}\xspace}
\newcommand{\VLA}{\textit{Karl G. Jansky Very Large Array}\xspace}
\newcommand{\Parkes}{\textit{Parkes Observatory}\xspace}
\newcommand{\MWA}{\textit{Murchison Widefield Array}\xspace}
\newcommand{\LOFAR}{\textit{LOw Frequency ARray}\xspace}
\newcommand{\ATA}{\textit{Allen Telescope Array}\xspace}
\newcommand{\GBT}{\textit{Green Bank Telescope}\xspace}
\newcommand{\APF}{\textit{Automated Planet Finder}\xspace}
\newcommand{\sgra}{Sgr\,A$^{*}$}
\newcommand{\subtot}[1]{{\textcolor{Cerulean}{#1}}}
\newcommand{\subsubtot}[1]{\textit{\textcolor{OliveGreen}{#1}}}
\newcommand{\tot}[1]{{\textcolor{NavyBlue}{#1}}}
\newcommand{\ie}{i.\,e.\,}
\newcommand{\eg}{e.\,g.\,}

\newcommand{\UCB}{Department of Astronomy,  University of California Berkeley, Berkeley CA 94720}
\newcommand{\SSL}{Space Sciences Laboratory, University of California, Berkeley, Berkeley CA 94720}
\newcommand{\SWIN}{Centre for Astrophysics \& Supercomputing, Swinburne University of Technology, Hawthorn, VIC 3122, Australia}
\newcommand{\GBTadd}{Green Bank Observatory,  West Virginia, 24944, USA}
\newcommand{\OXF}{Astronomy Department, University of Oxford, Keble Rd, Oxford, OX13RH, United Kingdom}
\newcommand{\NIJ}{Department of Astrophysics/IMAPP,Radboud University, Nijmegen, Netherlands}
\newcommand{\ATNF}{Australia Telescope National Facility, CSIRO, PO Box 76, Epping, NSW 1710, Australia}
\newcommand{\HOU}{Hellenic Open University, School of Science \& Technology, Parodos Aristotelous, Perivola Patron, Greece}
\newcommand{\USQ}{University of Southern Queensland, Toowoomba, QLD 4350, Australia}
\newcommand{\SETI}{SETI Institute, Mountain View, California}
\newcommand{\KZA}{University of Malta, Institute of Space Sciences and Astronomy}
\newcommand{\PWJD}{The Breakthrough Initiatives, NASA Research Park, Bld. 18, Moffett Field, CA, 94035, USA}
\newcommand{\BPF}{The Breakthrough Initiatives, NASA Research Park, Bld. 18, Moffett Field, CA, 94035, USA}
\newcommand{\cornell}{Cornell Center for Astrophysics and Planetary Science and Department of Astronomy, Cornell University, Ithaca, NY 14853, USA}
\newcommand{\PENN}{Department of Astronomy and Astrophysics, Pennsylvania State University, University Park PA 16802}
\newcommand{\MIT}{Massachusetts Institute of Technology, Cambridge, Massachusetts}
\newcommand{\NAU}{School of Informatics, Computing, and Cyber Systems, Northern Arizona University, Flagstaff, AZ, 86011}
\newcommand{\Curtin}{International Centre for Radio Astronomy Research, Curtin Institute of Radio Astronomy, Curtin University, Perth, WA 6845, Australia}

\correspondingauthor{Vishal Gajjar}
\email{vishalg@berkeley.edu}

\author[0000-0002-8604-106X]{Vishal Gajjar}
\affiliation{\UCB}

\author[0000-0002-6341-4548]{Karen I. Perez}
\affiliation{Department of Astronomy, Columbia University, 550 West 120th Street, New York, NY 10027, USA}

\author[0000-0003-2828-7720]{Andrew P. V. Siemion}
\affiliation{\UCB}
\affiliation{\NIJ}
\affiliation{\SETI}
\affiliation{\KZA}

\author[0000-0003-3197-2294]{Griffin Foster}
\affiliation{\UCB}

\author{Bryan Brzycki}
\affiliation{\UCB}

\author[0000-0002-2878-1502]{Shami Chatterjee}
\affiliation{\cornell}

\author{Yuhong Chen}
\affiliation{\UCB}

\author[0000-0002-4049-1882]{James M. Cordes}
\affiliation{\cornell}

\author[0000-0003-4823-129X]{Steve Croft}
\affiliation{\UCB}
\affiliation{\SETI}

\author[0000-0002-8071-6011]{Daniel Czech}
\affiliation{\UCB}

\author[0000-0003-3197-2294]{David DeBoer}
\affiliation{\UCB}

\author{Julia DeMarines}
\affiliation{\UCB}

\author{Jamie Drew}
\affiliation{\BPF}

\author[0000-0002-0826-6204]{Michael Gowanlock}
\affiliation{\NAU}

\author[0000-0002-0531-1073]{Howard Isaacson}
\affiliation{\UCB}
\affiliation{\USQ}

\author[0000-0003-1515-4857]{Brian C. Lacki}
\affiliation{\UCB}

\author{Matt Lebofsky}
\affiliation{\UCB}

\author{David H.\ E.\ MacMahon}
\affiliation{\UCB}

\author[0000-0003-0833-0541]{Ian S. Morrison}
\affiliation{\Curtin}

\author{Cherry Ng}
\affiliation{\UCB}

\author{Imke de Pater}
\affiliation{\UCB}

\author[0000-0003-2783-1608]{Danny C.\ Price}
\affiliation{\UCB}
\affiliation{\Curtin}

\author{Sofia Z. Sheikh}
\affiliation{\PENN}

\author[0000-0002-5389-7806]{Akshay Suresh}
\affiliation{\cornell}

\author{Claire Webb}
\affiliation{\MIT}

\author{S. Pete Worden}
\affiliation{\BPF}



\begin{abstract}

A line-of-sight towards the Galactic Center (GC) offers the largest number of potentially habitable systems of any direction in the sky. The Breakthrough Listen program is undertaking the most sensitive and deepest targeted SETI surveys towards the GC. Here, we outline our observing strategies with Robert C. Byrd Green Bank Telescope (GBT) and Parkes telescope to conduct 600 hours of deep observations across 0.7--93 GHz. We report preliminary results from our survey for ETI beacons across 1--8 GHz with 7.0 and 11.2 hours of observations with Parkes and GBT, respectively. With our narrowband drifting signal search, we were able to place meaningful constraints on ETI transmitters across 1--4 GHz and 3.9--8 GHz with EIRP limits of $\geq$4$\times$10$^{18}$\,W among 60 million stars and $\geq$5$\times$10$^{17}$\,W among half a million stars, respectively. For the first time, we were able to constrain the existence of artificially dispersed transient signals across 3.9--8 GHz with EIRP $\geq$1$\times$10$^{14}$\,W/Hz with a repetition period $\leq$4.3 hours. We also searched our 11.2 hours of deep observations of the GC and its surrounding region for Fast Radio Burst-like magnetars with the DM up to 5000 pc\,cm$^{-3}$ with maximum pulse widths up to 90\,ms at 6 GHz. We detected several hundred transient bursts from SGR\,J1745$-$2900, but did not detect any new transient burst with the peak luminosity limit across our observed band of $\geq$10$^{31}$ erg s$^{-1}$ and burst-rate of $\geq$0.23 burst-hr$^{-1}$. These limits are comparable to bright transient emission seen from other Galactic radio-loud magnetars, constraining their presence at the GC. 

\end{abstract}



\section{Introduction} 
\label{sec:intro}
The existence of intelligent life in the Universe is one of the most profound and fundamental questions posed to science. Recent discoveries of habitable exoplanets and their prevalence \citep{hsh+14,dc15, Bryson_2021} suggest that our Galaxy is likely to harbor life-bearing planets and perhaps technologically advanced civilizations.
Advances in new instruments such as the\textit{ James Webb Space Telescope} will likely provide some pathway to ``sniff'' atmospheres of several dozen exoplanets to seek indirect evidence of life by detecting biosignatures. However, in many instances such biosignatures are also expected to arise due to abiotic processes (see \citealt{wp14}) and may not provide substantial and direct evidence of life. Moreover, these surveys will be limited to only a few hundred light years, will require long observations, and are unlikely to answer the question of whether any life discovered is intelligent. Surveys are underway to seek evidence of ``technosignatures'' \citep{Tarter:2003p266} which can also provide indirect evidence of \gls{eti}. The \BLI\ (BLI) is a US \$100M 10-year effort to conduct the most sensitive, comprehensive, and intensive search for technosignatures on other worlds across a large fraction of the electromagnetic spectrum \citep{2017AcAau.139...98W,Isaacson:2017ib,gsc+19}.

Low-frequency electromagnetic waves, such as coherent radio waves, are prime candidates for such beacons as they are energetically cheap to produce, and can convey information at maximum speeds across vast interstellar distances. 
Technosignature searches at radio frequencies take place in a large, multidimensional parameter space. Two of the most challenging parameters of any technosignature search include (a) the location of such putative \gls{eti} transmitters and (b) the transmission frequency of such signals. Along with these, unknown signal characteristics such as strength, intermittency, polarization, modulation types, and other unknown characteristics also play a significant role in measuring completeness of any \gls{seti} survey \citep{Tarter:2003p266, wkl18}.  The \gls{bl} program mitigates some of these challenges by extending the search to a wide variety of targets across the entire electromagnetic spectrum accessible from existing ground-based observing facilities. The \gls{bl} program currently has dedicated time on three telescopes: the Robert C. Byrd Green Bank Telescope  (GBT;  \citealt{2018PASP..130d4502M}) and the \Parkes\ (Parkes; \citealt{Price:2018bv}) in the radio, and the \APF\ \citep[APF;][]{Lipman:2018wv} in the optical. Additionally, commensal observations will soon begin 
at the \MK\ radio telescope in South Africa. \cite{gsc+19} provides the current status of the above programs, as well as other observing facilities working alongside \gls{bl}. The primary targets of the \gls{bl} program include 1 million nearby stars, 100 nearby galaxies, and deep observations of the \gls{gc} and the Galactic plane \citep{Isaacson:2017ib}. \cite{Enriquez:2017} conducted one of the most sensitive surveys towards 692 nearby stars at 1.4\,GHz as part of the \gls{bl} program. More recently, \cite{price2020} expanded this search to 1327 stars and extended the survey to 3.7\,GHz. Further extending these surveys to more \gls{bl} targets, in this paper, we outline details of our comprehensive survey of the \gls{gc} with the BL program along with early results. 

\subsection{Beacon types and survey summary}
\label{sect:beacons}
There are two different ways of detecting evidence of \gls{eti} through radio technosignatures: detecting deliberate beacons, or eavesdropping on the leakage radiation which could be a by-product of extra-terrestrial technologies. However, it is much  harder to speculate on the type of leakage radiation and such putative leakage signals are also likely to be weaker. In this paper, we will focus on strong beacons deliberately transmitted by \gls{eti}. In the past, very few hours of observation on the \gls{gc} have actually been performed to search for these ETI beacons. A narrowband Doppler drifting beacon signal \citep{2004AAS...204.7504B,2012arXiv1211.6470H} is a prime candidate for deliberately transmitted technosignatures. \cite{SHOSTAK1985369} carried out one of the few dedicated surveys of the \gls{gc} for such narrowband signals in a 4-hour search with the Westerbork Synthesis Radio Telescope (WSRT) at a sensitivity threshold of $\sim$ $10^{18}$ W, reporting a non-detection. 

\cite{CoEk79} suggest that in addition to searching for signals concentrated in frequency space, searches for impulsive wide-band signals should also be conducted. Powerful wide-band radar systems are commonly used on Earth; they follow a dispersion relation similar to a pulse dispersed by the \gls{ism}. An advanced society may know about pulsars and the techniques required to detect them. They could generate a pulsar-like broadband radar signal readily detected by another technologically-capable society. Further, they could induce an artificial dispersion measure into the pulse to differentiate their transmitted signal from a pulsar. Such a signal serves as a second type of beacon readily identifiable as originating from non-astrophysical processes. \cite{CoEk79} further speculated that such broadband beacons will consist of extremely narrow ($\leq 4 \mu$s) in time pulses (non-physical) and conducted searches for naturally dispersed pulses towards many targets including the \gls{gc}. Although, Crab pulsar is known to produce $\mu$s structures at higher-frequencies \citep{Hankins_2007}. Thus, negative or artificial dispersion might be the only indicator of artificiality for such broadband signals. 

No other SETI survey with moderate sensitivity has been performed for the \gls{gc} region. Given the potential for discovery in the \gls{gc} region, our outlined survey will contiguously span the large fraction of visible radio window from ground-based facilities; i.e. from 1 GHz to 93 GHz. This constitutes the largest fractional bandwidth search for radio signals towards any source to date for SETI. In Section \ref{sect:motivations}, we highlight that the \gls{gc} is one of the most interesting locations to conduct a survey for technosignatures, along with Section \ref{sect:anci_science} outlining the ancillary science benefit. In Section \ref{sect:obs_stratergy} we outline our survey strategy for the Parkes telescope (Section \ref{sect:pks_suvery_stratergy}) and GBT (Section \ref{sect:gbt_survey_strategy}) which will conduct  observations across 0.7--3.8 GHz utilizing 352 hours and 3.9--93.0 GHz utilizing 280 hours, respectively. In Section \ref{sect:observations_report} we outline already carried out observations across 1--8 GHz from both these facilities. In Section \ref{sec:data_analysis} we detail our search for two beacon types. Section \ref{sect:narrow_seti} highlights our search for the above mentioned narrowband Doppler drifting signals, while Section \ref{sect:broad_seti} outlines our searches for four different types of broadband dispersed signals (originating from natural and non-astrophysical processes). Section \ref{sec:discussion} discusses the implications of our findings, and Section \ref{sec:conclusion} lists our final conclusions.  

\section{Galactic Center}
\label{sect:motivations}
\subsection{Highest number of line--of--sight targets}
\begin{figure}[!t]
\centering
    \includegraphics[width=0.45\textwidth, ]{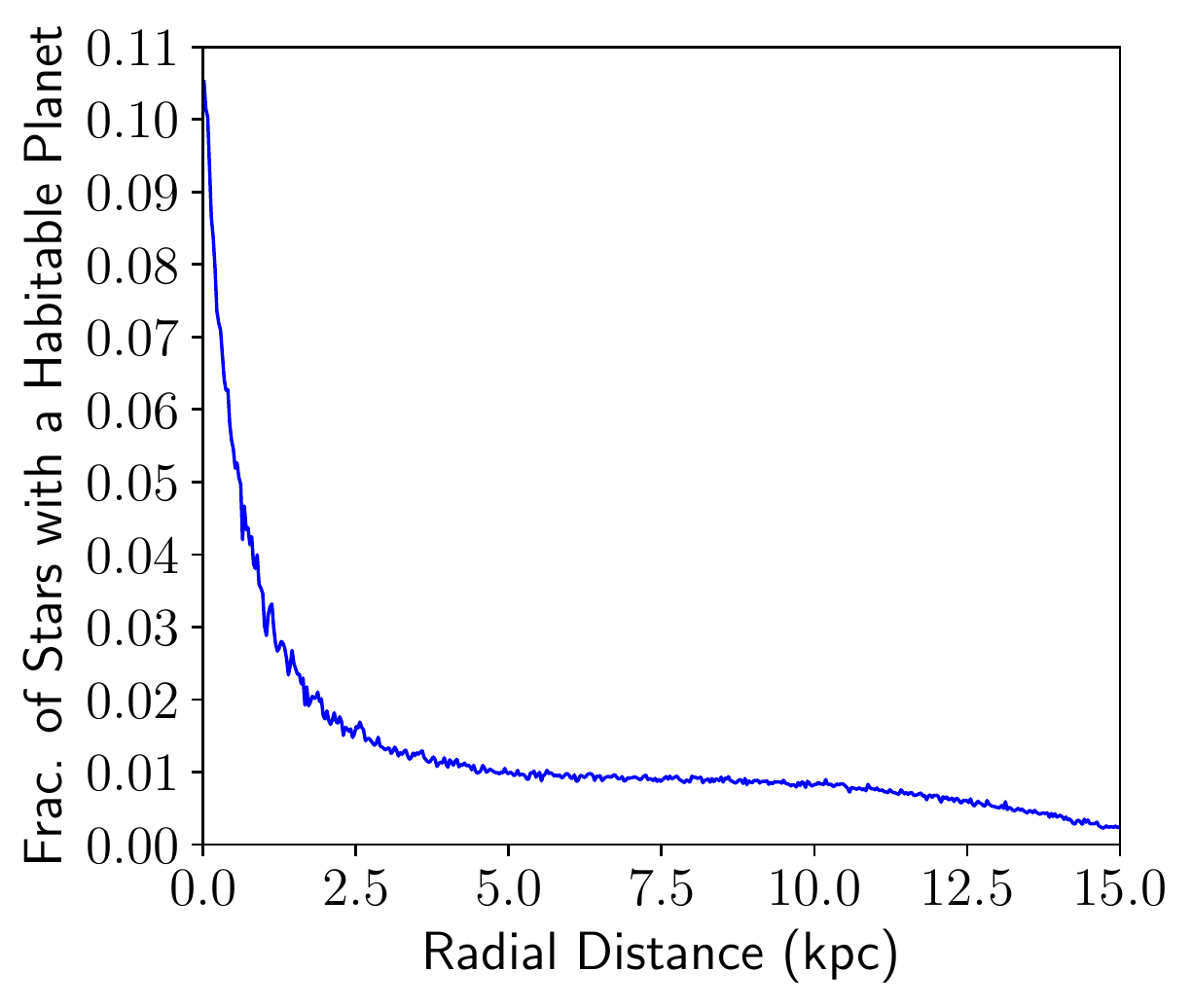}	
    \caption{The fraction of stars with a habitable planet as a function of galactocentric radius, $R$, using the most conservative model in \citet{gpm11}. The inner Galaxy hosts the greatest fraction of stars with a habitable planet. 
    }
   \label{fig:gowanlock2011_fraction_HP}
\end{figure}

The \gls{gc} has the largest concentration and highest number density of stars within the Galaxy. The line-of-sight toward the \gls{gc} offers the largest integrated Galactic star count of any direction in the sky. However, emergence of intelligence near the \gls{gc} depends on the probability of habitable planets along with their survival and stability due to nearby frequent supernovae (SNae) and other cataclysmic events
such as stellar flybys and flares from magnetars near the GC. \cite{lineweaver2004GHZ} also indicated that the inner region might be a less hospitable environment for habitability. In \citet{Torres_habitability}, it is found that, within $R=0.8$ kpc of the Galactic Center, stellar flybys are likely to disrupt Oort clouds, which would consequently result in the damage of planetary disks and systems. However, such flybys might deplete the Oort clouds on a fairly short time scale, thus making the planetary systems {\itshape more} habitable at longer times. 
\citet{gpm11} modeled the habitability of the Galaxy to host complex land-based life. \citet{Morrison:2015} assessed the propensity of the Milky Way to host intelligent life by including an additional timescale to the prior work of \cite{gpm11} that accounts for the transition from complex to intelligent life, which required 0.6 Gyr on Earth. Both these models suggest the inner Galaxy as the most likely place for the emergence of habitable worlds and intelligent life. However, \citet{gpm11} elected to model the Galactic disk only at $R\geq 2.5$ kpc, as the complicated formation history of the galactic bulge at $R<2.5$ makes it difficult to model habitability in the region. To understand the propensity of intelligent life in the very inner region of the Milky Way for our \gls{bl} survey, we update their model to account for the entire disk and include the $R<2.5$ kpc region. As is clearly evident from Figure \ref{fig:gowanlock2011_fraction_HP}, the fraction of stars with a habitable planet is greatest in the inner disk of the Galaxy. This is because there is much earlier planet formation in the region  than at the Solar neighborhood or the outskirts due to the inside-out formation history of the Galaxy. Despite the high SNae rate in the inner Galaxy, there are more opportunities for land-based, subsurface or oceanic complex life to emerge on planets in the region.The propensity for the emergence of intelligent life as introduced by \cite{Morrison:2015} closely follows the density of habitable worlds and thus also suggests the inner region of the Galaxy as the most promising region. In summary, with findings from \citet{gpm11}, \cite{Morrison:2015}, and our extension to the inner disk ($R<2.5$ kpc), if the emergence of life is commonplace within the galaxy, it follows that the inner Galaxy, especially regions around the \gls{gc}, would provide an ideal location to search for technosignatures in our Galaxy. 

\subsection{Space-faring societies}
The higher density of stars and habitable planets at the \gls{gc} will also be beneficial for the growth of more advanced societies.  Close proximity among the likely-inhabited worlds may accelerate development of interstellar communication and  travel, which can give rise to advanced space-faring societies.
\cite{NEWMAN_SAGAN} first discussed in detail diffusion of such advanced societies across the Milky Way. More recently, \cite{cfw+19} modeled the expansion of space-faring societies, and suggested that the high density of stars provides settlement fronts to move much quicker. Thus, if civilizations engage in settlements, planets in the inner region of the Galaxy are likely to be settled by advanced societies much earlier than other parts of the Milky Way. Similar arguments are used by \cite{DiStefano:2016} to suggest that globular clusters are likely places to harbor space-faring societies due to their close proximity to each other. Such advanced societies are likely to produce technosignatures which can be detected across large interstellar distances. 

\subsection{Schelling point}
In game theory, \cite{schelling_1960} suggested an approach by which different members of a group can derive a mutually-beneficial solution to a problem in the absence of direct communication. These arguments can be extended to propose a range of parameter spaces that ought to provide a maximum outcome in searches for technosignatures \citep{wright18}. For example, suggestions of searching near the hydrogen-line frequency by \cite{1959Natur.184..844C} and within the ``cosmic water-hole'' frequencies by \cite{NASA:2003p185} provide likely ``Schelling points'' for suitable frequencies to search for ETI beacons. In order to maximize the chances of detecting beacons across the entire sky, the optimal Schelling point for such a transmitter to exist would be the \gls{gc}. The \gls{gc} is a natural cynosure of the entire Milky Way Galaxy, and also suggested by Project Cyclops as the only likely direction for sensitive techosignature searches \citep{NASA:2003p185}\footnote{Although, this is due to our current limitations in surveying all-sky across all frequencies with compelling sensitivities.}. 

\cite{bbb10II} argued various motivations for an advanced society to transmit powerful beacons and highlights the importance of an energy budget. They argued that transmitting strong pulses as beacons with a certain duty cycle would be the most cost-effective way to build a beacon. \cite{bbb10II} suggested that advanced societies residing in our line-of-sight towards the \gls{gc} might choose to transmit such powerful beacons towards the \gls{gc} to get the maximum number of targets in the beam. Such transmitters might also direct back outwards away from the \gls{gc} because it is likely that societies like ours will be looking at the \gls{gc} due to the above mentioned Schelling point argument. They hence suggested a survey strategy to look at the GC on a roughly daily cadence over a year period. \cite{bbb10I} calculated the energy budget of such a strong beacon. As the required power for such a transmitter is proportional to the square of the distance to the target (i.e. $P_{transmitter}\propto{{D^{2}}}$), placing a transmitter at a location which can illuminate a large fraction of targets in the Milky Way with the least amount of energy would also favor the \gls{gc}. Recently, \cite{PhysRevD.103.023014} suggested that it is possible to extract energy from the spinning black hole at the GC. Advanced ETI transmitter might be able to obtain energy this way to remain active for billions of year. Thus, we propose that by placing a powerful transmitter either at the \gls{gc} or pointing towards and away from the \gls{gc} are the most cost effective ways, making it an ideal Schelling point for technosignature searches within the Milky Way. 

\subsection{Ancillary Science}
\label{sect:anci_science}

The \gls{gc} region is also an exciting observational target for a host of natural astrophysical phenomena, prominently including pulsars in close orbits around the central super-massive black hole, \sgra\ and other magnetars, or in new exotic systems such as a millisecond pulsar in a binary system with a black hole. 

\subsubsection{Pulsars}
Pulsars orbiting near the Galactic Center
\citep{1997ApJ...475..557C,PfahlLoeb2004}
could offer unprecedented insights into the surroundings of its super-massive black hole of mass $\sim 4 \times 10^6  M_{\odot}$ \citep{Ghez_BH_GC,Genzel_GC_BH}. In particular, the discovery of a pulsar on a $\sim 1$ yr orbit around \sgra\ will lead to high precision vetting of the theory of General Relativity in the strong field regime \citep{Liu2012}. So far, the nearest pulsars discovered are $10 - 15$\arcmin\ away, or approximately 30\,pc in projection \citep{2006MNRAS.373L...6J,2009ApJ...702L.177D}, despite the expectation of a large population of close-in neutron stars. Our survey data are already being utilized for deep searches for regular and millisecond pulsars which will likely provide some of the most stringent constraint on the presence of pulsar populations within the central 1\,pc region (Suresh et al.~2021, in prep.). 

\subsubsection{SGR J1745$-$2900}
The exciting discovery of a radio magnetar, SGR J1745$-$2900, at a projected distance of 0.1\,pc from the \gls{gc} \citep{Eatough2013ATel} was originally detected as an X-ray flare by the \textit{Swift} observatory \citep{Degenaar2013, Kennea2013}, and then shown to repeat with a spin period of 3.76\,s by \textit{NuSTAR} \citep{Mori2013}. Radio pulsations of this source were thereafter detected at frequencies ranging from 1.2 to 18.95\,GHz \citep{Spitler2014}. The estimated \gls{dm} for the source is $1778 \pm 3$\,pc\,cm$^{-3}$, and the \gls{rm}, \mbox{$- 66960\pm 50$\,rad\,m$^{-2}$}, is the highest \gls{rm} value for any known Galactic source. \citet{Eatough2013} suggested the \gls{rm} is consistent with a large magnetic field pervading the plasma surrounding the super-massive black hole. Surprisingly however, the observed scattered profiles were measured to have a scattering time of 1.3\,s at 1\,GHz, much less than predicted by the scattering models \citep{Spitler2014}.  \citet{Bower2014} obtained the first \gls{vlbi} image of the \gls{gc} magnetar, and measured an angular size of approximately \mbox{$\theta = 130$}~mas at 8.7\,GHz, similar to that of \sgra\ itself. Since SGR\,J1745$-$2900 is so close to the GC region (only 2\farcs4 from the \gls{gc}), even at our highest survey frequency of 93 GHz (10\arcsec\ beam with the GBT), it will be well within our central beam. Thus, our survey will observe SGR\,J1745$-$2900 across 0.7--93.0\,GHz. This will allow a detailed study of spectro-temporal properties of numerous bursts and evolution of the integrated profile. 

\subsubsection{Fast Transients}
\label{sect:fast_transients}
Fast Radio Bursts (FRBs) are highly dispersed, millisecond duration bursts of unknown origin. Even a decade since their discovery, they are one of the most exciting challenges for modern astrophysics. Due to the DM  excess beyond the Milky Way contribution and excellent interferometric localization of around a half-a-dozen sources, they are revealed to be extra-galactic in origin. There are a number of theories that have been proposed to explain the origin of FRBs. Among them, highly magnetized neutron stars such as magnetars have been favored as one of the most likely origins. This is further supported by the recent detection of Galactic magnetar SGR\,1935+2154 \citep{chime_sgr1935,stare2SGR}. This luminous Galactic FRB-like magnetar (with radio luminosity $\approx$10$^{38}$\,erg\,s$^{-1}$) was only detected as single pulses \citep{stare2SGR,chime_sgr1935,fastSGR1935}, and prolonged observations looking for regular pulsation through harmonic searches were unsuccessful \citep{surnisSGR1,surnisSGR2}. Finding more such Galactic radio-loud magnetars and studying their spectro-temporal properties would likely help us solve the origin mystery of FRBs. 

\cite{Dexter2014} suggested that the GC is likely to harbor a significant population of magnetars. Radio magnetars have been seen to show bright transient radio bursts \citep{crh+06} with emission across a wide range of frequencies as the \gls{gc} magnetar (SGR\,J1745$-$2900) has been detected all the way up to 225\,GHz \citep{tek+15}. Spectra of the radio-emitting magnetars are remarkably flat \citep{Camilo2007,Levin2010,tek+15}, different to normal pulsars that typically have steep flux spectra ($S_\nu \propto \nu^{-1.4}$). For example, \cite{crh+06} also reported bright transient radio pulses from XTE\,J1810$-$197 all the way up to 42\,GHz. This means that magnetars can be considerably brighter than pulsars at higher frequencies. Recently, \cite{gps+20} reported a detection of Swift\,J1819.0$-$0167 across 4--11\,GHz, which has also been suggested as a magnetar candidate with a rotation period of around 1.3\,s. Moreover, \citet{gps+20} reported that Swift\,J1819.0$-$1607 was seen to emit only single bright pulses from 4 -- 11\,GHz instead of regular weaker pulses. This suggests that bright radio pulses can be observed at higher-frequencies from magnetars. {\itshape Does the GC region harbour a population of FRB-like Galactic magnetars?} With our survey, we will be able to search for such bright pulses all the way from 0.7\,GHz to 93\,GHz originating from any putative recently active magnetars from the GC region. 

Other sources producing prominent transient emission include \gls{rrat} and pulsars producing giant pulses (GPs). We will be sensitive to such sources at lower frequencies from our survey; however, spectral indices of these sources are relatively steep,  making their detection  challenging across most of our observing bands. 

\section{Observation Strategy}
\label{sect:obs_stratergy}
Our strategy is to split these observations into two modes; shorter ON-OFF observations of the \gls{gc}$(0,0)$\footnote{As the beam size varies as a function of frequency, we refer to the central pointing on the \gls{gc} for each observed band as GC$(0,0)$.} and the surrounding region, and deep observations of just the \gls{gc}$(0,0)$ region. The shorter observations will be conducted with three ON-OFF cadences with 5-minute and 10-minute pointings for GBT and Parkes, respectively. For Parkes, we doubled the length of individual scans due to the significantly higher sky temperature (T$_{sky}$) at lower-frequencies towards the \gls{gc}. Such ON-OFF pointing techniques have previously been used in \gls{bl}  observations of nearby stars by \cite{Enriquez:2017} and \cite{price2020}. In order to keep optimum observing efficiency, for OFF-pointings, we will use pointings which are at least two \gls{hpbw} away from the same GC sampled region. 
 
Interstellar scintillation induces intermittency in otherwise steady signals and will have time scales dependent on the degree of scattering \citep{1991ApJ...376..123C, 1997ApJ...487..782C}.  Consequently, the preferred observing strategy for the \gls{gc}
is to observe a target at  multiple epochs.  Thus, in order to discriminate such highly scintillating signals, we aim to conduct longer scans towards the GC$(0,0)$ which will be divided into multiple epochs spread over a few days. Such deep observations will also help us search for transient signals from ETIs and other astrophysical sources with lower duty cycles. 

It should be noted that terrestrial interference makes identification of truly sky localized narrowband signals a challenging task, especially for such deep observations where we are not planning to conduct similar length OFF-pointing observations. At the GBT, all the receivers at frequencies higher than X-band ($>12$\,GHz) have multiple beams which will allow a natural discrimination of sky localized narrowband signals for our planned longer scans. For the \gls{uwl} receiver at Parkes, along with the C-band and X-band receivers at the GBT, we aim to utilize the effects of the ISM on such narrowband signals to discriminate them from terrestrial interference using state-of-the-art machine learning techniques. Such techniques are currently being explored by the \gls{bl} team \citep{bsc+20} and will be discussed in a future publication. 

\subsection {Parkes Telescope}
\label{sect:pks_suvery_stratergy}
\begin{figure}
    \centering
    \includegraphics[scale=0.32]{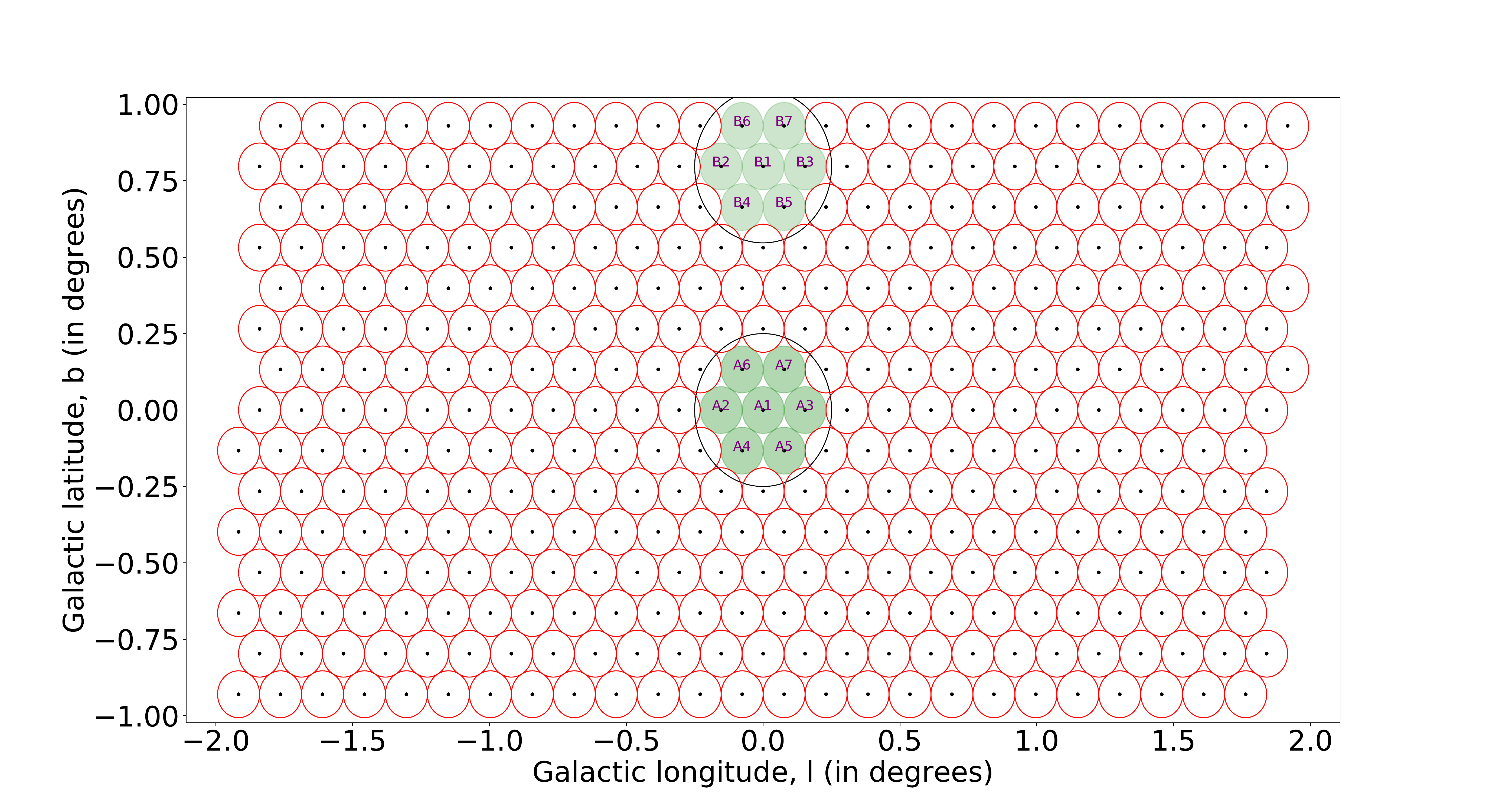}
    \caption{All 375 Parkes pointings planned to be surveyed with the \gls{bl} \gls{gc} survey. Each pointing, indicated with red solid circles, corresponds to the HPBW of the central frequency of the \gls{uwl} receiver. All of the  green pointings are already observed and reported in this paper. The black circle indicates the HPBW of 30\arcmin\ which corresponds to the lowest frequency of the \gls{uwl} receiver. The pointings inside the central 30\arcmin\ region are denoted A1--A7 while the corresponding OFF-source pointing are denoted B1--B7.  
    }
    \label{fig:Parkes_pointing}
\end{figure}

We will use 352 on-source hours to observe the GC$(0,0)$ and bulge using Parkes. Parkes will perform the lower frequency portion of the survey with the use of the newly installed \gls{uwl}. This  receiver covers 704\,MHz to 4.032\,GHz, with an average system temperature, $T_{\rm{sys}} = 22$\.K \citep{hobbs_UWL2020}. Because we will only be using one receiver and the \gls{hpbw} of Parkes is larger than that of the \gls{gbt}, the observing strategy will be to observe with approximately half the time spent on the \gls{gc} (168\,h) and the remaining time surveying the bulge.

At 700\,MHz, the \gls{hpbw} is approximately 30\arcmin. This defines the Galactic Center for the Parkes observations. At a distance of 8.2\,kpc (distance to the \gls{gc}; \citealt{Distance_to_gc}), this corresponds to a circular region with a radius of $\sim 35$\,pc. We plan to sample this region by using 7 beams at a distance of 9.2\arcmin\ from each other; this corresponds to the \gls{hpbw} at the mid-point of the frequency band for the \gls{uwl} (2.35\,GHz).  This is shown in Figure \ref{fig:Parkes_pointing}. We allocate approximately 24\,h per pointing. Each pointing will be observed multiple times in 2--3 hour increments, at a cadence of every 1--2 weeks, to identify transient and highly scintillating signals. Our search parameters include a range of $\pm 4$\,Hz\,s$^{-1}$, which is large enough to account for changes in the Doppler shifts due to Earth's spin, taking orbital velocity effects to be negligible, towards the GC on different days of observations.

The Galactic bulge will be surveyed for 184\,h. We define a region of
2\degr\ in Galactic latitude and 4\degr\ in Galactic longitude to cover
the extent of the bulge. This corresponds to 375 pointings including the above mentioned
\gls{gc} pointings. This will result in 30\,min per pointing, with three 10\,min-long scans per pointing.

\subsection{Green Bank Telescope}
\label{sect:gbt_survey_strategy}
The \gls{gbt} will be used for the high-frequency component of the survey starting from 4\, GHz (C-band) to 93\,GHz (W-band) (see Table \ref{tbl:gbt_rx}) with a total observing time of around 280\,h (excluding overhead). As mentioned earlier, a significant amount of observing time per receiver band will be dedicated for deep observations of the \gls{gc} $(0,0)$. The remaining time will be used to observe the region around the \gls{gc}. Many of the details in this section are taken from the GBT Proposer's Guide\footnote{https://science.nrao.edu/facilities/gbt/proposing/GBTpg.pdf} and Observing Guide\footnote{https://science.nrao.edu/facilities/gbt/observing/GBTog.pdf}. 

\begin{table}[h]
    \caption{Parkes low-frequency and GBT high-frequency receiver bands along with other necessary details. Columns list range of covered frequencies, number of beams, number of sub-bands, instantaneous bandwidth per sub-band, total bandwidth,  \gls{fwhm} beam size computed for the central frequency}, and \gls{sefd}, respectively. KFPA is a 7-beam receiver, however, we are only planing to use two of these beams with 4-GHz of band per beam for our survey.
    \centering
    \addtolength{\tabcolsep}{6.5pt}
    \begin{tabular}{l c c c c c c c}
    Receiver & Obs. Freq. & N$_{\textrm{beams}}$ & N$_{\textrm{bands}}$ & $\Delta \nu$/band & Total $\Delta \nu$ & Beam Size & SEFD \\
    & (GHz) & & & (GHz) & (GHz) & FWHM & (Jy) \\
    \hline
    \hline
    UWL & 0.704 - 4.032 & 1 & 1 & 4 & 4    & 9\farcm2 & 40 \\ 
    C  & 3.9 - 8     & 1  & 1  & 4    & 4    & 2\farcm5   & 10 \\
    X  & 8 - 11.6     & 1  & 1  & 2.4  & 2.4  & 1\farcm4   & 15 \\
    Ku & 12 - 15.4    & 2  & 1  & 3.5  & 3.5  & 54\arcsec   & 18 \\
    KFPA  & 18 - 27.5    & 2  & 2  & 8    & 9    & 32\arcsec   & 20 \\
    Ka & 26 -39.5     & 2  & 3  & 4    & 12   & 22\farcs6 & 25 \\
    Q  & 38.2 - 49.8  & 2  & 2  & 4    & 8    & 16\arcsec   & 60 \\
    W  & 67 - 93      & 2  & 4  & 6    & 24   & 10\arcsec   & 100 \\
    \hline
    \end{tabular}
    \label{tbl:gbt_rx}
\end{table}

\begin{figure}
\centering
    \subfloat[X-band]{
        \includegraphics[width=0.45\linewidth]{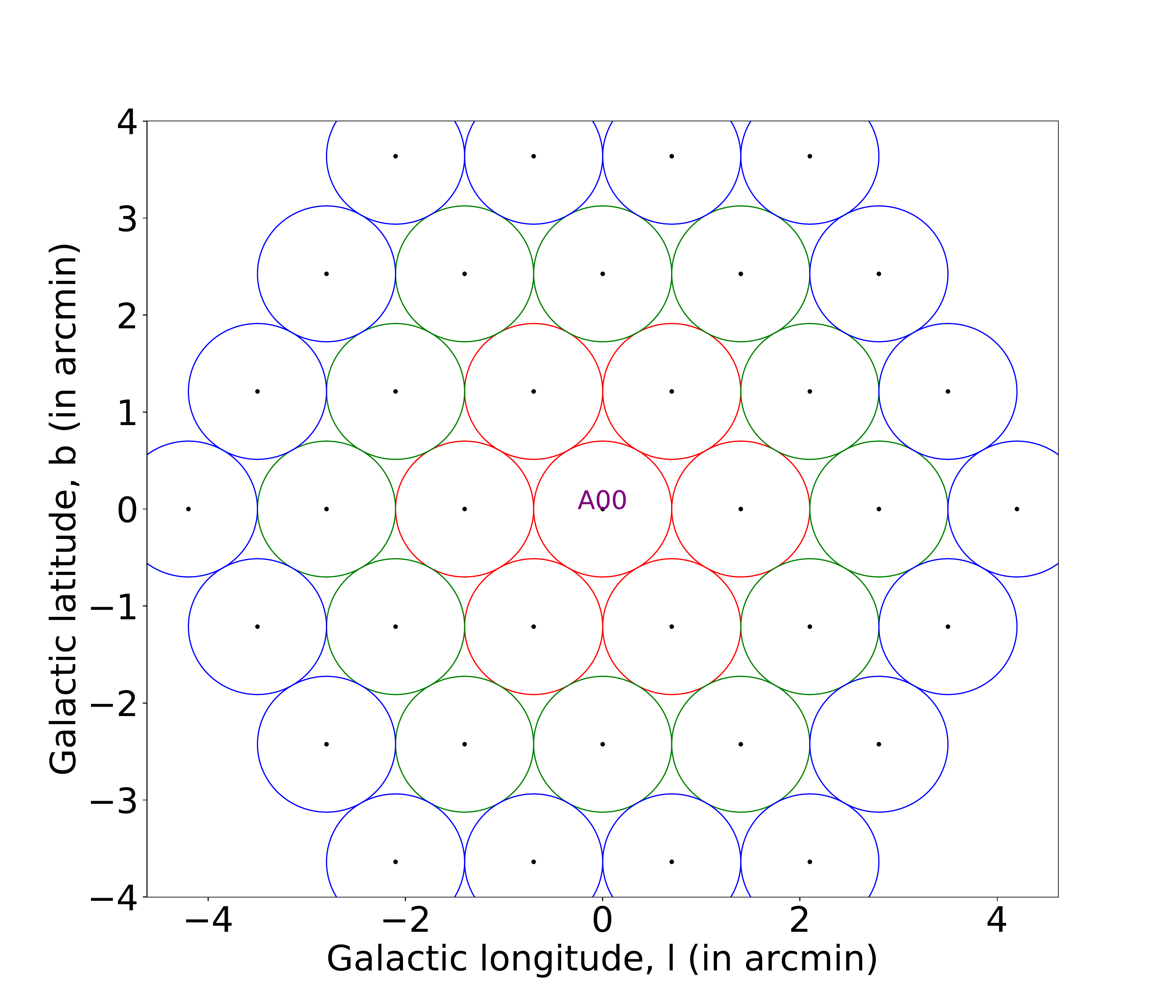}
        \label{fig:gbt_X_point}
    }
    \subfloat[Ku-band]{
        \includegraphics[width=.45\linewidth]{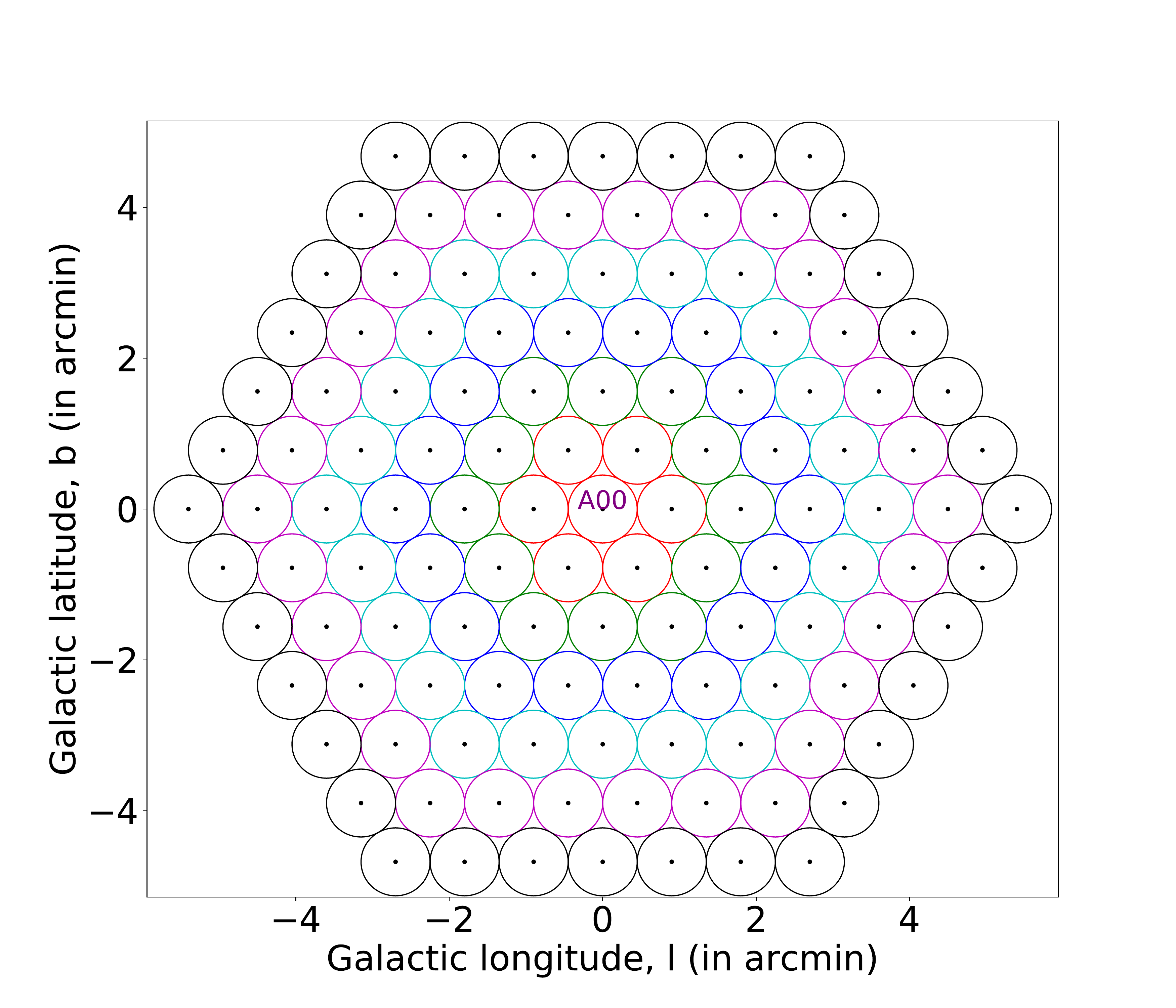}
        \label{fig:gbt_Ku_point}
    }
    \vspace{-0.4cm}
    \subfloat[KFPA-band]{
        \includegraphics[width=.45\linewidth]{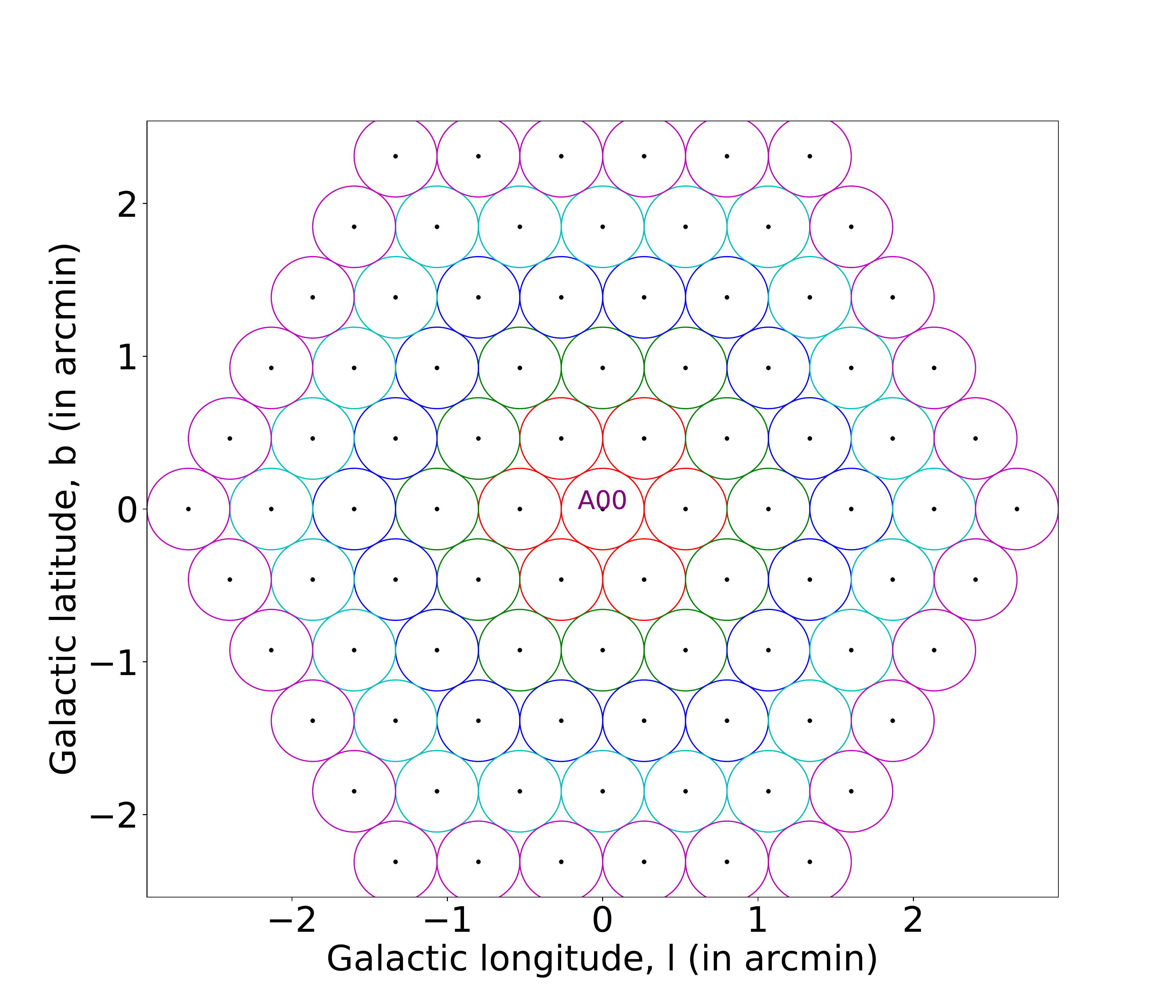}
        \label{fig:gbt_k2_point}
    }
    \subfloat[Ka-band]{
        \includegraphics[width=.45\linewidth]{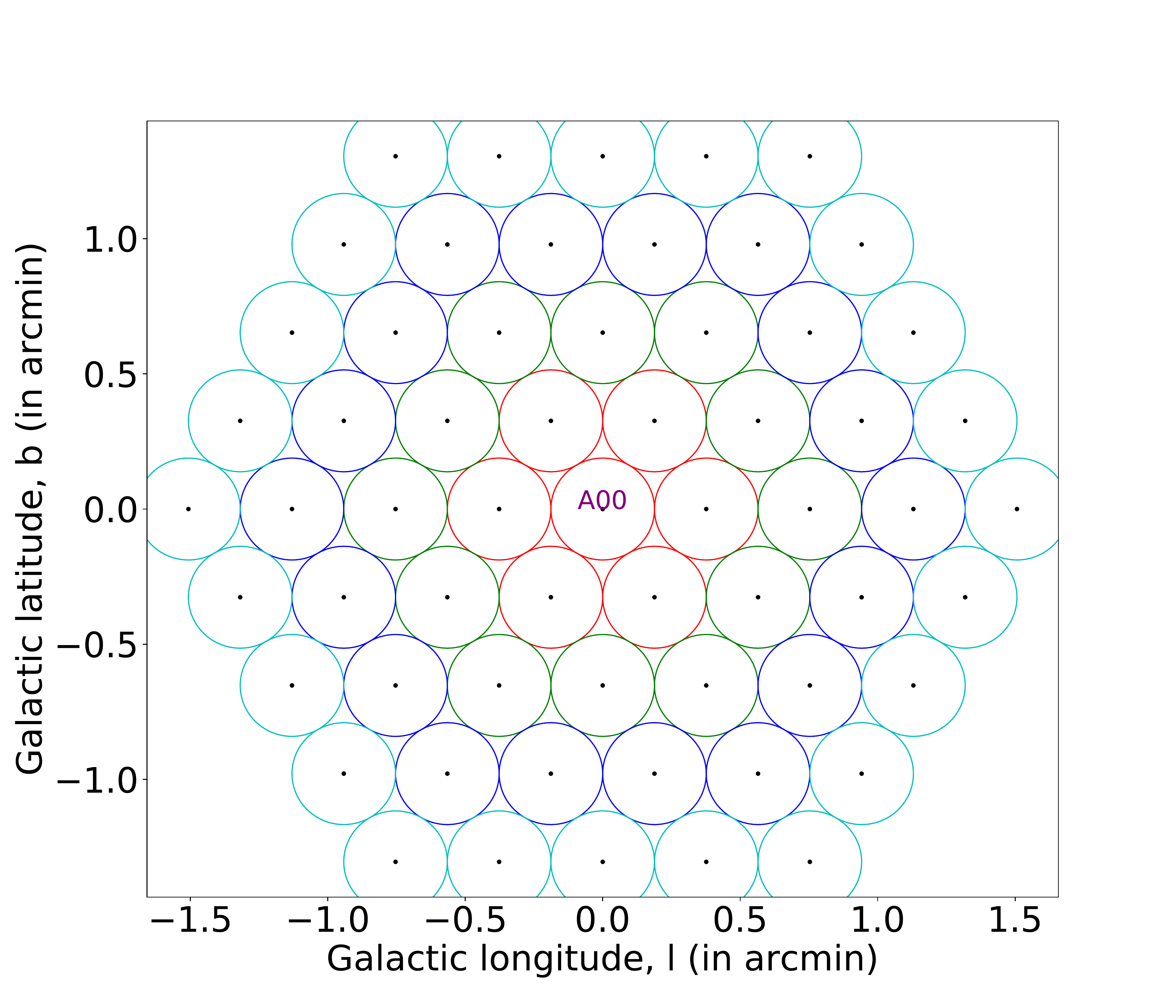}
        \label{fig:gbt_ka_point}
    }
    \vspace{-0.4cm}
    \subfloat[Q-band]{
        \includegraphics[width=.45\linewidth]{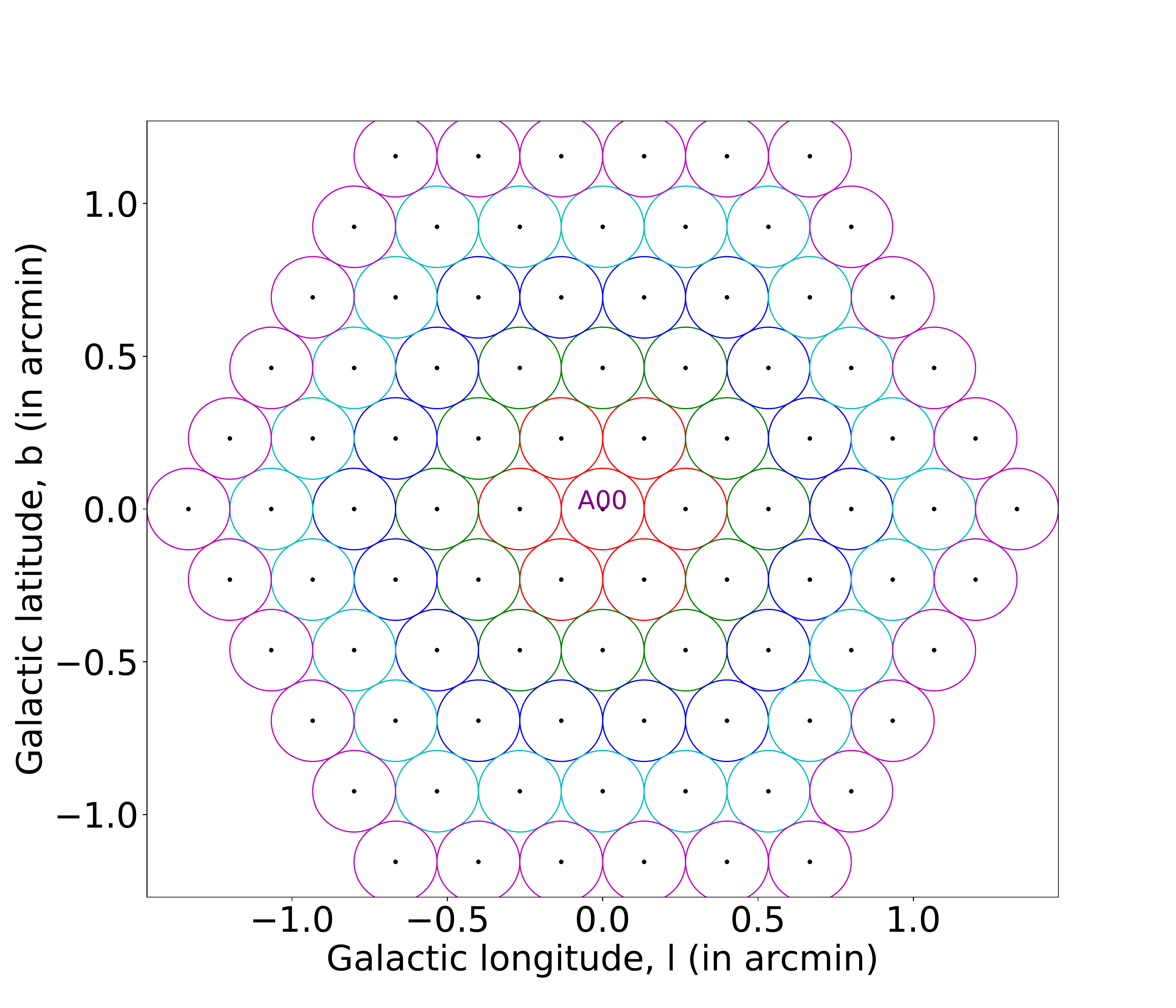}
        \label{fig:gbt_Q_point}
    }
    \subfloat[W-band]{
        \includegraphics[width=.45\linewidth]{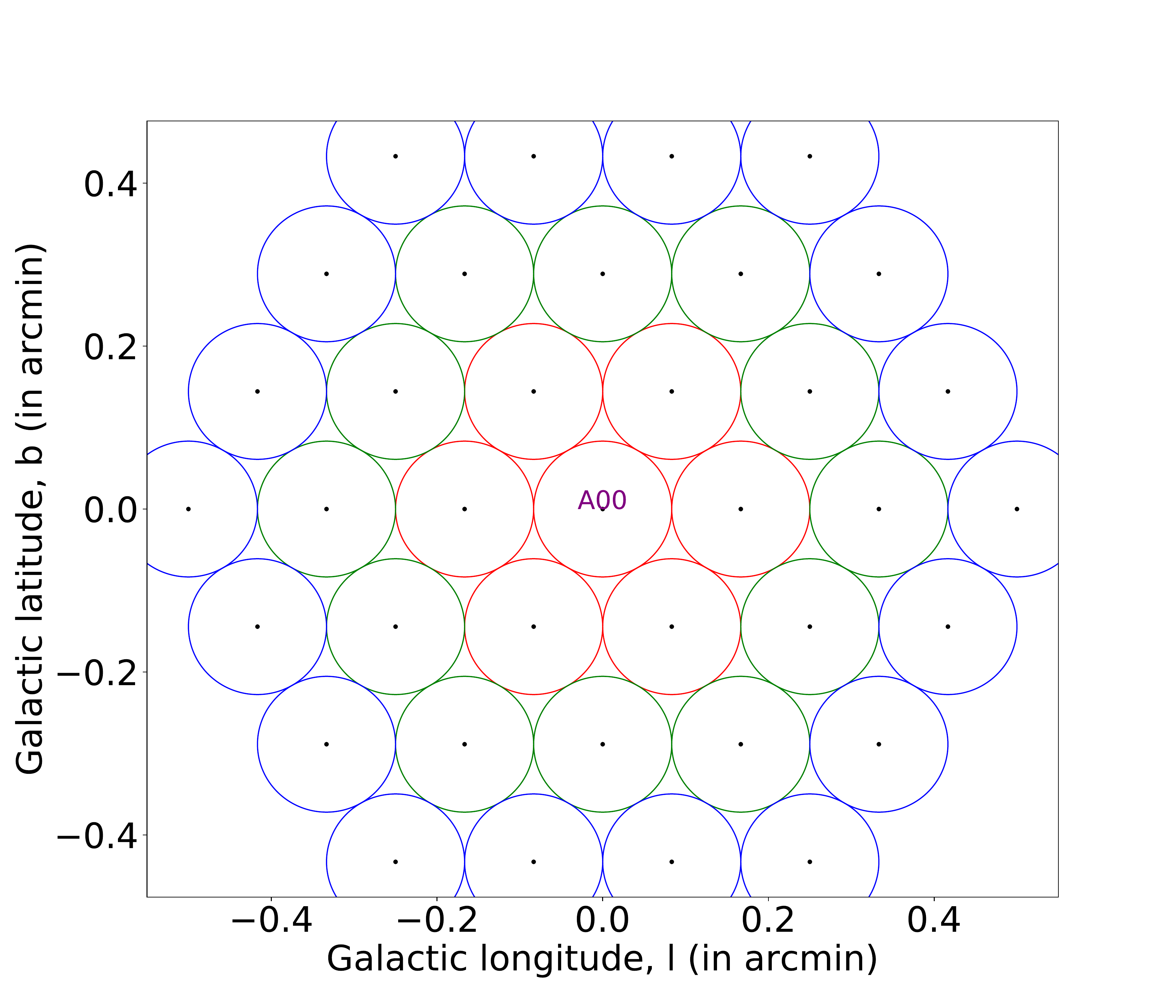}
        \label{fig:gbt_W_point}
    }
    \vspace{0.5cm}
    
    \caption{GBT pointings for X to W band. The middle pointing, labeled as A00, indicates deep integrations of the GC$(0,0)$. The rest of the beams are fully sampled at a fixed integration time
    around the \gls{gc}.
    }
    \label{fig:gbt_pointings}
\end{figure}

We would like to uniformly sample the \gls{gc} region with the complete observable \gls{gbt} band.  However, the \gls{gbt} beam size varies by a factor of 30 and the receiver \glspl{sefd} cover a factor of 10 in sensitivity from C-band to W-band. Also, the instantaneous bandwidth varies between receiver. As such, a compromise in survey area, bandwidth, and sensitivity must be made. The observation pointings and integration times are summarized in Figure \ref{fig:gbt_pointings} and Table \ref{tbl:gbt_rx_obs}. A deep integration, similar to the Parkes observing plan will be performed with each receiver of the GC$(0,0)$. The GC bulge around GC$(0,0)$, depending on the beam size and number of sub-bands to cover the full receiver bandwidth, will also be sampled at 15~minute integration times (three 5-minute scans). Further details of the receiver specific observing strategies are presented below. 


\begin{figure}
\centering
    {
        \includegraphics[width=0.6\linewidth]{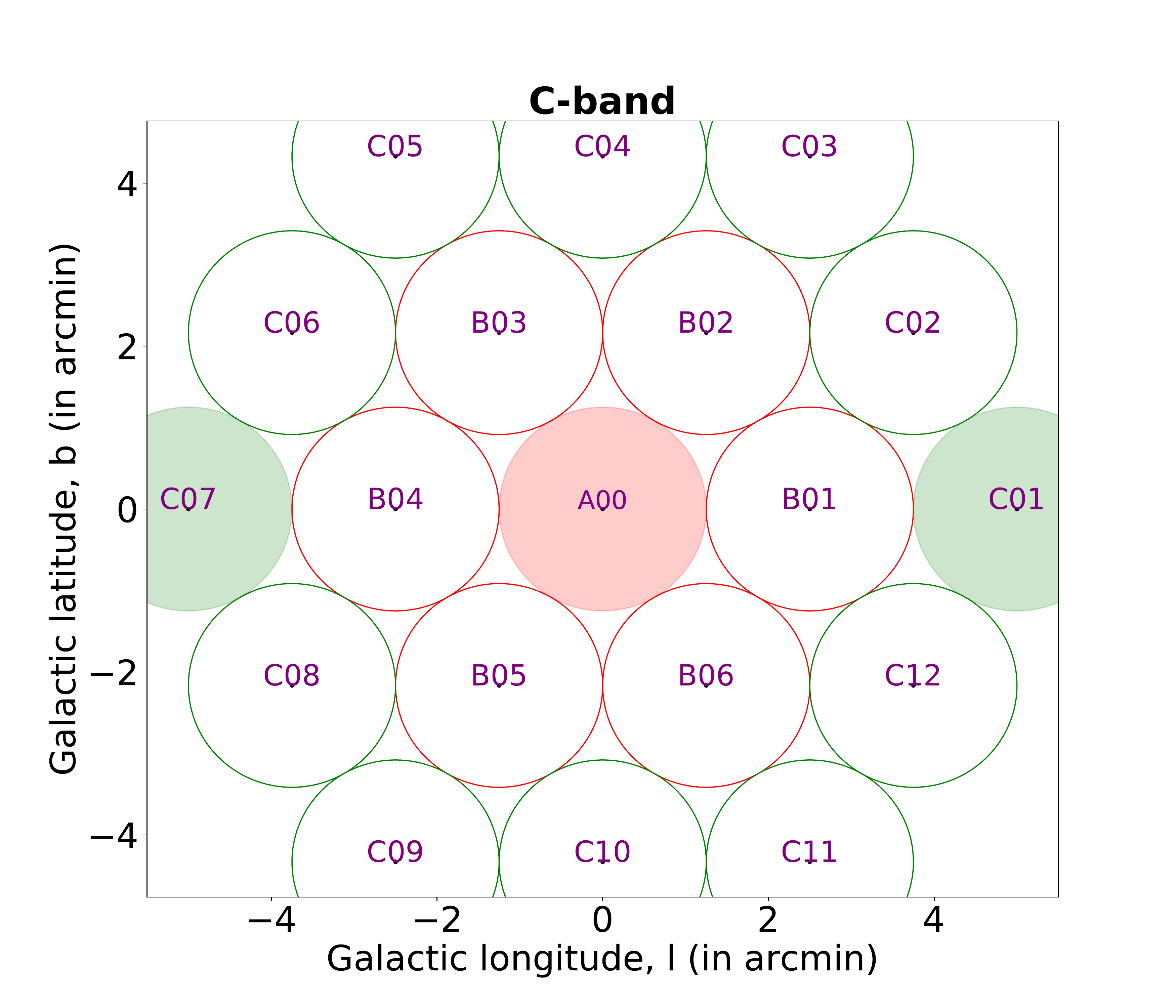}
        \label{fig:gbt_X_point}
    }
    \caption{Pointings at C-band for the \gls{bl} GC survey with the GBT. The central pointing, labeled as A00 and shown with light red fill, indicates the region of deep pointing of the GC$(0,0)$ at C-band. The rest of the pointings fully sample the $4 \arcmin \times 4\arcmin$ \gls{gc} bulge region. We highlight 2 pointings with green fill to demonstrate observation pairs with A00; in this case: A00-C01-C07-A00-C01-C07-A00-C01-C07. We should note that only for A00, C01, and C07 pointings we have used the approach of switching between three pointings. For all the other pointings, we have utilized pairs of two far away pointings to mimic standard ON-OFF observations.}
    \label{fig:gbt_pointings_Cband}
\end{figure}

\begin{table}[h]
    \centering
    \caption{Observing time and number of pointings for Parkes UWL and each \gls{gbt} receiver planned to be used in this \gls{gc} survey. The columns list the sampled region around the GC, available instantaneous bandwidth, number of deep pointing necessary to cover the full bandwidth at a given band and to fully sample GC$(0,0)$,  total observations time per each deep pointing, total number of bulge pointings, total observing hours for the bulge, and total observing hours for the full survey at a given band, respectively. The total observing hours for each receiver are calculated as $N_{\rm Int, deep}$ $\times$  $\tau_{\rm obs, deep}$ + $\tau_{\rm obs, bulge}$.
    }
    \addtolength{\tabcolsep}{5pt}
    \begin{tabular}{l c c c c c c c}
    Receiver & Sampled & Obs $\Delta \nu$ & Deep  &
    Obs Time & Bulge & Obs Time & Total Obs \\ 
    & Region & (GHz) & pointings, & per deep & pointings, & all bulge & (Hours) \\
    & & & $N_{\rm Int, deep}$ & pointing, & $N_{\rm Int, bulge}$ & pointings, & \\
    & & & & $\tau_{\rm obs, deep}$ & & $\tau_{\rm obs, bulge}$  & \\
    & & & & (hours) & & (hours)  & \\
    \hline
    \hline
    UWL & $2\degr \times 4\degr$ & 4 & 7 & 24 & 368 & 184 & 352 \\
    C  & $4\arcmin \times 4\arcmin$  & 4   &   1 & 10  & 18   &    4.5  & 14.5 \\
    X  & $4\arcmin \times 4\arcmin$  & 2.4 &   1 & 10  & 36   &    9    & 19   \\
    Ku & $4\arcmin \times 4\arcmin$  & 3.5 &   1 & 10  & 126  &    31.5 & 41.5 \\
    KFPA  & $2\arcmin \times 2\arcmin$  & 8   &   2 & 10  & 90   &    22.5 & 42.5 \\
    Ka & $2\arcmin \times 2\arcmin$  & 4   &   3 & 3   & 180  &    45   & 54   \\
    Q  & $2\arcmin \times 2\arcmin$  & 4   &   2 & 3   & 180  &    45   & 51   \\
    W  & $1\arcmin \times 1\arcmin$  & 6   &   4 & 2   & 144  &    36   & 44   \\
    \hline
    \end{tabular}
    \label{tbl:gbt_rx_obs}
\end{table}

\paragraph{C-Band (3.9--8 GHz)}
The C-band feed is dual polarizations, single beam receiver with 3.9\,GHz of instantaneous bandwidth. This feed has been used in previous \gls{bl} observations of the standard target list \citep{2017PASP..129e4501I}, stars located in the restricted Earth transit zone \citep{Sheikh_2020}, and FRB 121102 \citep{gaj18apj}. A 4\arcmin\ region around the \gls{gc} can be fully sampled with 19 pointings, including a deep pointing of the \gls{gc} (Figure \ref{fig:gbt_pointings_Cband}). The \gls{bl} digital backend (hereafter; BLDB \citealt{2018PASP..130d4502M}) will record baseband voltages across the entire 4\,GHz of the receiver bandwidth. Combining the deep pointing of GC$(0,0)$ for 10 hours with the standard ON-OFF pointings of the 4\arcmin\ bulge, we spent around 14 hours surveying the \gls{gc} region. Detailed analysis of these observations is presented in Sections \ref{sect:obs_cband} and \ref{sec:data_analysis}. 

\paragraph{X-Band (8.0--11.6 GHz)}
The X-band feed is a circularly polarized, single beam receiver with 2.4\,GHz of instantaneous bandwidth even though the total receiver bandwidth is 3.6\,GHz. We aim to sample a 4\arcmin\ region around the \gls{gc} with 37 pointings, including a deep 10\,h pointing of the GC$(0,0)$ (Figure \ref{fig:gbt_X_point}) with a total observing time of around 19\,h. 

\paragraph{Ku-Band (12.0--15.4 GHz)}
The Ku-band has two beams at a fixed separation ($5\farcm5$ in the cross-elevation direction), and each is circularly polarized.  The instantaneous bandwidth is 3.5\,GHz. Since the offset beam is likely to be outside the selected GC region for most of the GC pointings, it will be used to increase the sky coverage and to provide OFF-sources for GC region pointings. The BLDB is capable of recording 8~GHz of instantaneous bandwidth which will allows us to capture full receiver coverage simultaneously from both beams, which will eliminate the need for ON-OFF cadence scans. A 4\arcmin\ region around the \gls{gc} will be fully sampled with 127 pointings, including a deep pointing of the GC$(0,0)$ (Figure \ref{fig:gbt_Ku_point}) with a total observing time of 41.5\,h. 

\paragraph{K-Band Focal Plane Array (18.0--27.5\,GHz)}
The K-band receiver is a focal plane array (KFPA) of seven dual-polarization (circular) beams at fixed separations on the sky in a hexagonal pattern. The 2\arcmin\ region around the \gls{gc} can be fully sampled with 91 pointings as shown in Figure \ref{fig:gbt_k2_point} considering a \gls{hpbw} of 32\arcsec. There is a full seven beam mode at 1.8\,GHz instantaneous bandwidth per beam, and a dual-beam 8\,GHz total instantaneous bandwidth experimental mode. Since the \gls{bl} backend can capture 12\,GHz of instantaneous bandwidth, we will use the single central beam with 8\,GHz simultaneous bandwidth. For KFPA, we plan to use a similar ON-OFF observing strategy as that employed for the C-band and X-band. For the deep integration of the GC$(0,0)$ region, we plan to use the dual-beam mode of the KFPA with 4~GHz per beam. This doubles the number of independent pointings to cover the entire bandwidth for the deep GC$(0,0)$ pointing. Such a dual-beam approach is especially necessary for transient signals as dispersion delays at these frequencies are negligible which renders them indistinguishable from 
\gls{rfi}. The 90 pointings of 15\,min each from the GC bulge and 2 pointings (4\,GHz per pointing) of 10\,h each for GC$(0,0)$ gives a total observing time of 42.5\,h. 

\paragraph{Ka-Band (26.0--39.5 GHz)}
The Ka-band receiver consists of two beams, each with a single polarization (linear) at a fixed separation (78\arcsec\ in the cross-elevation direction). The beams are orthogonal in polarization. The band is split into three bands (26.0--31.0\,GHz, 30.5--37.0\,GHz, 36.0--39.5\,GHz), only one of which can be used at a time with an instantaneous bandwidth of 4\,GHz. Thus, we will use both beams with 4\,GHz per beam to survey the GC bulge and the GC$(0,0)$. 
Here, the second beam will provide OFF-source pointing for deep and standard pointings. Since three bands are required to sample the full receiver bandwidth, 60 pointings will be used per band for the bulge and a total of 3 deep observations of the GC$(0,0)$ (see Figure \ref{fig:gbt_ka_point}). The 180 pointings of 15\,min each from the GC bulge and 3 pointings (4\,GHz per pointing) of 3\,h each for GC$(0,0)$ gives a total observing time of 54\,h. 

\paragraph{Q-Band (38.2--49.8 GHz)}
Like Ka-band, Q-band also has a dual beam receiver, each dual-polarization (circular) at a fixed separation (57\farcs8 in the cross-elevation direction). The instantaneous bandwidth is 4\,GHz. Thus, with the BLDB we will record data from both beams. The offset beam will be used to increase the sky coverage and provide OFF-source pointings for the GC bulge and the GC$(0,0)$. The 2\arcmin\ region of the GC bulge is divided into 90 pointings (see Figure \ref{fig:gbt_Q_point}). We plan to spend 15\,min for bulge pointings and 3\,h for the deep observations of the GC$(0,0)$ region. Since two bands are required to sample the full receiver bandwidth, 90$\times$2 bulge pointings 
and 3$\times$2 deep observations provide us a total of 51\,h of observation. 

\paragraph{W-band (67--93 GHz)}
The W-band receiver also has two beams, each dual-polarization (linear) at a fixed separation (286\arcsec\ in the cross-elevation direction). Similar to Ka and Q-band, we will use the second beam to increase the sky coverage and obtain OFF-source pointings. We will sample a 1\arcmin\ region of the GC bulge with 37 pointings (see Figure \ref{fig:gbt_W_point}). We plan to spend the standard 15\,min for bulge pointings and 2\,h for the deep observations of the GC$(0,0)$ region. The band is split into four separate bands (67--74\,GHz, 73--80\,GHz, 79--86\,GHz, and 85--93.3\,GHz), only one of which can be used at a time. Thus, 36$\times$4 pointings for the bulge along with 2$\times$4 pointings for the GC$(0,0)$ will be completed in 44\,h. 

\section{Observations}
\label{sect:observations_report}
In this section, we discuss preliminary observations from the GBT (11.2\,h) and Parkes (7.0\,h). Table~\ref{tab:observations} includes a list of pointings and observing dates that are discussed in the following sections. 

\begin{table}[h]
\caption{Details of the observations analyzed in this paper from the GBT and Parkes. Columns list observing date, start time of the session, observed fields, calibrator, test pulsar, and total observing time for each session, respectively.}
    \centering
    \begin{tabular}{c c c c c c}
    \hline
    \hline
    \multicolumn{6}{c}{GBT} \\
    \hline
    Date & Start MJD & Fields & Calibrator & Test pulsars & Total On Source Time (min) \\ 
    \hline
    2019--08--07 & 58702.20313657 & A00,C01,C07 & 3C295 & B0355+54, J1744--1134 & 40 \\
    2019--08--09 & 58704.99252314 & A00,B01--B06,C01--C12 & 3C286 & B1133+16, J1744--1134 & 285 \\
    2019--09--07 & 58733.98109953 & A00                   & 3C286 & B2021+51 & 28 \\
    2019--09--08 & 58734.95761574 & A00                   &  $-$    &    $-$     & 60 \\
    2019--09--11 & 58737.95781249 & A00                   & 3C286 & B2021+51 & 258 \\ 
    
    \hline
    \hline
    \multicolumn{6}{c}{Parkes} \\
    \hline
    Date & Start MJD & Fields & Calibrator & Test pulsars & Total On Source Time (min) \\ 
    \hline
    2020--04--10 & 58949.58363425 & A1, B1        &  $-$ &     $-$       & 60    \\ 
    2020--04--29 & 58968.61862268 & A2, B2  & 0407-658 & J1141-6545 & 60 \\
    2020--04--23 & 58962.55708333 & A3,B3, A4, B4 & 1613-586  &  J1141-6545 & 120  \\
    2020--04--24/25 & 58963.56557870 & A5,B5,A6,B6, A7, B7 & 0407-658  &  J1141-6545 & 180  \\
    \hline
    \end{tabular}
    \label{tab:observations}
\end{table}

\subsection{GBT}
\label{sect:obs_cband}
Here, we report observations on the \gls{gc} carried out in five sessions between 2019 August 07 to 2019 September 11 using the C-band receiver at the GBT. As indicated in Section \ref{sect:obs_stratergy}, these observations were conducted towards 19 pointings (see Figure \ref{fig:gbt_pointings_Cband}) for three 5-min scans each. We also conducted deep observations of around 6.5\,h towards the GC$(0,0)$ marked as pointing A00. We labeled the surrounding hexagon pointings as B and C ( D, E, F, and G for higher bands), numbering the beams counterclockwise starting immediately to the right of A00. In order to eliminate false positives for the narrowband searches due to terrestrial \gls{rfi}, we carried out observations in a sequence which allowed two HPBW separations between consecutive pointing centers. The shorter scans towards A00 were followed by two shorter scans of outer ring pointings, i.e. C01 and C07. The other pointings were observed in the following pairs; B01--B04, B02--B05, B03--B06, C02--C04, C03--C05, C06--C08, C09--C11, and C10--C12 (see Figure \ref{fig:gbt_pointings_Cband}). We also recorded noise-diode scans on the flux calibrators and a strong pulsar to measure our sensitivity and configuration during most sessions (see Table \ref{tab:observations}). 

The observations, also listed in Table \ref{tab:observations}, were conducted using the BLDB. The BLDB provides a unique flexibility to capture and record baseband raw-voltages across 12\,GHz of bandwidth with 64 compute nodes at 32-bit resolution, making it one of the most powerful backends at the GBT. These compute nodes are clustered in 8 banks with each bank hosting 8 compute nodes. We also have 10 storage nodes with a total recording capacity of 8.2 PB for long term storage. For our observations, the BLDB re-purposed the existing VEGAS backend (Versatile Greenbank Astronomical Spectrometer; \citealt{vegas_2015}) to tap into the GBT analog downconversion streams with four 1500\,MHz wide tunable passbands covering the entire C-band receiver frequency response, spanning 3.9--8.0\,GHz. This interface provided coarsely channelized raw digitized voltages into 512 polyphase channels--with 2.97\,MHz of channel resolution. The recording of these raw voltages are carried out across an array of 32 compute nodes in four banks with each compute node configured to record 187.5\,MHz of the bandwidth. We configured our backend with central on-sky frequencies of 4312.5, 5437.5, 6562.5, and 7687.5\,MHz for the four overlapping bands. These frequencies were chosen to allow signal at the edges of each passband’s intermediate frequency filter to overlap the adjacent passband by 2$\times$187.5\,MHz, allowing overlap of two compute nodes between banks. This resulted in a total 4875.0\,MHz of bandwidth overlapping the entire C-band receiver instantaneous bandwidth of 4100\,MHz. 

\subsection{Parkes}
We report observations of the GC carried out in four sessions between 2020 April 10 to 2020 April 25 using the \gls{uwl} receiver at the Parkes telescope. We conducted observations towards 14 pointings, with three scans each for 10\,min per scan (Figure \ref{fig:Parkes_pointing}). We observed the GC$(0,0)$, labeled as A1, and its immediate surrounding region (A2-A7), using OFF-pointings one-beam width away (B1-B7) to avoid detecting false positives as described above in Section \ref{sect:obs_cband}. The pairs for our 14 pointings are as follows: [A1-B1], [A2-B2], ..., and [A7-B7]. The calibrators and test pulsars used to verify system integrity for most sessions can be found in Table \ref{tab:observations}.

Observations were recorded on the 26 \gls{bl} compute nodes that comprise the \gls{bl} data recorder system at Parkes. \cite{Price:2018bv} provides a detail summary of the \gls{bl} data recorder and storage systems at Parkes. As mentioned previously, for these observations we used the \gls{uwl} receiver \citep{hobbs_UWL2020}. The \gls{uwl} digital systems digitize incoming voltages from the receiver at 16 bit resolution (for each pol), channelize the data streams into 26 coarse channels of width 128\,MHz, and then output over high-speed Ethernet (total data rate $\sim$213 Gb/s). The BL data recorder \citep{Price:2018bv} captures these data (one 128\,MHz subband per compute node) and produces spectral products as 32-bit floats (see Table\,\ref{tab:formats}). A copy of the data are also sent (via Ethernet multicast) to the primary telescope digital processor, Medusa \citep{hobbs_UWL2020}.


\subsection{Data Products}
\label{sect:data_products}
\begin{table}[h]
\caption{Summary of data products produced from the standard \gls{bl} reduction pipeline and various signal types possible to search. Signal searches which are already carried out and discussed in this paper are marked with bold-faced texts. All these data products from our survey will be publicly available for the community.}
    \centering
    \begin{tabular}{c c c c}
    Data product    & Frequency Resolution  & Temporal Resolution & Possible Signal Searches\\
    \hline
    \hline
   High-spectral resolution &   $\sim$ 3\,Hz &  $\sim$ 18\,s  & narrowband drifting signals \\
   Mid-spectral resolution  &   $\sim$ 3\,kHz &  $\sim$ 1\,s  & Spectral line; wide-band pulses from ETI \\
   Mid-temporal resolution &    $\sim$ 350\,kHz &  $\sim$ 349 $\mu$s & natural and artificially dispersed transients \\
   High-temporal resolution\footnote{These data products were only produced for the GBT. } &   $\sim$ 91\,kHz  &  $\sim$ 43\,$\mu$s & \makecell{
   Pulsars,  MSPs, FRBs, \\ Wide-band pulses from ETI, Pulsars, FRBs} \\
   \hline
    \end{tabular}
    \label{tab:formats}
\end{table}{}

The \gls{bl} standard reduction pipeline at both the facilities reduces the collected raw-voltage data products into various temporal and spectral resolution products for offline processing. A detailed summary of this \gls{bl} reduction pipeline and discussion on various products is provided by  \cite{leb19_bl_data_format}. These formats are shown in Table \ref{tab:formats}. As discussed in Section \ref{sect:broad_seti}, the \gls{gc} region has a large electron density which is likely to induce large dispersion and scattering for any putative transient signals. At lower frequencies, lower than 4\,GHz with Parkes, these effects are dominating (with also increased noise temperature from the GC) and thus does not allow sensitive searches for transient signals. Thus, in this paper, we only searched for transient signals (naturally and artificially dispersed) above 4\,GHz with the GBT. Moreover, at the GBT, we also produced further high-temporal resolution products which will be useful for the community to search for pulsars and MSPs above 4\,GHz. We plan to publicly release all of our data collected from these observations via the Breakthrough Listen Open Data Portal\footnote{\url{https://breakthroughinitiatives.org/opendatasearch}}. 

\section{Analysis}
\label{sec:data_analysis}
In this section, we discuss our search for two different classes of signals using the data collected from the Parkes with the \gls{uwl} and with the GBT using the C-band receiver. 

\subsection{Narrow-band drifting signal searches}
\label{sect:narrow_seti}
\begin{table}[hb]
\caption{Narrowband search parameters for the \gls{bl} \gls{gc} survey}
    \centering
    \begin{tabular}{c|c}
        \hline
        \hline
        Parameter & Range \\
        \hline
        {S/N}$_{\rm min}$ threshold & 20 \\
        Drift Rate search & $\pm 4$\,Hz\,s$^{-1}$ \\
        GBT Frequency Range  & 3.9 --8\,GHz \\
        GBT trial drift rates & 860 steps \\
        Parkes Frequency Range & 1 -- 4\,GHz \\
        Parkes trial drift rates & 1720 steps \\
        filter 2 & candidates seen in at least 1 ON and no OFFs\\
        filter 3 & candidates seen in all 3 ONs and no OFFs \\
        \hline
    \end{tabular}
    \label{tab:turboSETI_parameter}
\end{table}

Narrowband signals have been proposed to be prime candidates for a deliberately transmitted beacon across large interstellar distances \citep{1959Natur.184..844C}. We will not discuss the merit of such signals any further here and refer to the earlier literature (see \cite{Tarter:2003p266} for a review). Due to the relative velocity between the transmitter and observer, such narrowband signals exhibit Doppler drifting across the observed band. If the narrowband signal at a fixed frequency is originating from a rotating object, the non-linear velocity difference will cause the observed frequency of this signal to change non-linearly. However, for shorter observing lengths we can approximate such Doppler drifting as a linear chirp. We used turboSETI \citep{Enriquez:2017} to search for such linearly chirped narrowband signals from the high-spectral resolution data products. Here, we used chunks of 2.9 MHz of zero-drift time-summed spectrum to determine the standard deviation ($\sigma$). We used a drift rate search in the range of $\pm4$\,Hz\,s$^{-1}$ with 860 and 1720 steps for the GBT and Parkes telescopes, respectively. For each trial drift rate step, we produced a time-averaged high-spectral resolution dechirp spectrum. Fine channels from this 2.9 MHz chunk which were found to be above ${\rm 20}\sigma$ were identified as narrowband drifting signals, or "hits".


Such a range is sufficient to detect drifting introduced by any transmitter located on typical Earth size planets including highly eccentric near Earth objects \citep{Sheikh_2019}. \citet{Sheikh_2019} recommend that ideally drift rates in a range of $\pm200$\,Hz\, s$^{-1}$ should be used to search for such narrowband signals. This range covers observed drift rates from transmitters possibly located on numerous exotic objects such as gaseous planets and planets in highly eccentric obits. However, because searching through such a wide range of drift rates is computationally expensive, here we limit our drift rate search in the range of $\pm4$\,Hz\,s$^{-1}$, with the caveat that this is by no means a full search of all possible narrowband drifting signals contained in all the observations analyzed in this paper. For example, ongoing efforts to improve our narrowband searching code, turboSETI, have led us to find an improvement in sensitivity which will be tackled in future papers. Due to our specific frequency and time resolution of high-spectral resolution products, signals with higher drift rate will spread across multiple nearby channels within a single spectra, incurring loss in sensitivity equivalent to {\rm1/N}, where N is the number of frequency channels over which a single signal is spread (see \cite{mpg+21} for a detailed discussion). However, it is possible to improve sensitivity by a factor of $\sqrt{N}$ for such high drifting signals by adding adjacent frequency bins or by using a varying width moving boxcar \citep{price2020}.

\begin{figure}[h]
  \centering
  \subfloat[]{
  \includegraphics[width=0.51\textwidth]{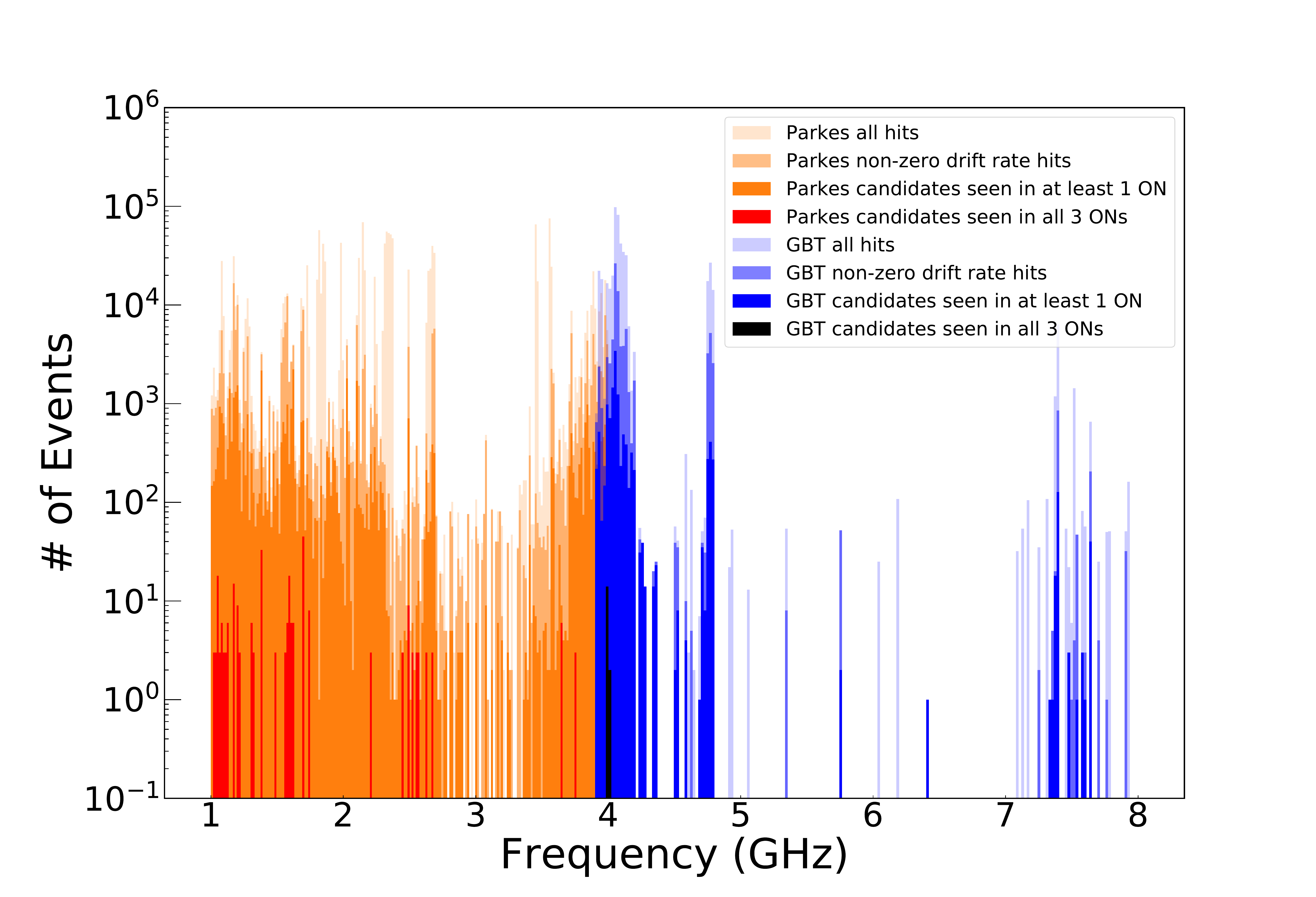}
  \label{fig:Freq_Hist}} 
  \hspace*{-0.75cm}
  \subfloat[]{
  \includegraphics[width=0.51\textwidth]{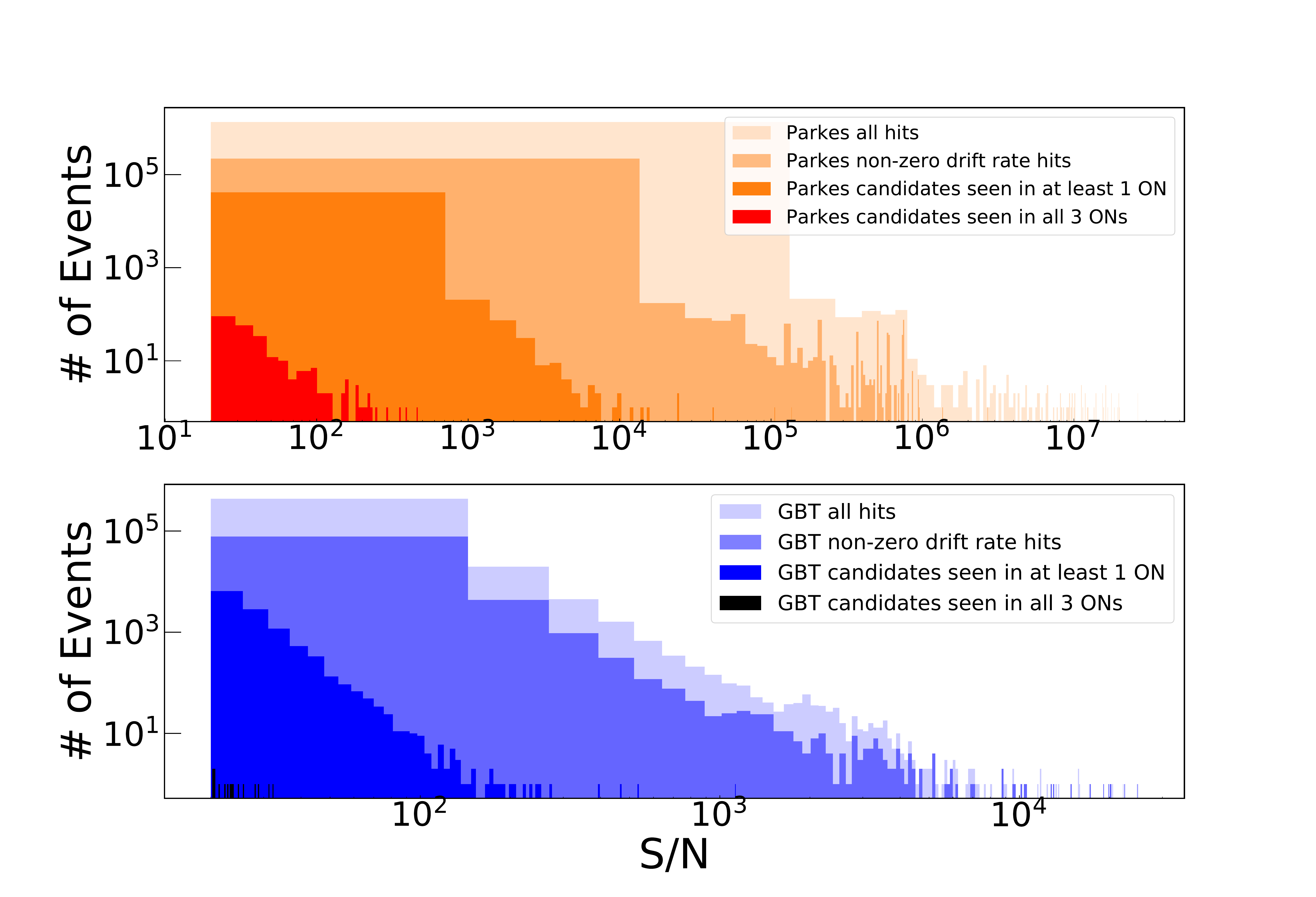}
  \label{fig:SNR_Hist}}
  \caption{Distribution of `hits' obtained from the narrowband drifting signal searches conducted across 1 to 8\,GHz as a part of the BL-GC survey. {\itshape Left:} Histogram shows the distribution of frequencies for all events detected in the GBT C-band and Parkes \gls{uwl} data analyzed in this paper.  {\itshape Right:} Histogram of SNR for the same events. We label these as ``all hits", where we use 200 bins for frequency, 50 bins for drift rate, and 200 bins for SNR. Hits shown in lighter to darker orange are from Parkes while hits shown in lighter to darker blue represents hits from the GBT for various filter cuts. For the frequency distribution of hits, the number of bins and their widths chosen are kept constant across Parkes and GBT data. However, because we only place a minimum threshold for SNR of 20, and there is no maximum limit, the width of the bins for SNR histogram is not restrained by a set maximum bound, but rather by the maximum value found within each filter. The final filter 3 hits are depicted by red bins for Parkes and black bins for GBT.   
}
\label{fig:freq_hist_SNR_hist_comb}
\end{figure}

\begin{figure}[h]
  \centering
  \subfloat[]{
  \includegraphics[width=0.51\textwidth]{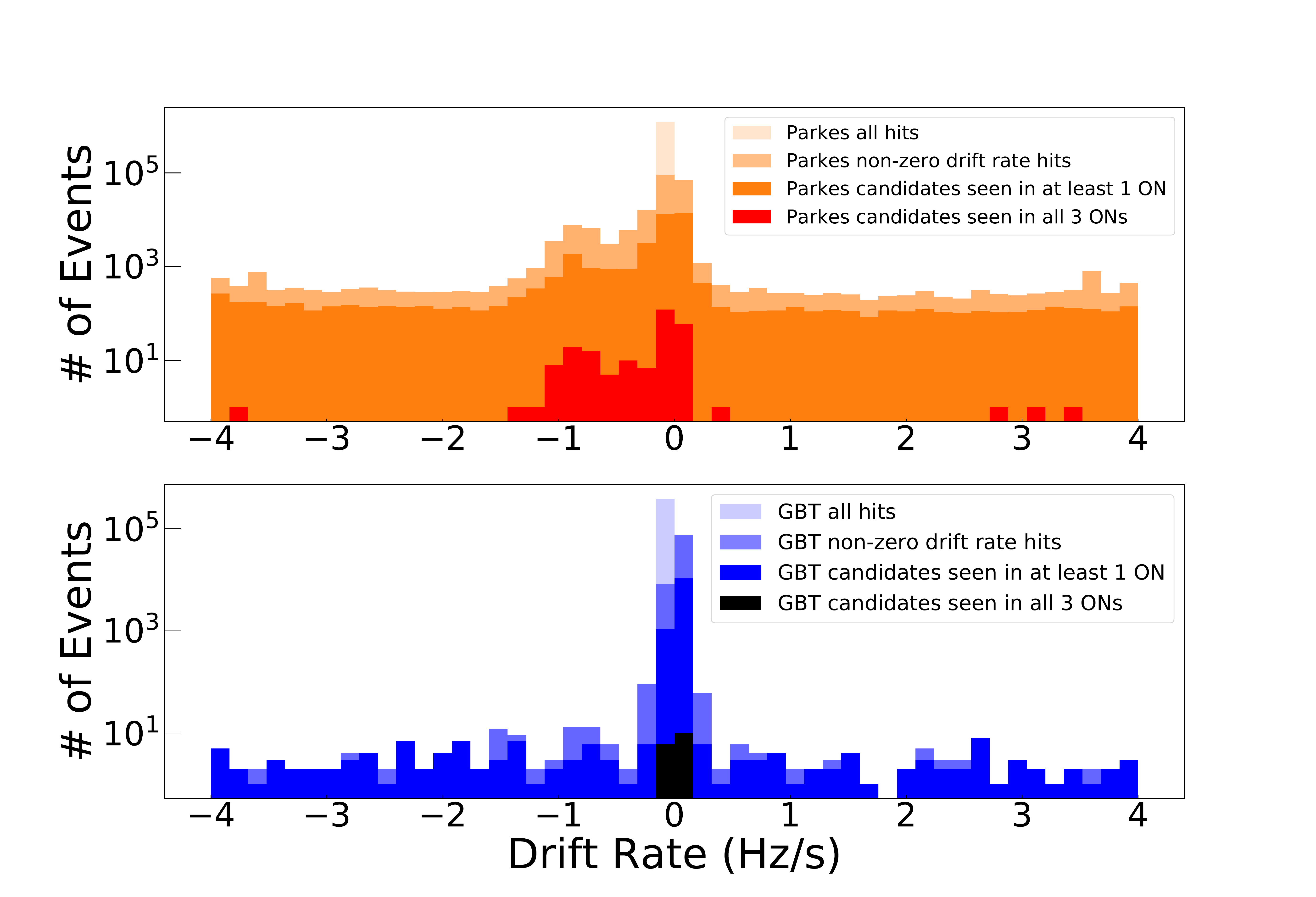}
  \label{fig:Drift_Hist}} 
  \hspace*{-0.75cm}
  \subfloat[]{
  \includegraphics[width=0.51\textwidth]{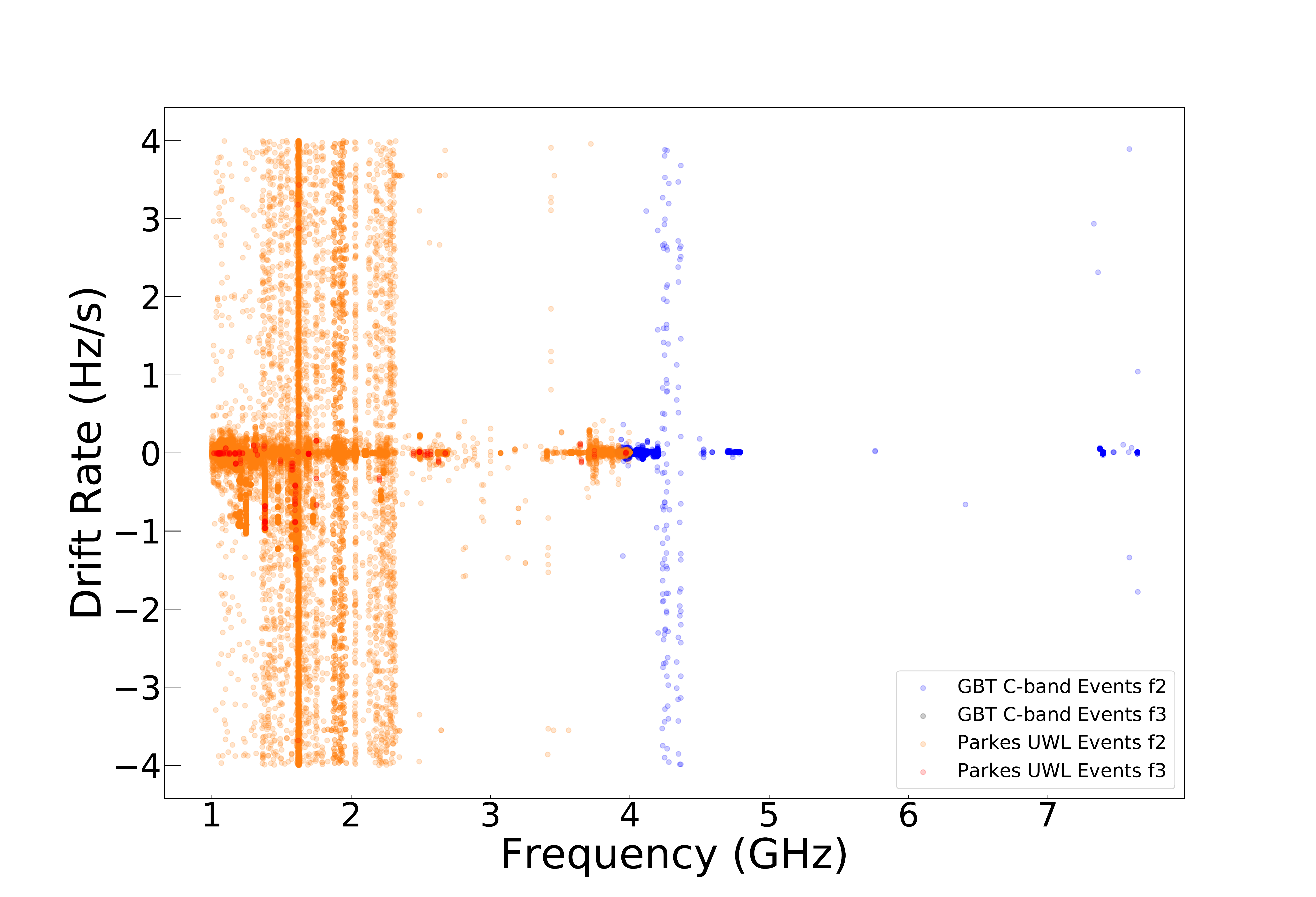}
  \label{fig:Drift_rate_Dist_scatter}}
  \caption{Distribution of `hits' obtained from the narrowband drifting signal searches conducted across 1 to 8 GHz as a part of the BL-GC survey.  {\itshape Left:} Histogram shows the distribution of drift rates for all events detected in the GBT C-band and Parkes \gls{uwl} data analyzed in this paper. {\itshape Right:} The scatter plot shows the distribution of drift rates across frequencies for the same events. The color labeling is similar to Figure \ref{fig:freq_hist_SNR_hist_comb}.}
  \label{fig:Drift_hist_Drift_scatter_comb}
\end{figure}

Using the search parameters listed in Table \ref{tab:turboSETI_parameter}, we obtained a list of hits for each ON-OFF pair, as defined in Table \ref{tab:obs_gbt_pks_coord}, where each pair consists of a total of 6 observations -- 3 ON and 3 OFF. We defined a ``hit" as a single strong narrowband signal in an observation which is a fine-channel above the set threshold. We found a total of 1,354,519 and 463,068 raw hits from Parkes and GBT observations, respectively. This hit density is relatively lower compared to the hit density obtained at L-band from \cite{price2020} as C-band is relatively cleaner from interference and we have used a relatively higher threshold of 20-sigma. The distribution of these hits in frequency, SNR, and drift-rate are shown in Figures \ref{fig:Freq_Hist}, \ref{fig:SNR_Hist}, and \ref{fig:Drift_Hist}, respectively. Due to relatively lower observing frequencies with Parkes, we observed around double the number of total candidates with roughly similar length of observations. We removed all the hits with a zero drift rate to eliminate as much \gls{rfi} as possible, since any narrowband signals deliberately transmitted by ETIs would likely have some Doppler drift due to the relative velocity differences between the transmitter and the receiver. All the remaining hits, a total of 304,658 candidates combined from Parkes and GBT, are shown in Figures \ref{fig:freq_hist_SNR_hist_comb} and \ref{fig:Drift_hist_Drift_scatter_comb} as non-zero drift rate hits. 

After collating hits from the three ON-OFF pairs, we remove any hits for which at least one of the OFF observations has a hit in the range 
\begin{equation}
  \nu_{bound, OFF} = \nu \pm |\dot{\nu}| \times 2(\Delta{T}).   
\end{equation}
Here, $\nu$ is the detected frequency of the hit, $\dot{\nu}$ is the maximum drift-rate, and $\Delta{T}$ is the length of observations. Any remaining hits were considered an event, where we defined an ``event" as a strong narrowband signal that was associated with a hit in at least one ON-observation and no OFF-observations. We defined these as ``filter 2'' events, which totaled to 53,702 candidates combined from Parkes and GBT, and they are shown in Figures \ref{fig:freq_hist_SNR_hist_comb} and \ref{fig:Drift_hist_Drift_scatter_comb}. Finally, we filtered out hits by removing all those that did not appear in all three ON-observations, and appeared in any OFF-observation. We defined them as ``filter 3'' events which totaled to only 249 candidates. These are also shown in Figures \ref{fig:freq_hist_SNR_hist_comb} and \ref{fig:Drift_hist_Drift_scatter_comb}. 

\begin{figure}[!h]
\centering
    \subfloat[]{
        \includegraphics[width=0.3\linewidth]{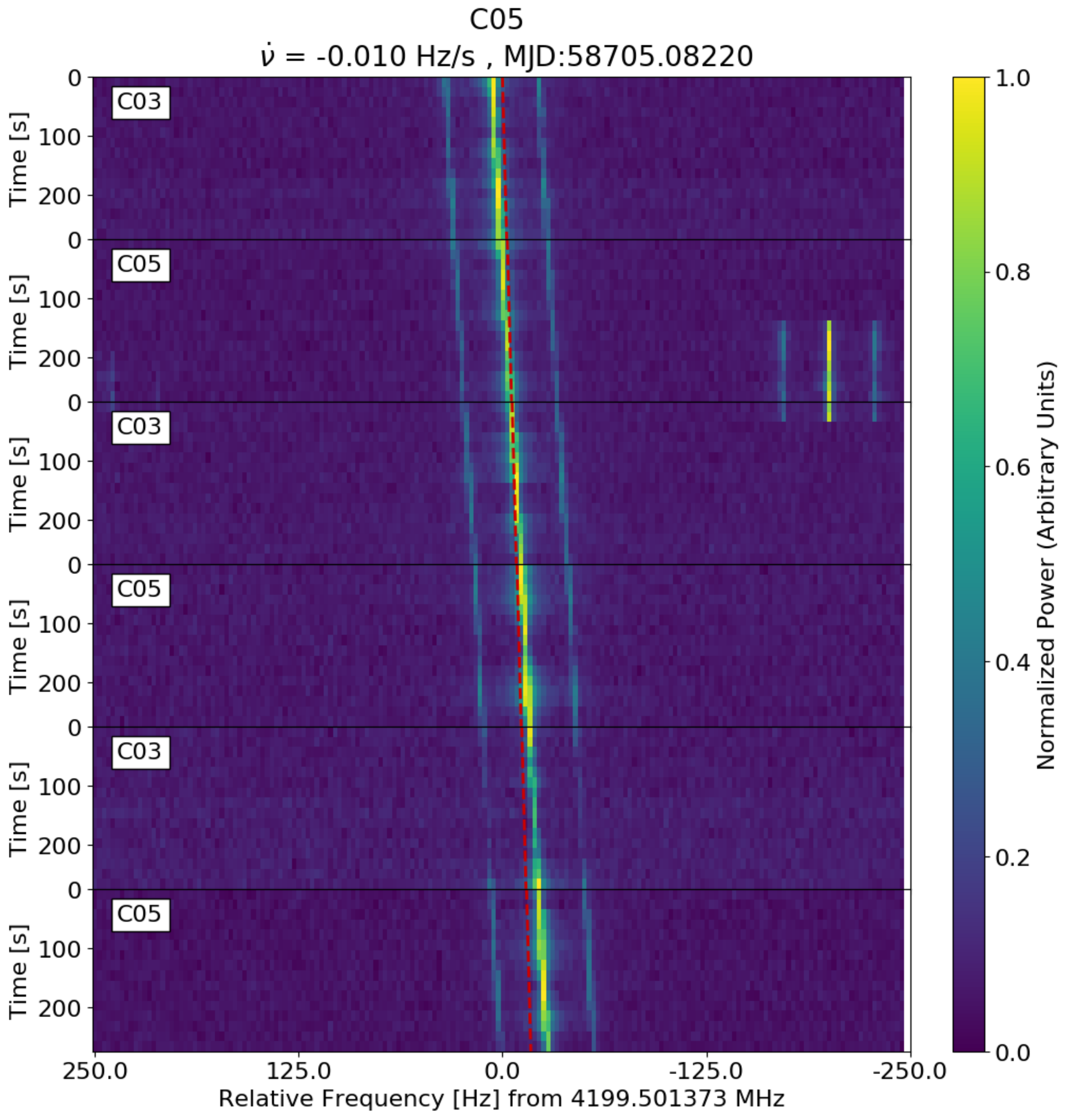}
        \label{fig:cand_1}
    }
    \subfloat[]{
        \includegraphics[width=.3\linewidth]{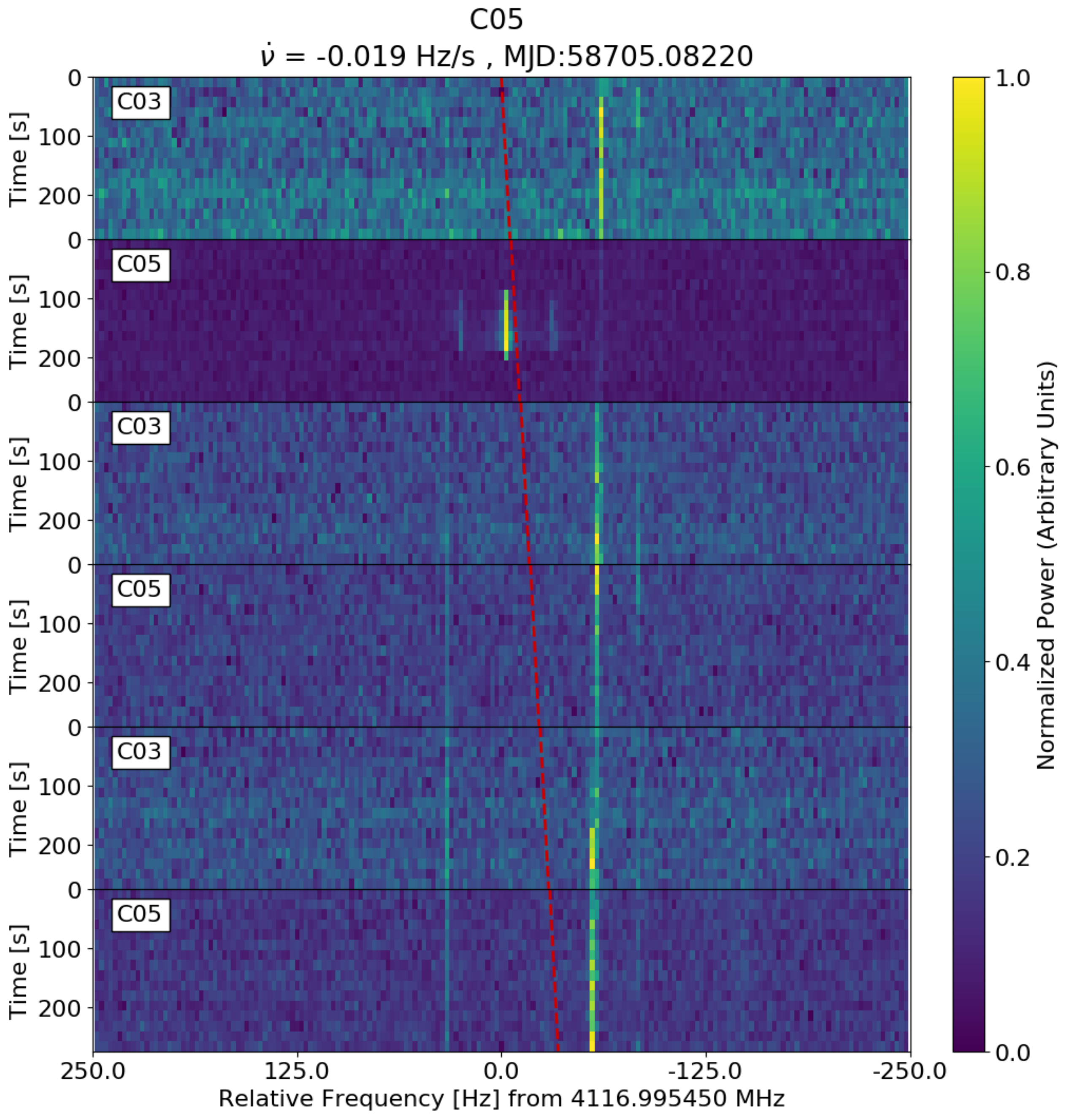}
        \label{fig:cand_2}
    }
    \subfloat[]{
        \includegraphics[width=.3\linewidth]{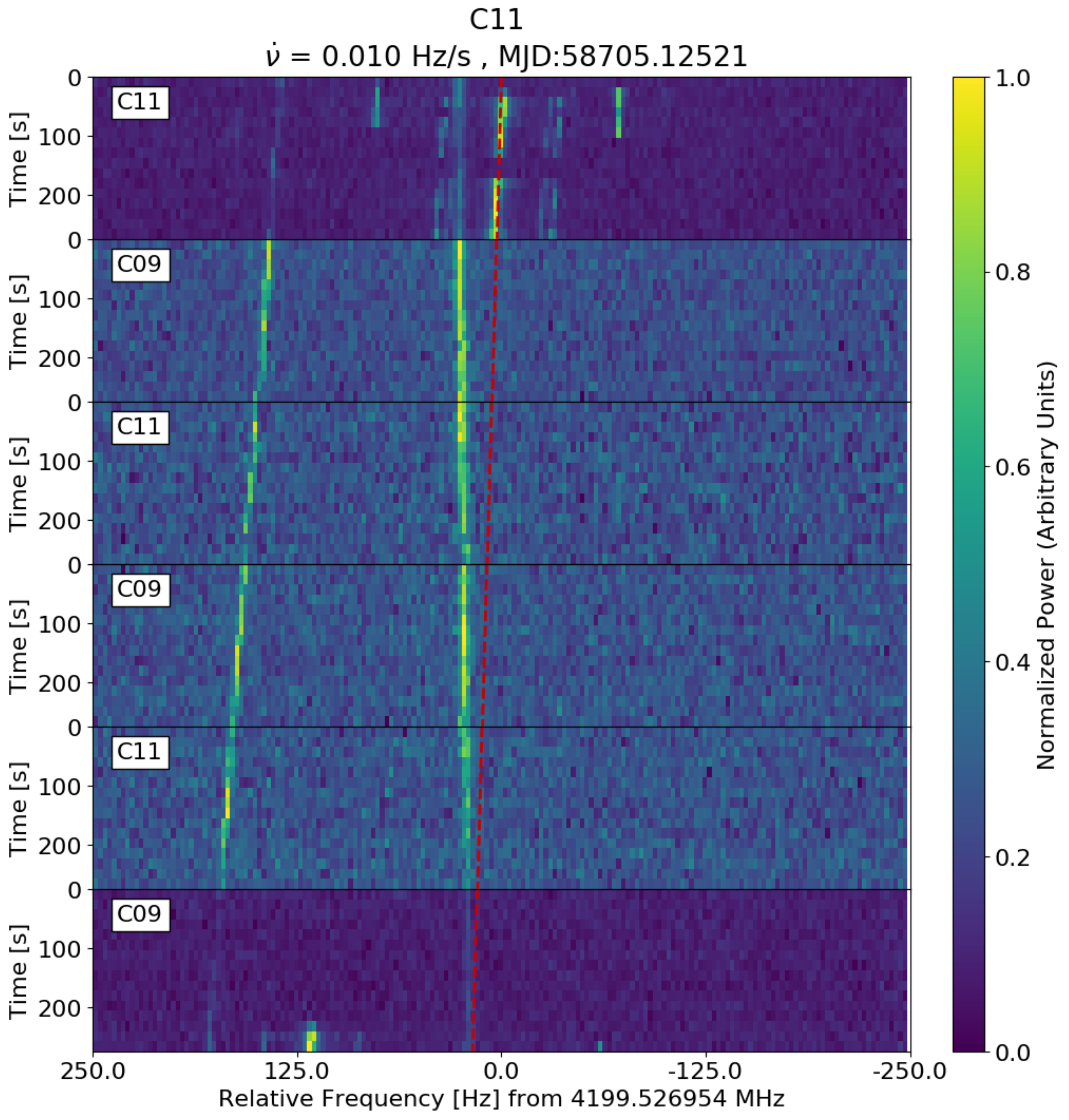}
        \label{fig:cand_3}
    }
    \hspace{0mm}
    \subfloat[]{
        \includegraphics[width=.3\linewidth]{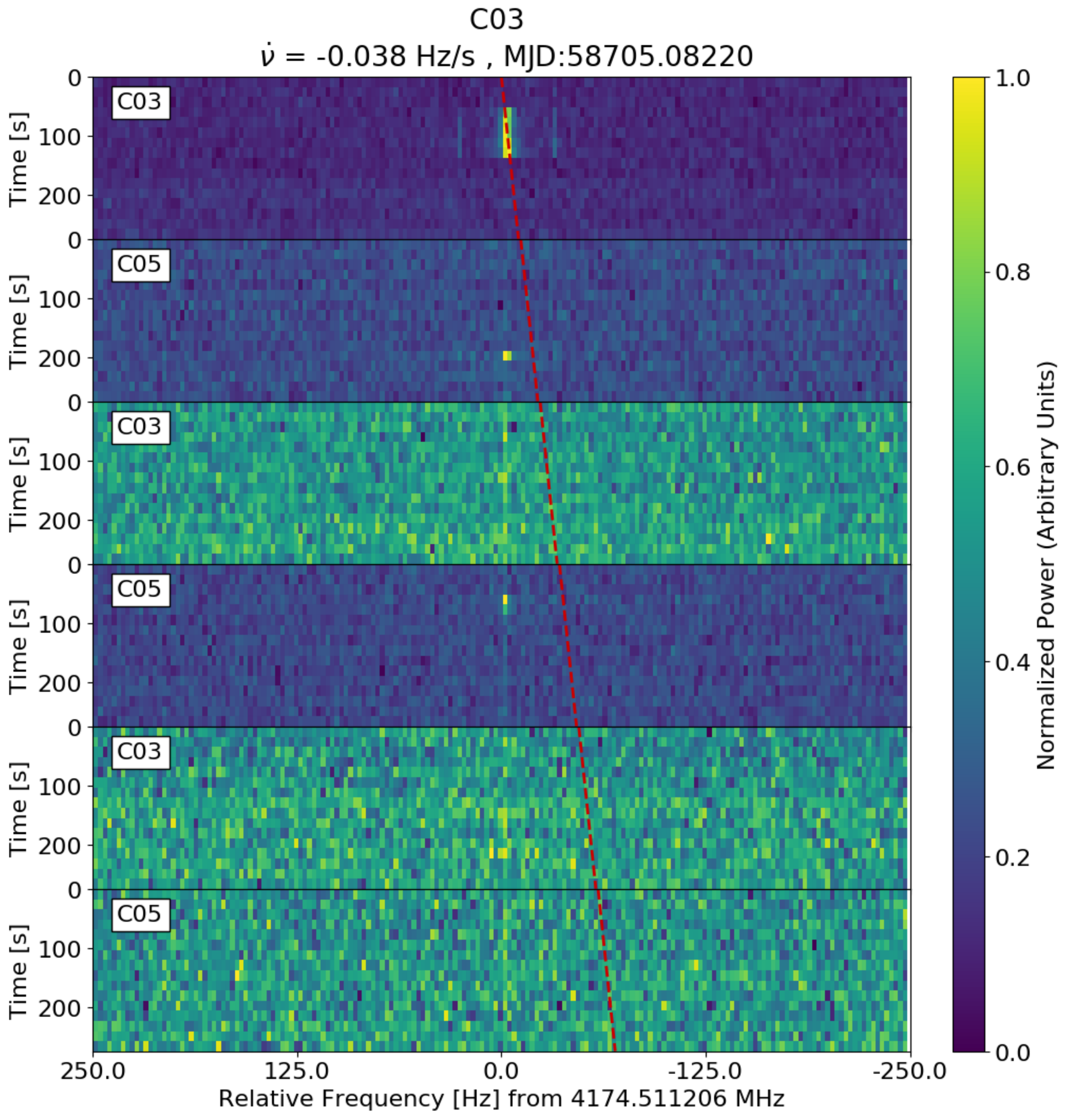}
        \label{fig:cand_4}
    }
    \subfloat[]{
        \includegraphics[width=0.3\linewidth]{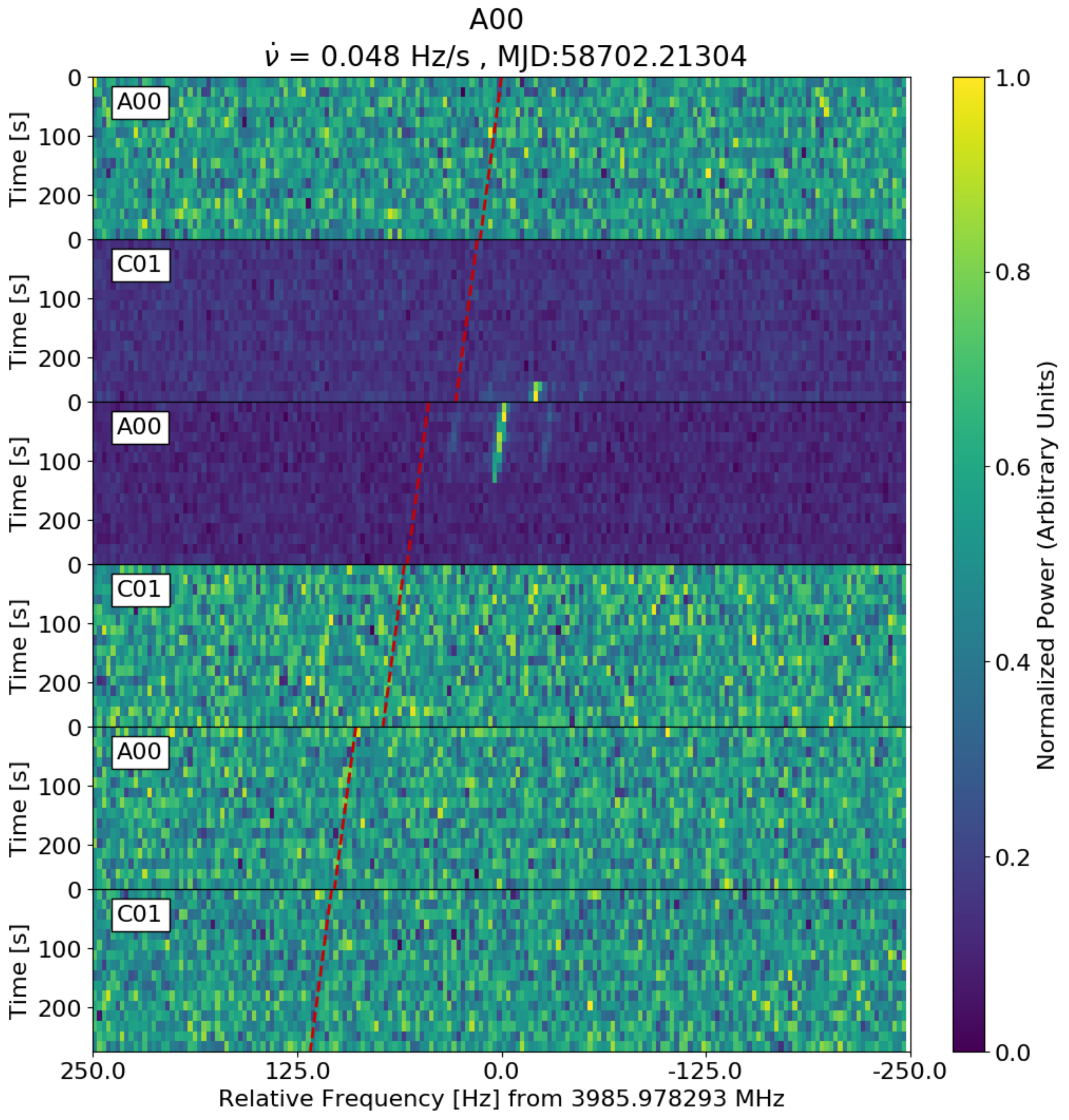}
        \label{fig:cand_5}
    }
     \subfloat[]{
        \includegraphics[width=.3\linewidth]{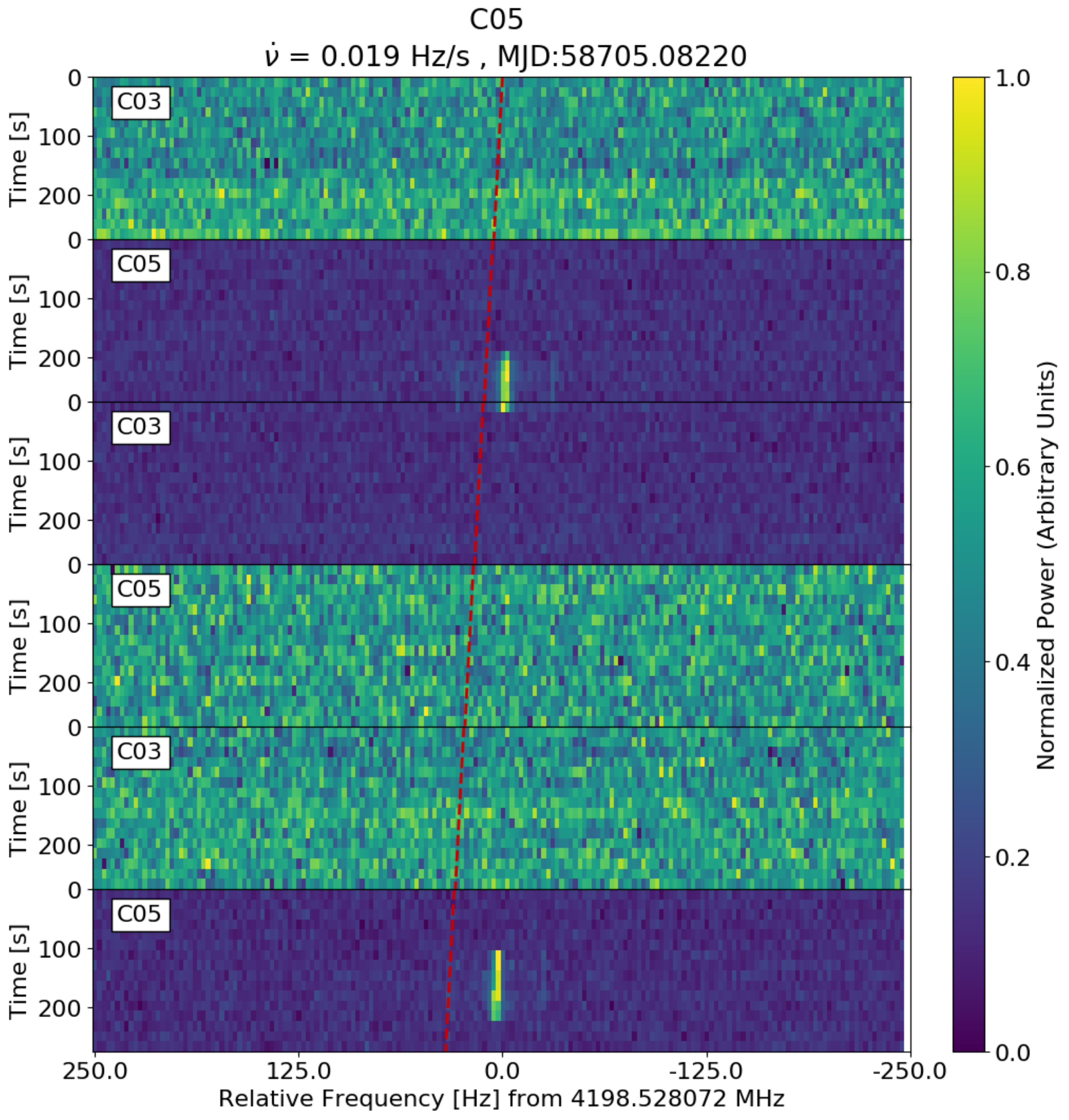}
        \label{fig:cand_6}
    }
    \hspace{0mm}
    \subfloat[]{
        \includegraphics[width=0.3\linewidth]{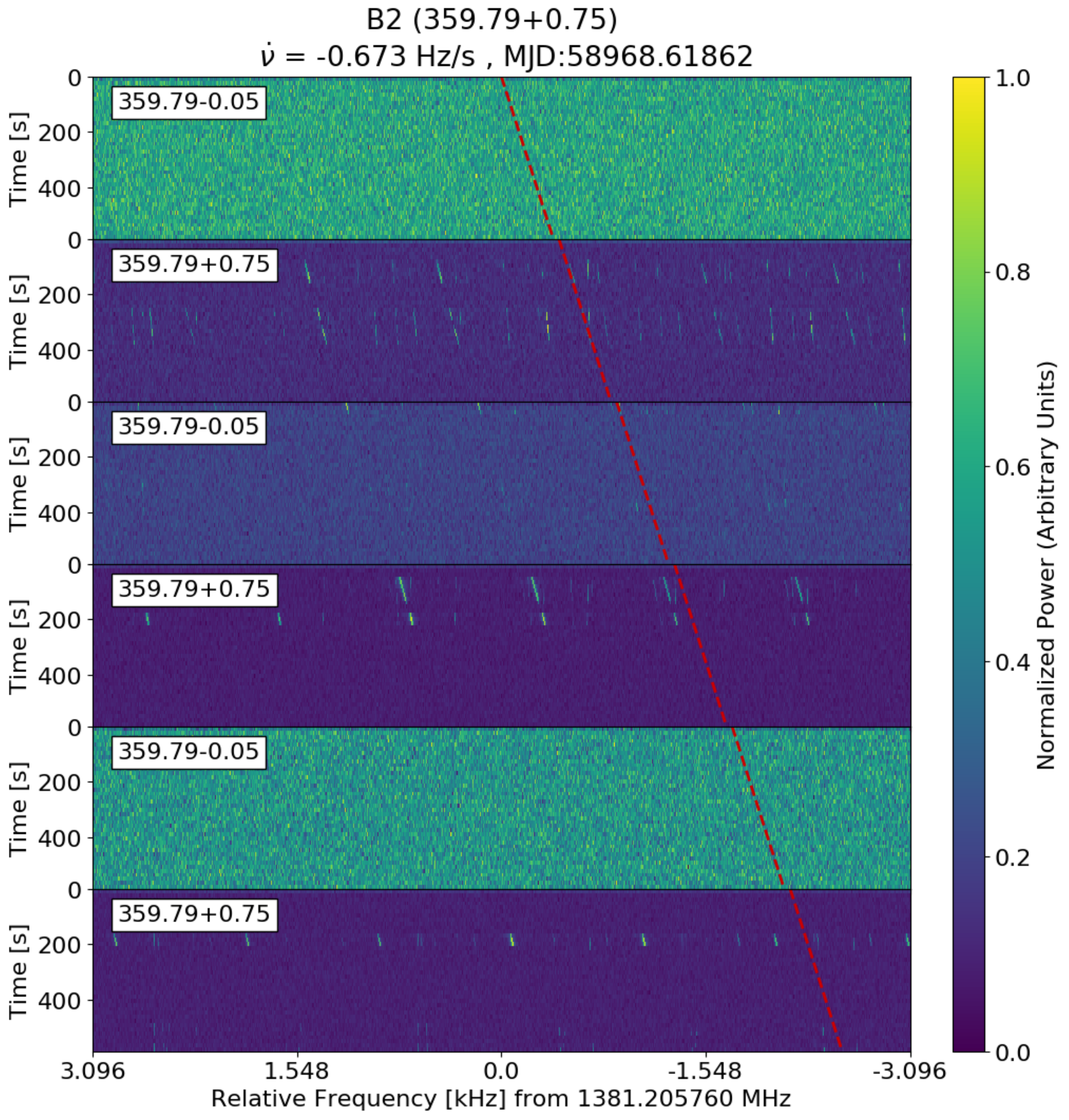}
        \label{fig:Navstar_sat}
    }
    \subfloat[]{
        \includegraphics[width=.3\linewidth]{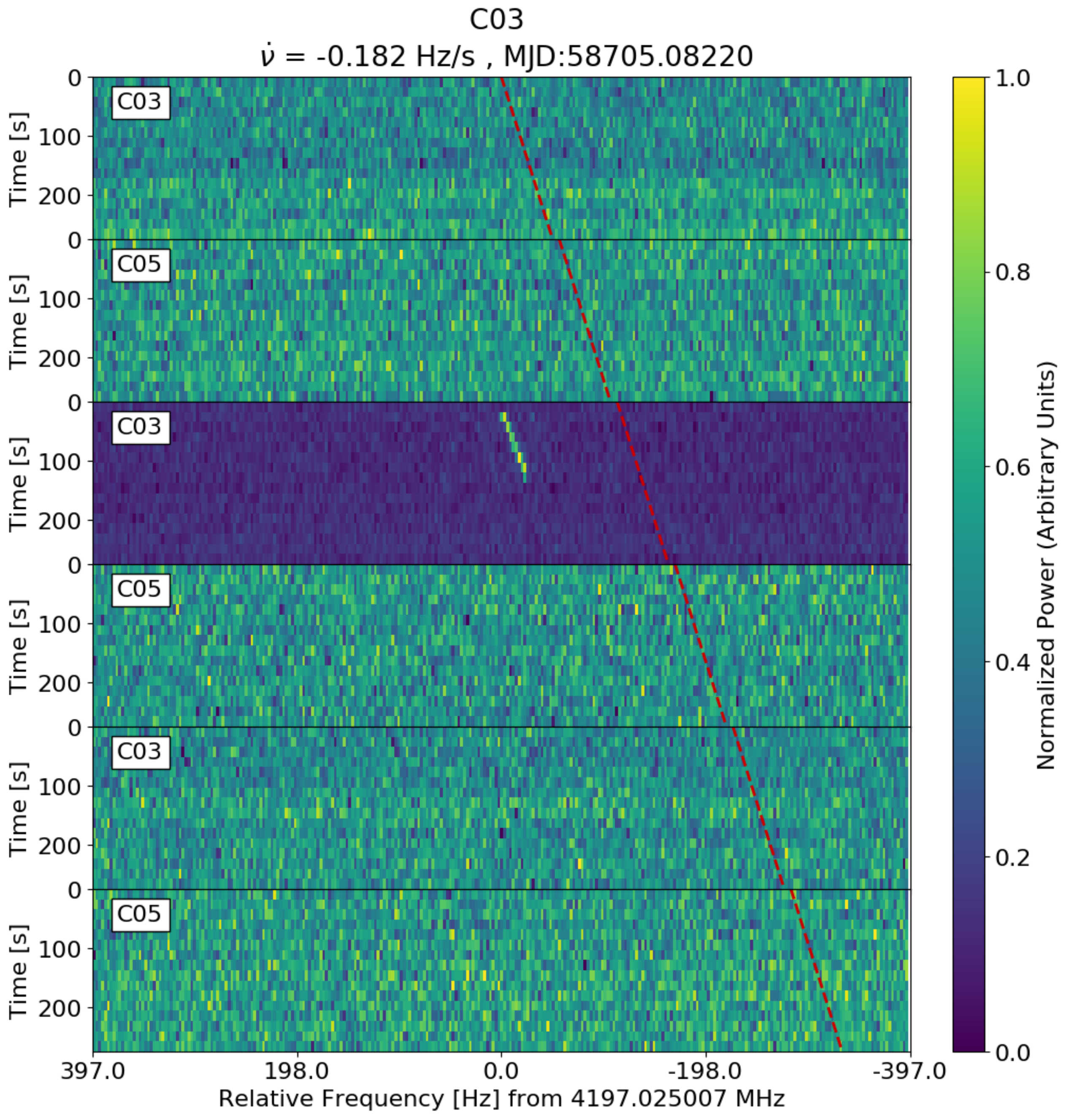}
        \label{fig:cand_7}
    }
    \subfloat[]{
        \includegraphics[width=.3\linewidth]{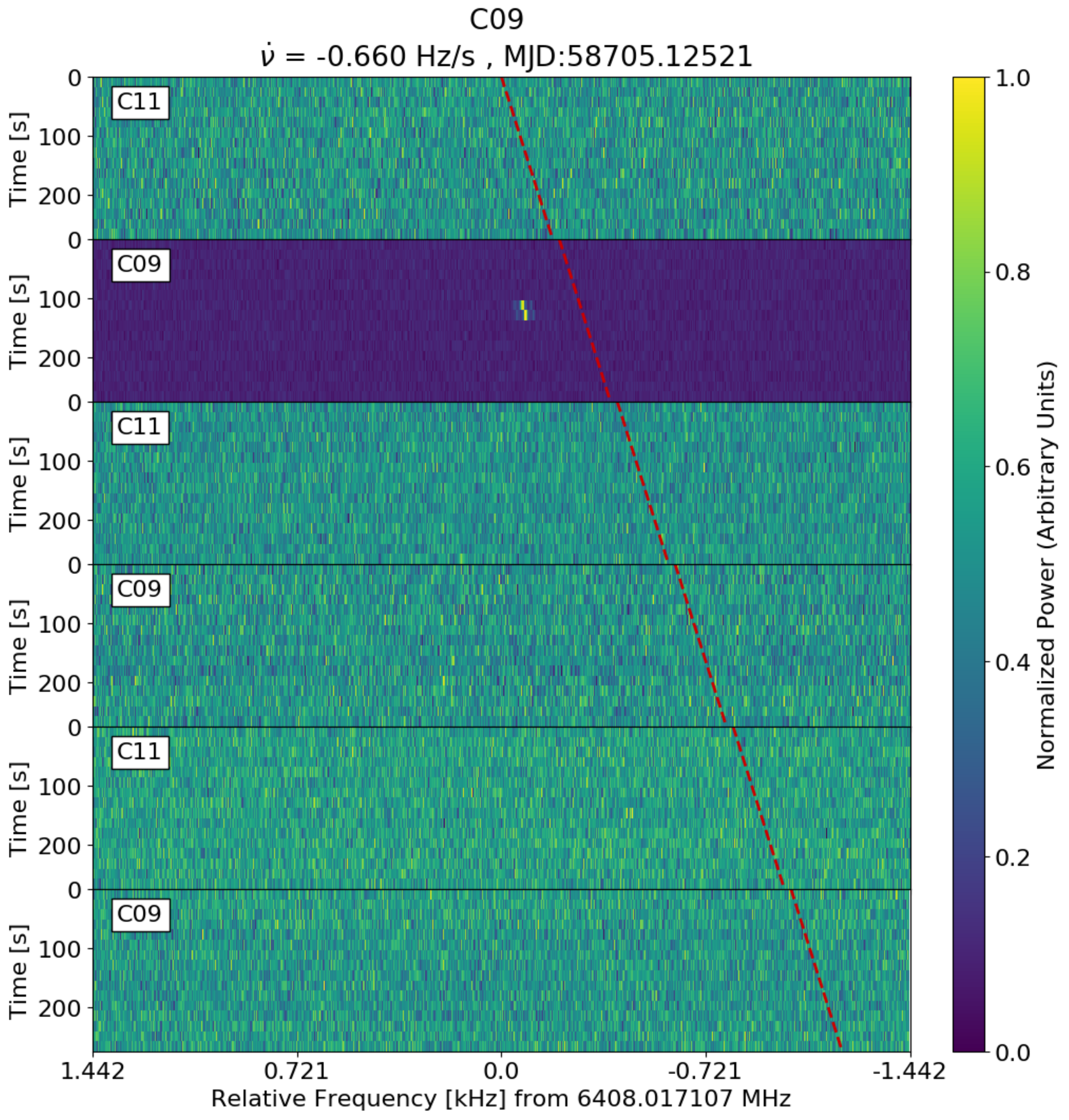}
        \label{fig:cand_8}
    }
    
    \vspace{0.5cm}
    \caption{Examples of narrowband drifting signals found from the BL GC survey conducted across 1 to 8\,GHz from Parkes and GBT. For each plot, dynamic spectra are shown for six observing ON-OFF pairs with the top panel showing the first scan in the pair. The field name for the associated candidate is shown at the top of each plot. The center frequency of each event and measured drift rate are also listed at the bottom and top of each plot, respectively. The overlaid red dashed line shows the detected drift rate obtained from the first ON observation in which the signal first appears with a slight offset in frequency for visualization; however, longer observations are needed for an exact drift rate. Apart from (g), all the examples are for candidates found from the GBT C-band observations.} 
    \label{fig:turboSETI_cands}
\end{figure}

We plotted all of our event candidates for both filter 2 and 3 which totaled to around $5.4 \times 10^4$ candidates, and vetted them by visual inspection. Any strong narrowband signals would appear in all 3 of the ON observations, but none of the OFF observations. Table \ref{tab:obs_gbt_pks_coord} shows the number of event candidates found for each ON-source within each pair for both filter 2 and 3. We ran this process for each pair of observations: 14 Parkes \gls{uwl} pointings and 19 GBT C-band pointings. Because the OFF-observation for each ON-observation is strategically placed inside the range of all our pointings, we are able to use this to our advantage by easily making double usage of each observation pair. For example, for the pair B01-B04 in C-band, we can run turboSETI twice and look for narrowband drifting signals in two different areas: once using B01 as the ON-source, and a second time using B04 as the ON-source. 

It is important to note that all of our pairs have a total of 6 observations; however, due to a split in our observation session for the GBT data, we only use 4 observations total (or 2 per pointing) for the pairs C07-A00 and C10-C12. Thus, an excess in the number of candidates for these pointings can be seen in Table \ref{tab:obs_gbt_pks_coord}. The rest of the pointings with high numbers of events are attributed to the large amount of \gls{rfi} found in our data, as can be seen in \ref{fig:Drift_rate_Dist_scatter} where many events were found to have drift rates close to zero. We did come across a few interesting Doppler-drifting candidates within this analysis; however, most of them can be attributed to \gls{rfi} since they appear in at least one of the OFF observations. Some examples of such interesting candidates are shown in Figure \ref{fig:turboSETI_cands}. Further details on these candidates are discussed in Section \ref{sect:discussion_candidates}. 

\subsection{Transient search}
\label{sect:broad_seti}

\begin{table}[hb]
\caption{Transient search parameters for the \gls{bl} \gls{gc} survey. Here, DM ranges refer to both natural and three types of artificial DMs.}
    \centering
    \begin{tabular}{c|c}
         \hline
            \hline
            Parameter  & Range  \\
            \hline
            DM (pc-cm$^{-3}$)  & -5000 to +5000 \\
            Maximum sensitivity loss due to DM steps & $\leq$15\% \\ 
            {S/N}$_{\rm min}$ threshold for pDM & 6 \\
            {S/N}$_{\rm min}$ threshold for aDM & 10 \\
            Width for pDM & 0.38 to 97 ms - with steps 0.38$\times$2$^{n}$; with n$\in$[0,8] \\
            Width for aDM & 0.7 to 194 ms - with steps 0.7$\times$2$^{n}$; with n$\in$[0,8] \\
            \hline
    \end{tabular}
     \label{tab:spandak_parameter}
\end{table}

\begin{figure}[h]
    \centering
    \includegraphics[scale=0.6]{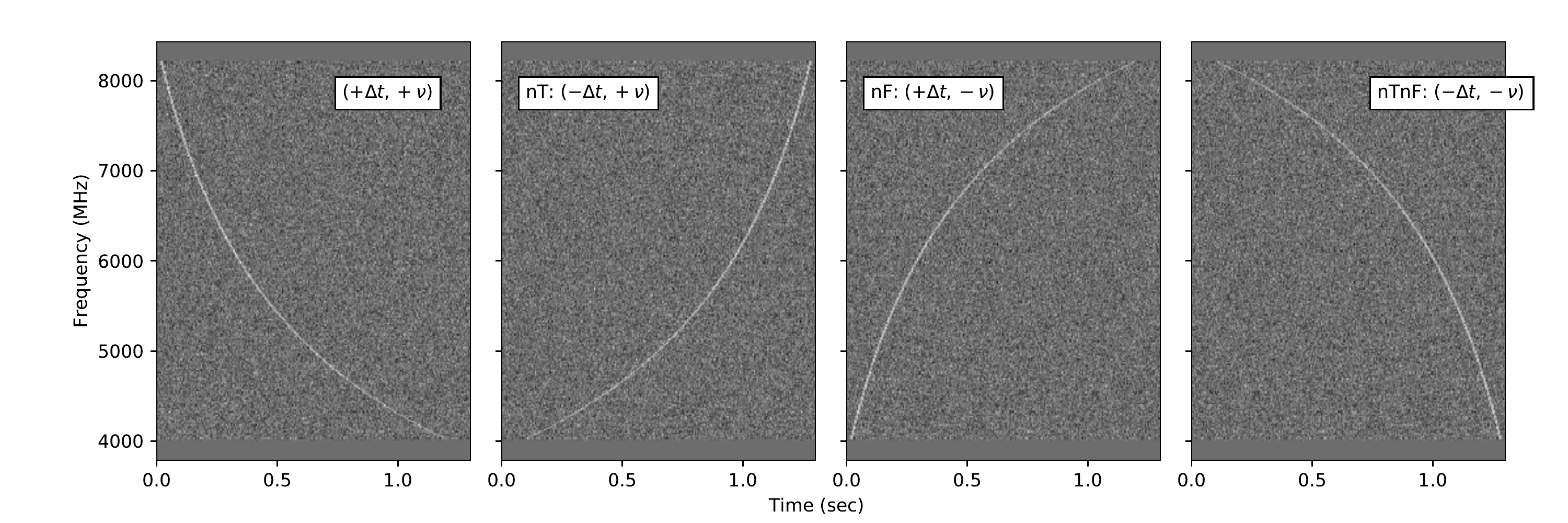}
    \caption{Four different types of transient signals we searched using the SPANDAK pipeline. Panels show dynamic spectra of dispersed pulses as function of time and frequency. The left most panel shows the 
    naturally dispersed pulse, occurring due to ionized interstellar medium. The other three panels are artificially-dispersed pulses which are not known to occur in nature and can be used by ETIs as beacons.}
    \label{fig:neg_DM}
\end{figure}

As suggested in Section \ref{sect:beacons}, ETI inhabiting the Galactic Center region might transmit powerful artificially-dispersed broadband pulses from an isotropic and/or a rotating beam transmitting all across the Galaxy as a beacon. This would require a similar amount of energy as the isotropic narrowband transmitter we searched for in the previous section. Any broadband signal passing through the interstellar medium experiences dispersion delay $\Delta{t}$ due to the free electrons, at the observed frequency of $\nu$ compared to the infinite frequency as, 
\begin{equation}
\Delta{t}(DM,\nu)~\propto~\frac{DM}{\nu^2}. 
\end{equation}
Here, we can express the arrival time ($\Delta{t}$) and corresponding observed frequency ($\nu$) in quadrants ($\pm\Delta{t},\pm\nu$), also shown in Figure \ref{fig:neg_DM}. It should be noted that the negative time and negative frequency only represent relative values for our given observation band in terms of respective channel and time sample indexes. For example, a signal with the noted arrival time of $\Delta{t}=-1$ is only in reference to arrival at $\Delta{t}=0$ for a signal without any (artificial) dispersion delay. Here, the quadrant 1 (i.e. $+\Delta{t}$ and $+\nu$) transient signals are the ones which we refer to as naturally dispersed pulses which are known to originate due to astrophysical transients such as FRB-like magnetars near the GC. An advanced communicating ETI civilization might choose to artificially disperse a broadband pulse which is not known to occur in nature. These other three classes of transient signals, located in the remaining three quadrants, are nonphysical and can be considered prime candidates for artificially-dispersed broadband pulses which are easily distinguishable from the standard natural-dispersed signals. Here, we refer to these artificially-dispersed signals as nT ($-\Delta{t}$, $+\nu$), nF ($+\Delta{t}$, $-\nu$), and nTnF ($-\Delta{t}$, $-\nu$). Such signals have been speculated as negative DM signals  \citep{Siemion:2010p6845,vonKorff:2010p3275,Harp:2018apj,li_gajjar_2020} but no comprehensive searches have ever been performed to the best of our knowledge. Thus, our analysis is going to be the first of its kind which targets such artificially-dispersed transient signals. 

In order to search these four types of broadband transient signals (shown in Figure \ref{fig:neg_DM}), we have deployed a state-of-the-art GPU-accelerated and ML-assisted broadband signal detection pipeline named SPANDAK. This pipeline is an extension of the pipeline used for the detection of -- so far the only highest radio frequency -- bursts from FRB \citep{gaj18apj}. More details about the SPANDAK pipeline will be discussed in future publications (Gajjar et al.\ 2021 in prep.). As mentioned in Section \ref{sect:data_products}, we only carried out these transient searches for our GBT data since scattering and dispersion losses at lower frequencies render such searches less sensitive. Some of the search parameters for our GBT C-band transient search are listed in Table \ref{tab:spandak_parameter}. For simplicity, natural DM candidates will be referred to as pDM (i.e. positive DM) candidates and three other types of transient signals -- nT, nF, and nTnF -- will collectively be referred to as artificial DM (i.e. aDM) candidates for the remainder of the discussion.

\begin{figure}[h]
    \centering
    \subfloat[]{
    \includegraphics[scale=0.3]{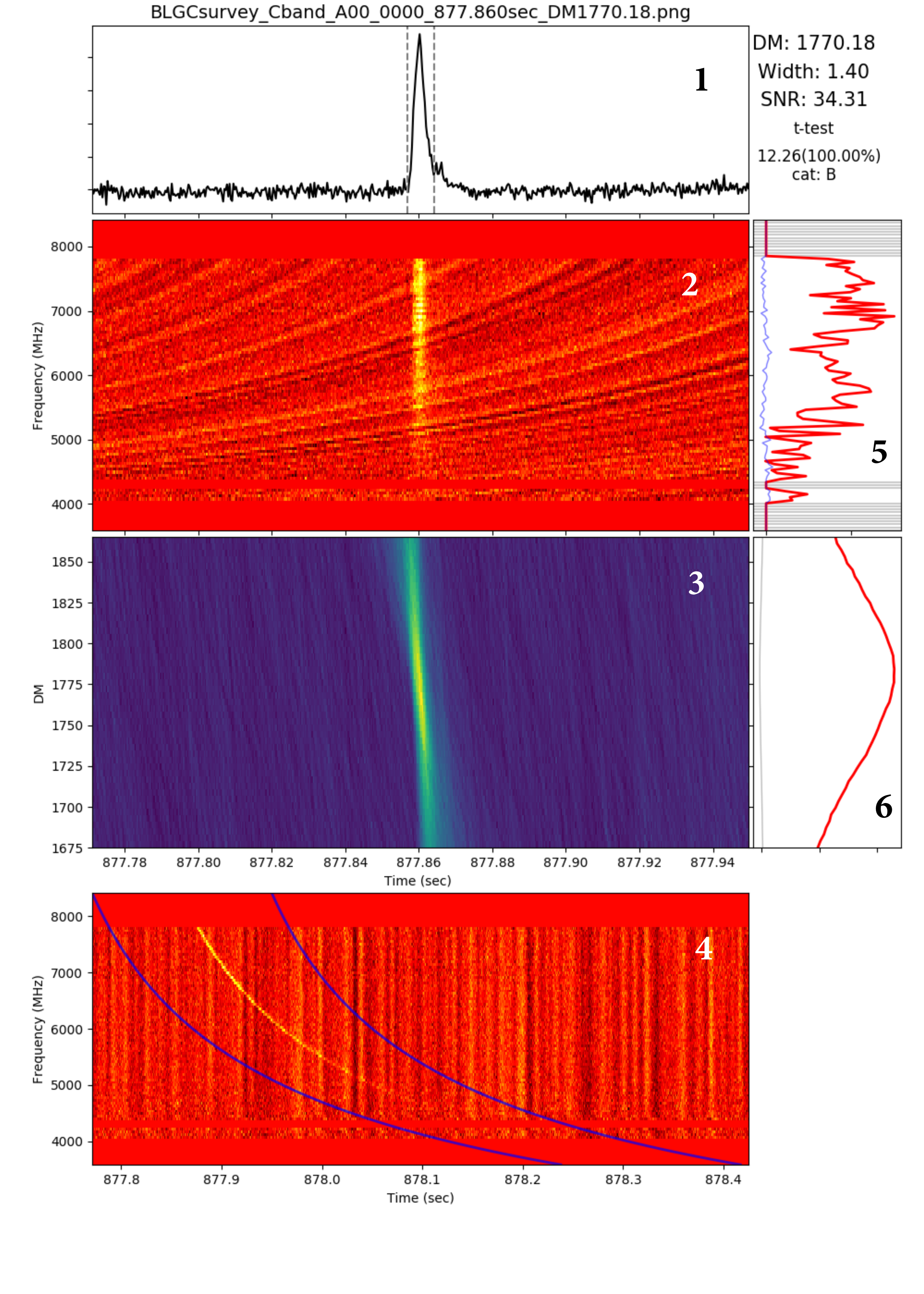}
    }
    \subfloat[]{
    \includegraphics[scale=0.3]{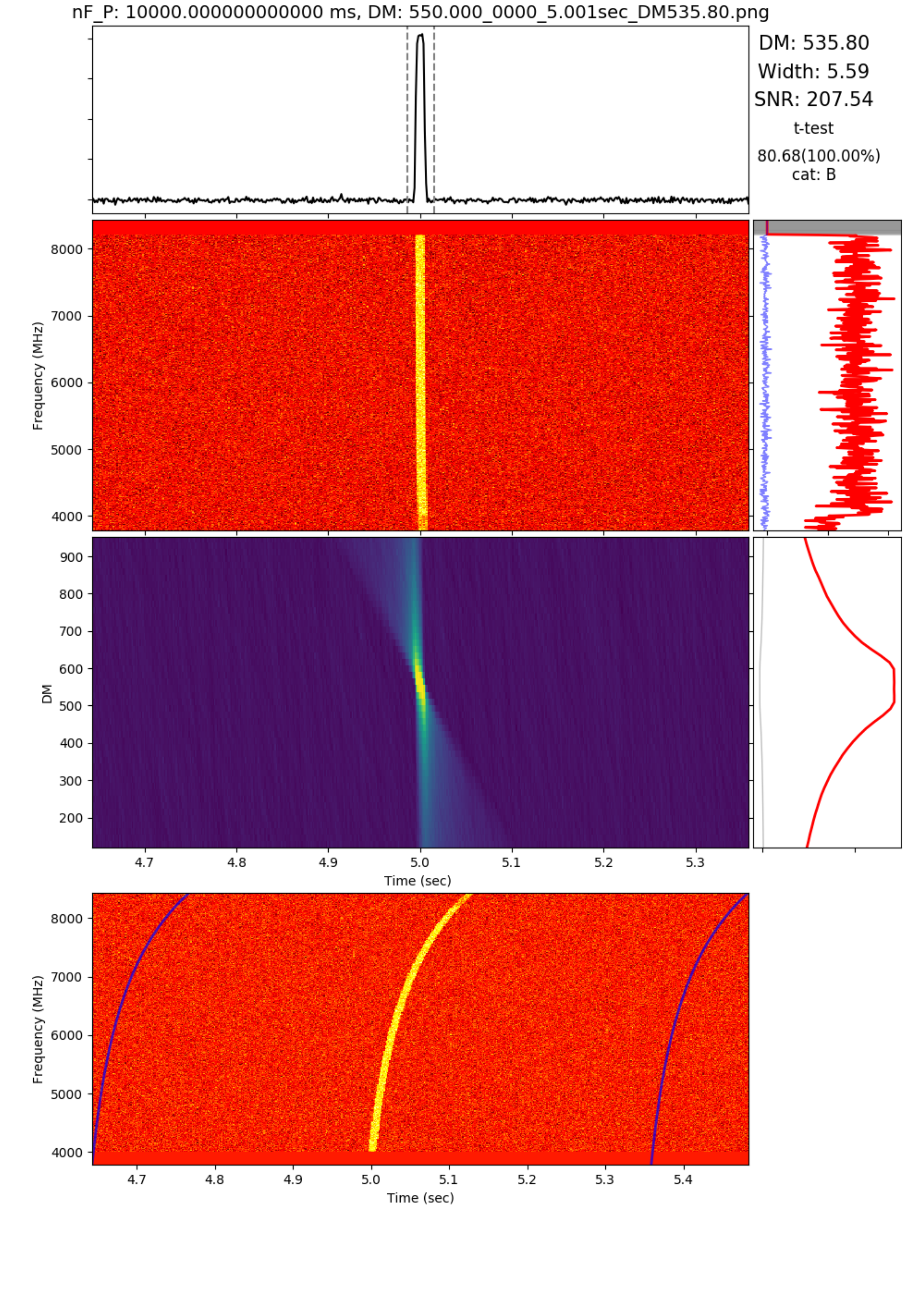}
    }
    \caption{Example plots from the SPANDAK pipeline. {\itshape Left:} A single pulse detected from SGR\,J1745--2900 from our observations of the GC$(0,0)$ across 3.9 to 8\,GHz from the GBT. Panel 1 shows a dedispersed single pulse with an on-pulse region marked by two dotted lines along with the necessary diagnostic information next to it showing DM (pc-cm$^{-3}$), S/N, and width of the detected candidate in milliseconds. Panel 2 shows dedispersed dynamic spectra showing a broadband pulse across 3.9 to 8\,GHz with on-pulse and off-pulse spectra shown in red and gray lines in panel 5, respectively. Panel 3 shows the DM-vs-time plot, with on-pulse DM-vs-SNR shown in panel 6. 
    Panel 4 shows the original data with the detected dispersed pulse along with two blue lines for visual guidance. Panels 1, 2, and 3 share the same time-axis. The color in all three panels are normalized intensities in arbitrary units. 
    {\itshape Right:} A simulated pulse detected from the SPANDAK pipeline showing nF-aDM pulse.}
    \label{fig:spandak_example_plot}
\end{figure}

\begin{figure}[h!]
    \centering
    \subfloat[]{
    \includegraphics[scale=0.5]{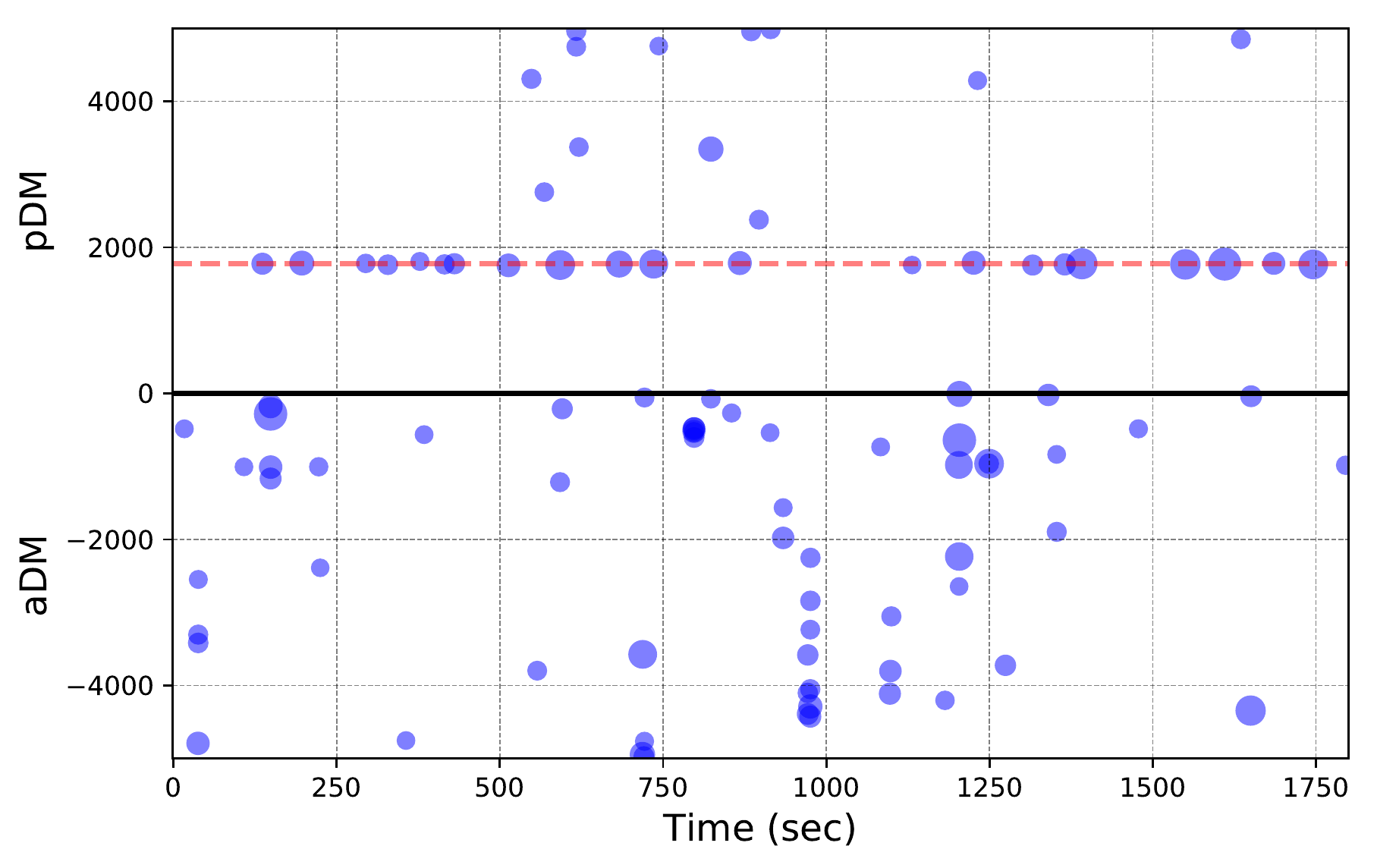}
    \label{fig:DM_vs_Time}
    }
    \subfloat[]{
    \includegraphics[scale=0.5]{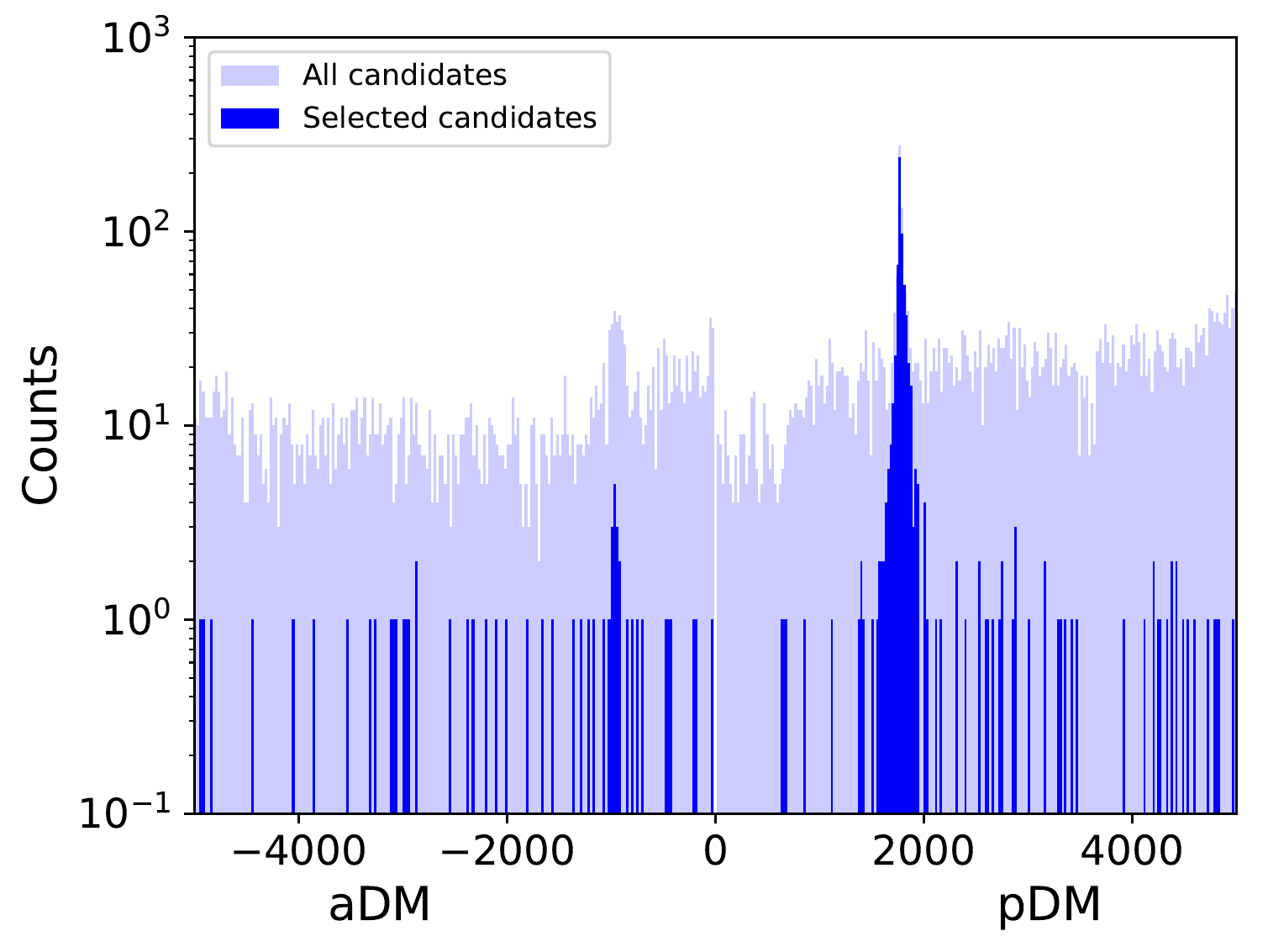}
    \label{fig:DM_hist}
    }
    \caption{Distribution of pDM (pc-cm$^{-3}$) and aDM (pc-cm$^{-3}$) candidates from the SPANDAK pipeline from the GBT C-band observations. {\itshape Left:} Distribution of pDM and aDM candidates from one of the 30-min long scans towards the A00 field. The size of the point represents the relative S/N. The red dotted line indicates the DM of 1778\,pc\,cm$^{-3}$. {\itshape Right:} Histogram of all the detected candidates as a function of pDM and aDM from all fields. A clear peak near the DM of 1778\,pc\,cm$^{-3}$ is apparent due to large number of single pulses detected from SGR\,J1745--2900. For clarity, only candidates with S/N$>$10 were used for both these plots.
    }
    \label{fig:all_spandak_cand}
\end{figure}

\subsubsection{Positive DM signal searches}
\begin{figure}[h]
    \centering
    \subfloat[]{
    \includegraphics[scale=0.25]{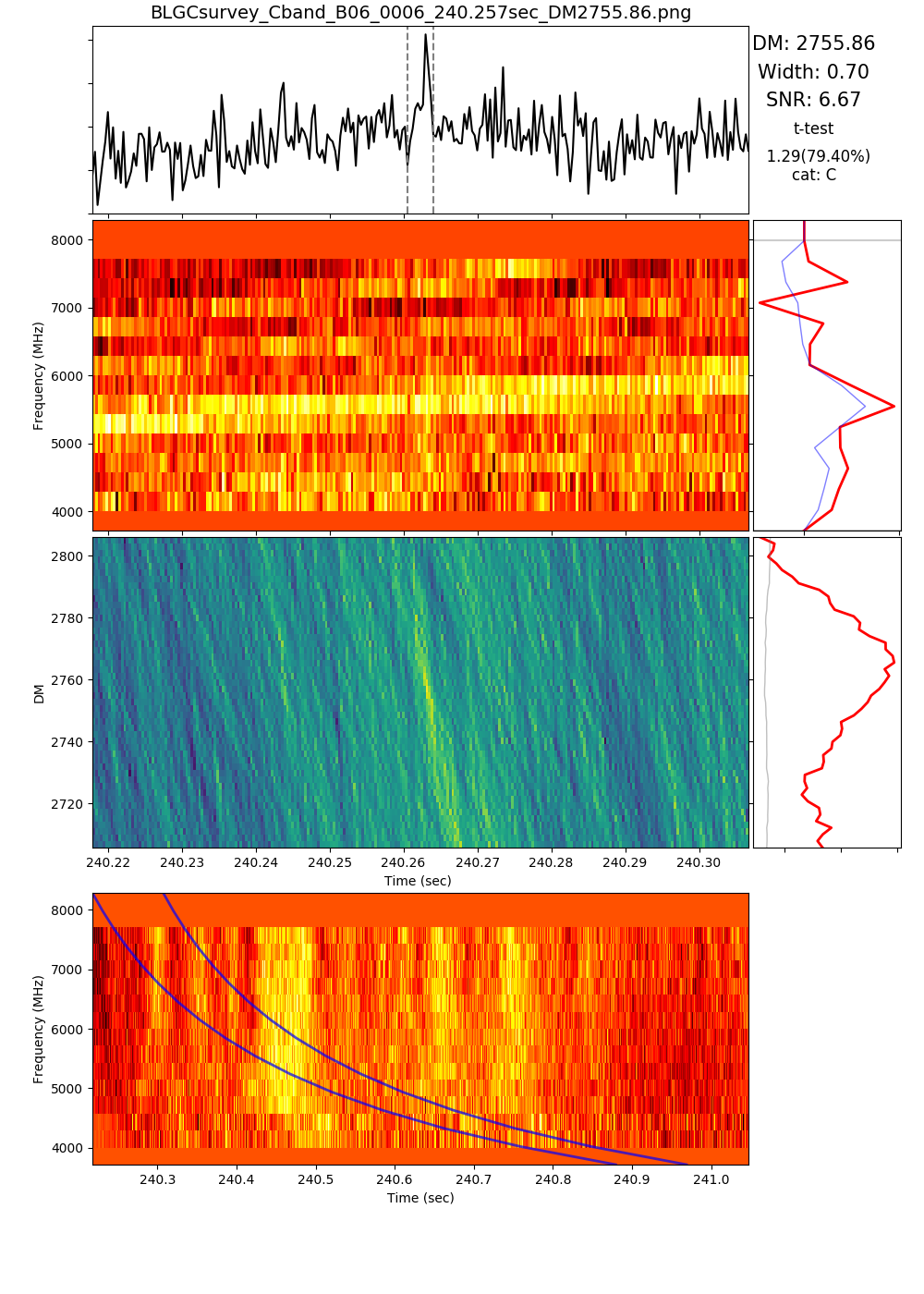}
    \label{fig:pDM_cand}
    }
    \subfloat[]{
    \includegraphics[scale=0.25]{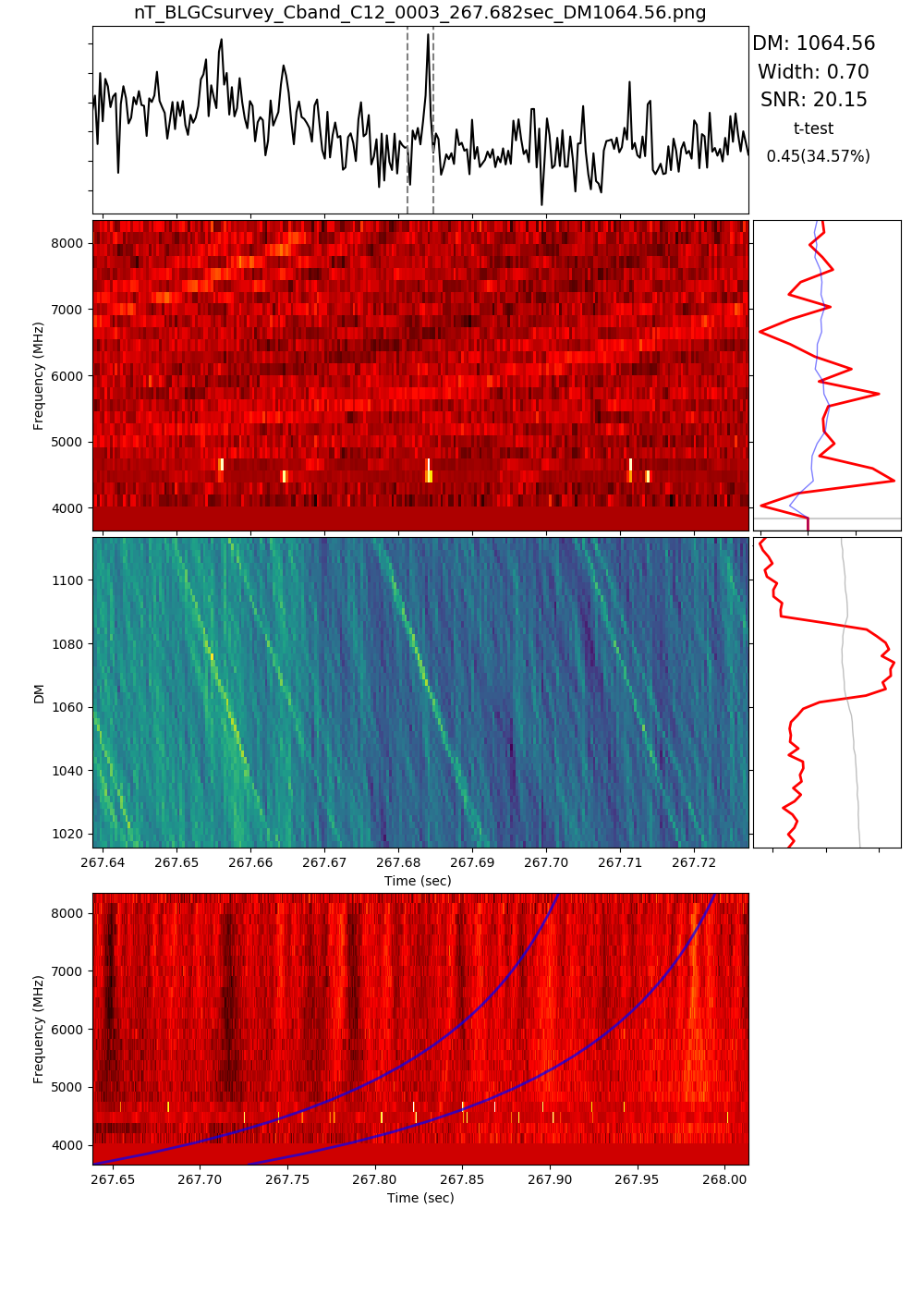}
    \label{fig:aDM_cand1}
    }
    \subfloat[]{
    \includegraphics[scale=0.25]{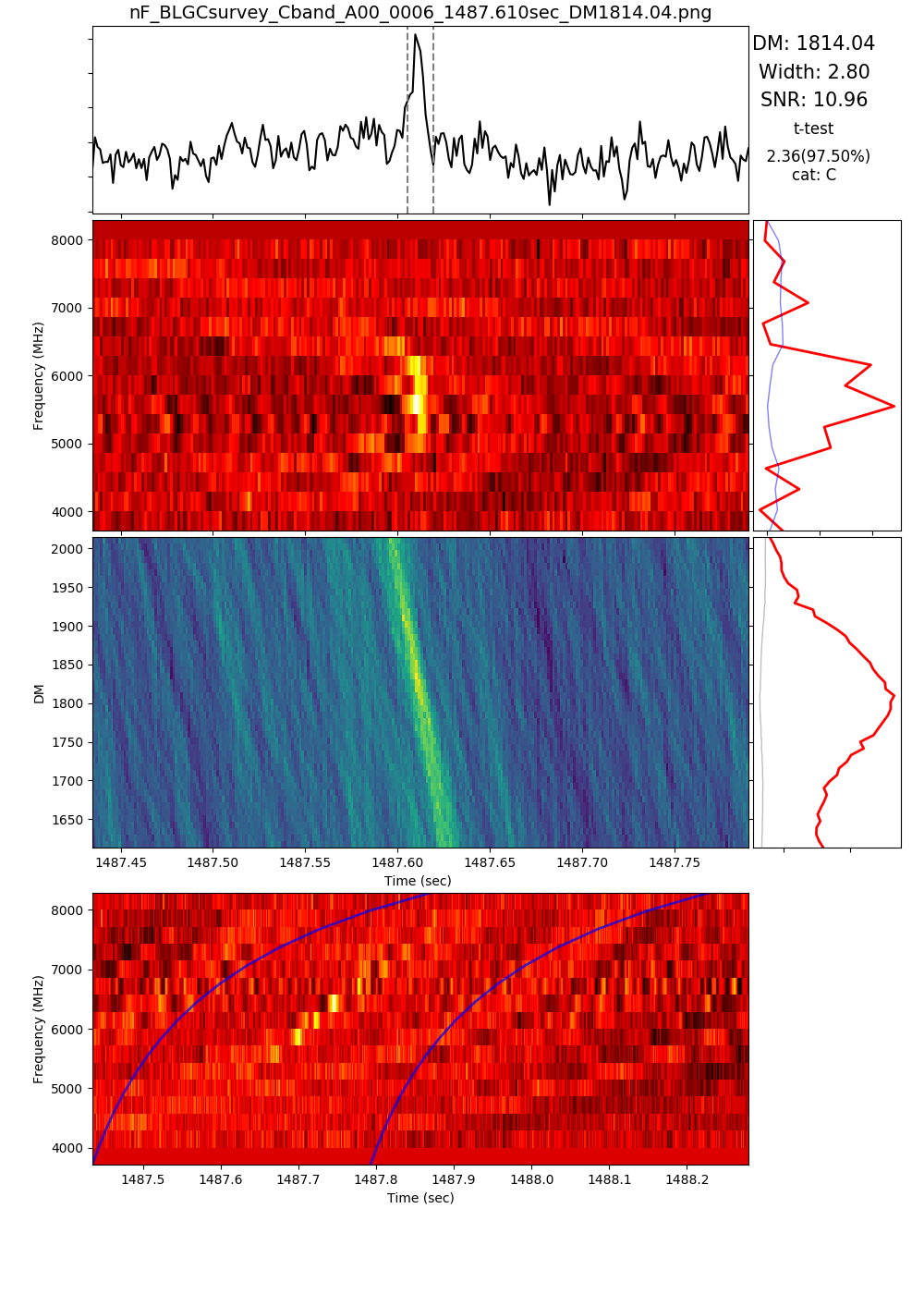}
    \label{fig:aDM_cand2}
    }
    \caption{Tentative pDM and aDM candidates from the GC survey conducted across 3.9 -- 8\,GHz from the GBT found using the SPANDAK pipeline. {\itshape Left:} pDM candidate found towards B06 field at the DM of 2755\,pc\,cm$^{-3}$. After careful examination, we were unable to verify the validity of this candidate. {\itshape Center:} One of the nT-aDM candidates found towards C12 field at the aDM of -1065\,pc\,cm$^{-3}$. We detected several such candidates as can be seen as a marginal excess in candidates around -1000 \,pc\,cm$^{-3}$ in Figure \ref{fig:all_spandak_cand} due to this peculiar interference {\itshape Right:} One of the nF-aDM candidate found towards A00 field at the aDM of -1814\,pc\,cm$^{-3}$.}
    \label{fig:my_label}
\end{figure}

The \gls{gc} region is likely to harbor an FRB-like magnetar population as suggested in Section \ref{sect:fast_transients}. We searched for such transients using SPANDAK with the parameters listed in Table \ref{tab:spandak_parameter} on the mid-temporal resolution data product mentioned in Table \ref{tab:formats}. We first collated 2 nearby frequency channels and reduced the frequency resolution by a factor of 2 to get 6656 channels as SPANDAK has a limitation of processing only up to 8k channels. We searched up to a DM of 5000\,pc\,cm$^{-3}$. With our final temporal and frequency resolution products, the maximum channel smearing we will get for the highest DM is $\sim 0.38$\,ms. This is slightly higher than our sampling time of the data products (0.349\,ms) used for this analysis. With this lower bound on the possible widths, we searched for bursts with widths up to 90\,ms as mentioned in Table \ref{tab:spandak_parameter}. We found a total 4467 pDM candidates from all the pointings shown in Figure \ref{fig:gbt_pointings_Cband}. A summary of number of candidates observed from each of our C-band pointings are listed in Table \ref{tab:obs_gbt_pks_coord}. For each of the off-GC pointings (B and C rings) we roughly observed 80 to 100 candidates which is consistent with our lower {S/N}$_{\rm min}$ threshold and interference environment. As can also be seen from Table \ref{tab:obs_gbt_pks_coord}, around 70\% of the total candidates were found toward the GC(0,0) region. Figure \ref{fig:spandak_example_plot} shows an example output plot from the SPANDAK pipeline. We visually inspected such plots for 4467 candidates and separated 667 candidates for further inspection. Figures \ref{fig:DM_vs_Time} and \ref{fig:DM_hist} show the distribution of all observed candidates from one of the 30-min scans and the distribution of 667 candidates from our entire observations with a first visual inspection cut. Out of these candidates, we identified 603 candidates from the Galactic center magnetar, SGR\,J1745-2900, as is clearly evident from Figure \ref{fig:DM_hist} with a large fraction of these candidates around the DM of 1778\,pc\,cm$^{-3}$. The rest of the 63 candidates were viewed carefully by changing the number of sub-bands, number of time bins, and adjusting the detected DM to verify if the detected signal was real. Figure \ref{fig:pDM_cand} shows one such example of an interesting candidate which was later rejected. After this careful inspection, we were unable to verify the validity of these 63 remaining candidates and were unable to find any new pDM transient signals from our observations originating from astrophysical processes. 

\hspace{1cm}

\subsubsection{Artificial DM signal searches}
The interstellar medium heavily impacts any signal being transmitted close to the GC. We refer to the Galactic density model from \cite{YMW16}, and considered that the largest DM one is likely to encounter towards the GC is on the order of 4000\,pc\,cm$^{-3}$. It should be noted that density model by \cite{YMW16} may sometimes over-predict the DM in the plane of the Galaxy. Another Galactic density model by \cite{Cordes2002} (NE2001) may be more accurate; however, towards the GC, it also predicts a similar high DM. All three types of aDM signals get impacted differently. For example, signals with nT-aDM are indistinguishable from naturally dispersed signals if they are transmitted originally with the aDM less than the DM contribution from the Galaxy. Hence, when we refer to the DM limit for these signals, original transmitted aDM would likely be $\sim 4000 - 9000$\,pc\,cm$^{-3}$. For nF-aDM and nTnF-DM types of signals, the dispersion due to interstellar medium heavily impacts their shape and would render them indistinguishable from interference. Thus, in order to recover their original aDM shape, we dedisperse all mid-time resolution data products at the DM of 4000\,pc\,cm$^{-3}$ before searching for nF-aDM and nTnF-aDM bursts. With this dispersion, we had to collate frequency resolution by a factor of 4 (3k channels) compared to the original mid-temporal resolution products. Thus, due to further channel smearing, our minimum burst width will be on the order of 0.7\,ms for artificial transients across 4 to 8 GHz.

For the aDM signal searches, we searched twice the maximum pulse widths compared to our pDM search as indicated in Table \ref{tab:spandak_parameter}. We found a total 2276 candidates combined from all three different aDM types with a {S/N}$_{\rm min}$ cut of 10. Table \ref{tab:obs_gbt_pks_coord} summarizes a number of candidates found from each of the pointings. Figure \ref{fig:DM_hist} shows a histogram of these candidates. From our initial inspection of SPANDAK plots (such as shown in Figure \ref{fig:spandak_example_plot}), we found 56 candidates for further inspection. Figure \ref{fig:aDM_cand1} and \ref{fig:aDM_cand2} show examples of two such interesting aDM candidates. Further discussion on these candidates are outlined in Section \ref{sect:discussion_candidates}. 

\begin{table}[h]
    \centering
    \caption{Summary of BL Galactic center survey conducted from the Parkes and GBT telescopes across 0.7 to 8\,GHz. The right ascension, declination, and Galactic coordinates for each ON-source pointing are listed in columns 3-6. Columns~7 and 8 show the number of candidates detected by turboSETI for each ON-source, for each filter as described in Section \ref{sect:narrow_seti}. The ON-sources for which 4 observations were used instead of the standard 6 are labeled with a $^\dagger$; this only applies to our turboSETI analysis. The last two rows show the number of transient candidates found for both naturally (pDM) and artificially (aDM) dispersed signals. The transient analysis used all three 5-min long observations per pointing, with the exception of A00, for which we included twelve 30-min long searches as well. These values are labeled with a $^\bigstar$, and are combined candidates from these deep and shorter observations towards the A00 field. The broadband analysis was not extended to the fields observed from the Parkes telescope due to excess scattering and channel smearing losses.}
    \begin{tabular}{l c c c c c c c c c}
    \hline
    \hline
    \multicolumn{10}{c}{Parkes: 1.0 to 4.0 GHz} \\
    \hline
    Pair & ON-source & RA & DEC & l & b & f2 & f3 \\ 
    & pointing & (h:m:s) & (d:m:s) & (deg) & (deg) & Events & Events \\
    \hline
    A1-B1 & A1 & 17:45:40.04 & -29:00:28.10 & 359.94 & -0.05 & 291 & 0 \\
    & B1 & 17:42:34.15 & -28:35:25.90 & 359.94 & 0.75 & 9,301 & 138 \\
    A2-B2 & A2 & 17:45:18.09 & -29:08:19.17 & 359.79 & -0.05 & 1,424 & 0 \\
    & B2 & 17:42:12.06 & -28:43:15.04 & 359.79 & 0.75 & 7,149 & 36 \\
    A3-B3 & A3 & 17:46:01.93 & -28:52:36.81 & 0.10 & -0.05 & 1,012 & 3 \\
    & B3 & 17:42:56.19 & -28:27:36.52 & 0.10 & 0.75 & 3,446 & 27 \\
    A4-B4 & A4 & 17:46:00.21 & -29:08:32.63 & 359.87 & -0.18 & 1,116 & 0 \\
    & B4 & 17:42:54.00 & -28:43:32.13 & 359.87 & 0.62 & 3,384 & 15 \\
    A5-B5 & A5 & 17:46:22.11 & -29:00:41.14 & 0.02 & -0.18 & 1,036 & 0 \\
    & B5 & 17:43:16.04 & -28:35:42.55 & 0.02 & 0.62 & 4,935 & 12 \\
    A6-B6 & B6 & 17:44:57.98 & -29:00:14.25 & 359.87 & 0.09 & 786 & 0\\
    & B6 & 17:41:52.26 & -28:35:08.43 & 359.87 & 0.88 & 2,436 & 9 \\
    A7-B7 & A7 & 17:45:19.92 & -28:52:23.38 & 0.02 & 0.09 & 802 & 0 \\
    & B7 & 17:42:14.35 & -28:27:19.50 & 0.02 & 0.88 & 4,727 & 9 \\
    
    \hline
    \hline
    \multicolumn{10}{c}{GBT: 3.9 to 8 GHz} \\
    \hline
    Pair & ON-source & RA & DEC & l & b & f2 & f3 & pDM & aDM \\ 
    & pointing & (h:m:s) & (d:m:s) & (deg) & (deg) & Events & Events & Candidates & Candidates\\
    \hline
    A00-C01 & A00 & 17:45:40.04 & -29:00:28.10 & 359.944 & -0.046 & 2 & 0 & see below & see below \\
    & C01 & 17:45:51.95 & -28:56:11.99 & 0.028 & -0.046 & 3,129 & 3 & 43 & 21\\
    C07-A00 & C$07^{\dagger}$ & 17:45:28.12 & -29:04:44.14 & 359.861 & -0.046 & 1,687 & 0 & 51 & 29 \\
    & A$00^{\dagger}$ & 17:45:40.04 & -29:00:28.10 & 359.944 & -0.046 & 1 & 0 &  see below & see below \\
    B01-B04 & B01 & 17:45:45.99 & -28:58:20.05 & 359.986 & -0.046 & 1 & 0 & 52 & 112 \\
    & B04 & 17:45:34.08 & -29:02:36.13 & 359.906 & -0.046 & 203 & 0 & 80 & 150 \\
    B02-B05 & B02 & 17:45:34.57 & -28:58:16.40 & 359.965 & -0.010 & 75 & 0 & 113 & 240 \\
    & B05 & 17:45:45.52 & -29:02:39.78 & 359.923 & -0.082 & 53 & 0 & 84 & 140 \\
    B03-B06 & B03 & 17:45:28.61 & -29:00:24.42 & 359.923 & -0.010 & 64 & 0 & 83 & 90 \\
    & B06 & 17:45:51.47 & -29:00:31.72 & 359.965 & -0.082 & 38 & 0 & 80 & 100 \\
    C02-C04 & C02 & 17:45:40.52 & -28:56:08.37 & 0.007 & -0.010 & 271 & 0 & 73 & 55 \\
    & C04 & 17:45:23.14 & -28:58:12.69 & 359.944 & 0.026 & 457 & 0 & 35 & 58 \\
    C03-C05 & C03 & 17:45:29.10 & -28:56:04.69 & 359.986 & 0.026 & 291 & 0 &100 & 65 \\
    & C05 & 17:45:17.18 & -29:00:20.67 & 359.903 & 0.026 & 2,038 & 0 & 86 & 27\\
    C08-C06 & C08 & 17:45:39.56 & -29:04:47.83 & 359.882 & -0.082 & 57 & 0 & 73 & 15 \\
    & C06 & 17:45:22.65 & -29:02:32.42 & 359.882 & -0.010 & 582 & 0 & 53 & 31 \\
    C11-C09 & C11 & 17:46:02.90 & -29:00:35.29 & 359.986 & -0.118 & 1,605 & 0 & 51 & 23 \\
    & C09 & 17:45:51.00 & -29:04:51.46 & 359.903 & -0.118 & 118 & 0 & 46 & 81 \\
    C10-C12 & C$10^{\dagger}$ & 17:45:56.95 & -29:02:43.38 & 359.944 & -0.118 & 282 & 16 & 51 & 20 \\
    & C$12^{\dagger}$ & 17:45:57.42 & -28:58:23.65 & 0.007 & -0.082 & 917 & 0 & 109 & 59 \\
    Deep & A$00^{\dagger}$ & 17:45:40.04 & -29:00:28.10 & 359.944 & -0.046 & - & - &  $3190^{\bigstar}$ & $955^{\bigstar}$ \\
    \hline
    \end{tabular}
    \label{tab:obs_gbt_pks_coord}
\end{table}

\section{Discussion}
\label{sec:discussion}

\subsection{Candidate signals from ETI}
\label{sect:discussion_candidates}
We carried out searches for two different types of beacons which are likely to originate from (a) a transmitter placed near the \gls{gc} illuminating the entire Galaxy and/or (b) any star in the lines-of-sight of our pointings towards the \gls{gc}. 
\subsubsection{Narrowband drifting candidates}
As mentioned earlier, we carried out three pairs of ON-OFF observations and compared narrowband drifting signals across ON and OFF pointings for all pairs. Some straightforward examples of candidates classified as \gls{rfi} can be seen in Figure \ref{fig:turboSETI_cands}. A clear \gls{rfi} signal can be seen in Figures \ref{fig:cand_1} - \ref{fig:cand_3}, where we see the signal appearing in all six observations. In these three examples, a clear and sometimes intermittent triplet pattern (central carrier with two side-bands) can be seen which is likely to be a terrestrial communication signal. If the drift rate of this signal was significantly non-zero and was not present in the off scans, this would be an incredibly intriguing signal. Signals shown in Figures \ref{fig:cand_1} and \ref{fig:cand_3} also appear to have similar frequency (4199.5\,MHz) despite coming from different pointings. A slightly dimmer \gls{rfi} signal can also be seen in Figure \ref{fig:cand_4} for ON and OFF observations. Such weaker signals are less likely to be detected above our {S/N}$_{\rm min}$ threshold and likely to be rejected automatically without careful visual inspection. Examples shown in Figures \ref{fig:cand_5} and \ref{fig:cand_6} are interesting candidates where a signal is seen in one and two ON-source pointings, respectively. However, upon careful inspection it can be seen that the same signal is also visible at the beginning of the OFF pointing. Additionally, the drift rate of these signals are close to zero and thus, they are likely to originate from terrestrial sources. 

Although most of the examples shown in Figure \ref{fig:turboSETI_cands} were detected with the GBT data, an interesting example was found in the A2-B2 (359.79-0.05 \& 359.79+0.75) pair with the Parkes \gls{uwl} observations. Figure \ref{fig:Navstar_sat} shows several closely spaced strong narrowband drifting signals in the $2^{nd}$, $4^{th}$, and $6^{th}$ panels. This strong signal appears as a candidate event several times throughout our candidate list, both using filter 2 and 3 across a few other pointings. Since this is an interesting candidate passing our final filter 3, we carefully compared the detected frequency of the candidate against the satellite database and their transmission frequencies. Indeed, we find that the frequency at which these signals are found (1381\,MHz) is also the frequency at which the GPS satellite Navstar transmits. Thus, due to its detection across other pointings and association of the detection frequency with a known satellite transmission, we are unable to consider this as a likely candidate of extra-terrestrial origin. Nevertheless, it is a good depiction of what we expect a potential ETI signal to look like: non-zero drifting and appearing in all 3 ON-observations but none of the OFF. The two remaining events from Figure \ref{fig:turboSETI_cands}: \ref{fig:cand_7} and \ref{fig:cand_8} are the most interesting. The candidate shown in Figure \ref{fig:cand_7} is found at 4.197\,GHz where a large concentration of hits are seen across other pointings as shown in Figure \ref{fig:Freq_Hist} and thus can still be rejected as \gls{rfi}. However, the candidate shown in Figure \ref{fig:cand_8} occurring at 6.408\,GHz is unique and is seen only once in our entire set of observations (see Figure \ref{fig:Freq_Hist}). Since we did not find it to repeat during the other ON pointings, we still discard it as \gls{rfi} due to some intermittent terrestrial source or coincidental passing of a low earth orbit satellite across our beam during that single 5-min observation pointing towards C09. In summary, we did not find any strong evidence of narrowband drifting signals based on our criteria. Thus, we can reject the presence of strong beacons transmitting between 1 to 8 GHz with a 100\% duty cycle located either at the \gls{gc} or from other line-of-sight stars. However, we should point out that there are a few caveats on these constraints we obtained on the existence of such signals. For example, we only searched a small fraction of drift-rate parameter space ($\pm$4 Hz/sec) and with a reduced sensitivity towards higher drift rate (see \citealt{mpg+21}). Also, our detection pipeline may have an imperfect recovery rate and we might miss a strong signal in our data. We will address completeness of analysis pipeline in future publications (Perez et al. 2021 in prep).

\subsubsection{Broadband candidates}
We carried out a search for three different types of aDM candidates for the first time. After eliminating candidates originating due to \gls{rfi}, we shortlisted 56 candidates. We carefully further inspected these selected candidates by changing the number of frequency channels, number of bins, and adjusting aDM. Most of these candidates appear to occur due to peculiar interference which aligns so as to mimic the aDM dispersion pattern. For example,  in Figure \ref{fig:aDM_cand1}, there is a comb of interference present between 4.2 to 4.4\,GHz. These comb of narrow frequency interference coincidentally happen to align with the nT-aDM dispersion curve. Figure \ref{fig:aDM_cand2} shows another example of a nF-aDM candidate found towards the A00 field at the aDM of -1814\,pc\,cm$^{-3}$. This candidate is likely due to interference at zero DM, which due to its limited spectral coverage, appears to partially follow the nF dispersion curve. Overall, we did not find any real artificially dispersed candidates originating from a likely transmitter placed at the \gls{gc} or from stars close to the \gls{gc}. Since we only dedisperse the data products to a DM of 4000\,pc\,cm$^{-3}$ to compensate for their likely origin at the \gls{gc}, with our current analysis we are unable to verify the existence of such signals transmitted from other stars along the line-of-sight towards the \gls{gc}. In the future, we plan to dedisperse our data products at multiple DMs before searching aDM signals to compensate for their Galactic dispersion and likely origin from line-of-sight stars spread across 8.2\,kpc. 

\subsection{Number of stars surveyed}
\label{sect:number_of_stars}
In Section \ref{sect:motivations}, we highlighted that a line-of-sight towards the \gls{gc} encounters the largest number of habitable stars compared to any other direction in the sky. In this section, we make an order-of-magnitude estimation on the number of stars surveyed based on the stellar number density (stars/pc$^{3}$) in the radial direction of our survey. \cite{gpm11} outlined various stellar number density models and found that the \cite{Carroll2007} estimation of the number density profile appears to match the observed stellar density in the Solar neighbourhood. This number density is represented as a function of radial distance from the \gls{gc}, R, and vertical height from the Galactic midplane, ${\rm Z}$, as,

\begin{equation}
  \rm  n(Z,R) ~=~ n_0~(e^{-Z/Z_{\rm thin}}~+~0.085e^{-Z/Z_{\rm thick}})e^{-{\rm R/h_R}} ~ {\rm stars/pc^3}. 
\end{equation}

Here, n$_0$ is the normalization factor which we consider from \cite{gpm11} as 5.502 stars/pc$^3$. The constants Z$_{\rm thin}$=350 pc, Z$_{\rm thick}$=1000 pc, and h$_R$=2.25 kpc are thin disk scale height, thick disk scale height, and radial scale, respectively. As can be seen, although stellar density falls off very quickly in the ${\rm Z}$ direction, for smaller angular scales, it can be considered to have constant stellar density. For example, at the distance of 8.5 kpc with our survey region of $4\arcmin \times 4\arcmin$, the diameter of the region probed would be around 10\,pc, for which the number density reduces by only 2.5\% (from 5.96 to 5.81 stars/pc$^3$). This is not a significant change; thus, we use the midplane density and take it as a constant for the ${\rm Z}$ scale height. 

To calculate the total number of stars, we considered a cone in the Galactic midplane towards the \gls{gc} with the Sun at the apex. The angle of the cone changes as a function of our observing frequencies and proposed target region. We considered stars up to the distance of 8.5 kpc from Earth which includes the central Galactic bulge which has the largest concentration of stars. For our current $4\arcmin \times 4\arcmin$ region, if we count the number of stars from Earth to the GC, we will encounter $\sim 6 \times 10^5$ total stars. This number will decrease as we go to higher-frequencies with the GBT due to our reduced survey regions (see Table \ref{tbl:gbt_rx_obs}). For our already conducted survey of the Parkes region which covers two $30\arcmin \times 30\arcmin$ fields (see Figure \ref{fig:Parkes_pointing}), we estimated that we surveyed around 60 million stars. The large concentration of these total surveyed stars are likely to reside close to the GC region. As indicated in Figure \ref{fig:gowanlock2011_fraction_HP}, a large fraction of these stars, especially the ones that are closest to the \gls{gc}, may host habitable planets and thus our survey provides some of the best constraints on the two types of beacons from the large number of habitable systems. We should highlight that these are only an order-of-magnitude estimates and gauging the exact numbers, especially for Parkes ($2\degr \times 4\degr$), is a non-trivial task due to the validity of our assumed stellar number density model and small angle approximations. 

\subsection{Survey sensitivity}
\label{sect:survey_sensitivity}
We have conducted searches for two different signal types, and limits can be placed on our detection sensitivity to both these signal types. For narrowband drifting signals, the minimum detectable flux can be given as, 

\begin{equation}
    S_{\rm min,narrow}~=~ \frac{{S/N}_{\rm min}}{\beta} \frac{S_{\rm sys}}{\delta\nu_t}\sqrt{\frac{\delta{\nu}}{n_p \tau_{\rm obs}}}. 
\end{equation}

Here, S$_{sys}$ is the system equivalent flux density, $\tau_{\rm obs}$ represents integration time, n$_{p}$ is the number of polarizations, $\delta\nu_t$ and $\delta\nu$ are bandwidths of transmitted and received narrowband signals, respectively. For our survey we assumed a transmitted bandwidth of 1 Hz\footnote{ Interstellar scattering causes  broadening of a narrow spectral line  \citep{1991ApJ...376..123C}. For transmitters near the Sun the broadening is $\ll 1$~Hz and thus unimportant but the strong scattering toward the \gls{gc} can make this a significant effect. Thus, it is likely that transmitter bandwidth will be larger which will increase our sensitivity but for simplicity we have assumed it to be 1 Hz.}. {S/N}$_{\rm min}$ represents a desired threshold and $\beta$ is the \emph{dechirping efficiency}. As mentioned in Section \ref{sect:narrow_seti}, due to our frequency ($\delta{\nu}\sim3$\,Hz) and temporal ($\delta{t}\sim$18\,sec) resolutions, high drift rate signals get spread across multiple channels which reduces our sensitivity. This dechirping efficiency can also be expressed as follows for our survey.

\begin{equation}
    \beta = 
\begin{cases}
    1 & |\dot{\nu}| \leq 0.16 ~ Hz/sec\\
    \frac{3}{|\dot{\nu}|\times\,18}  & |\dot{\nu}| > 0.16 ~ Hz/sec.
\end{cases}
\end{equation}

Here, $\dot{\nu}$ is the trial drift rate under consideration. Our dechirping efficiency reaches its maximum for drift rates $\leq$ 0.16 Hz/sec and gradually declines beyond this limit. As mentioned in Section \ref{sect:narrow_seti}, future narrowband searches will improve this efficiency by collapsing nearby channels or by using a moving boxcar of corresponding width for the number of channels a given signal is likely to have spread across \citep{price2020}.

For the broadband pDM and aDM transient signals, the minimum detectable flux can be expressed as, 

\begin{equation}
S_{\rm min, transient} = {S/N}_{\rm min} \frac{S_{\rm sys} \beta}{\sqrt{n_{\textrm{p}} \; \Delta \tau_{\rm pulse}\; \Delta \nu}}
\label{eq:radiometer_wide_single}
\end{equation}

Here, $\Delta\tau_{\rm pulse}$ is the pulse width of pDM and aDM candidates, $\Delta\nu$ is the total instantaneous bandwidth, and $\beta$ is the loss due to digitization. For pDM candidates we used a {S/N}$_{\rm min}$ threshold of 6 and for aDM candidates we used a {S/N}$_{\rm min}$ threshold of 10. For simplicity, we considered scatter-independent pulse widths. A more accurate width estimation based on the scatter broadening and corresponding sensitivity limitation are discussed further in Section \ref{sect:sgr1745sp}. For both the signal types (narrowband and broadband), the S$_{sys}$ can also be represented as, 

\begin{equation}
    S_{\rm sys}~=~\frac{2{k}(T_{\rm sys} + T_{\rm GC})}{{A_{\rm eff}}}. 
\end{equation}

Here, T$_{\rm sys}$ is the system temperature at typical weather conditions\footnote{Table 2.2 in \url{https://science.nrao.edu/facilities/gbt/observing/GBTog.pdf}}, $k$ is the Boltzmann constant, and A$_{\rm eff}$ is the effective aperture of the receiving antenna. The \gls{gc} region contains a significant amount of background flux, especially with the lower frequency receivers (up to Ku-band). The excess noise contribution, which is expressed as T$_{\rm GC}$, will make a significant additional contribution to the S$_{\rm sys}$ \citep{2006MNRAS.373L...6J,Macquart2010}, and need to be included in the sensitivity calculations. \citet{Law_2008} conducted a detailed continuum survey of the \gls{gc} region using the GBT and computed a calibrated map of the region. \citet{Rajwade_2017}. used these measurements and derived the frequency dependence of this excess noise from the \gls{gc} as, 

\begin{equation}
    T_{\rm GC} ~=~ \frac{568}{\nu_{\rm GHz}^{1.13}} ~ {\rm K}.
\end{equation}

\begin{figure}[!tbp]
  \centering
  \subfloat[]{
  \includegraphics[scale=0.13]{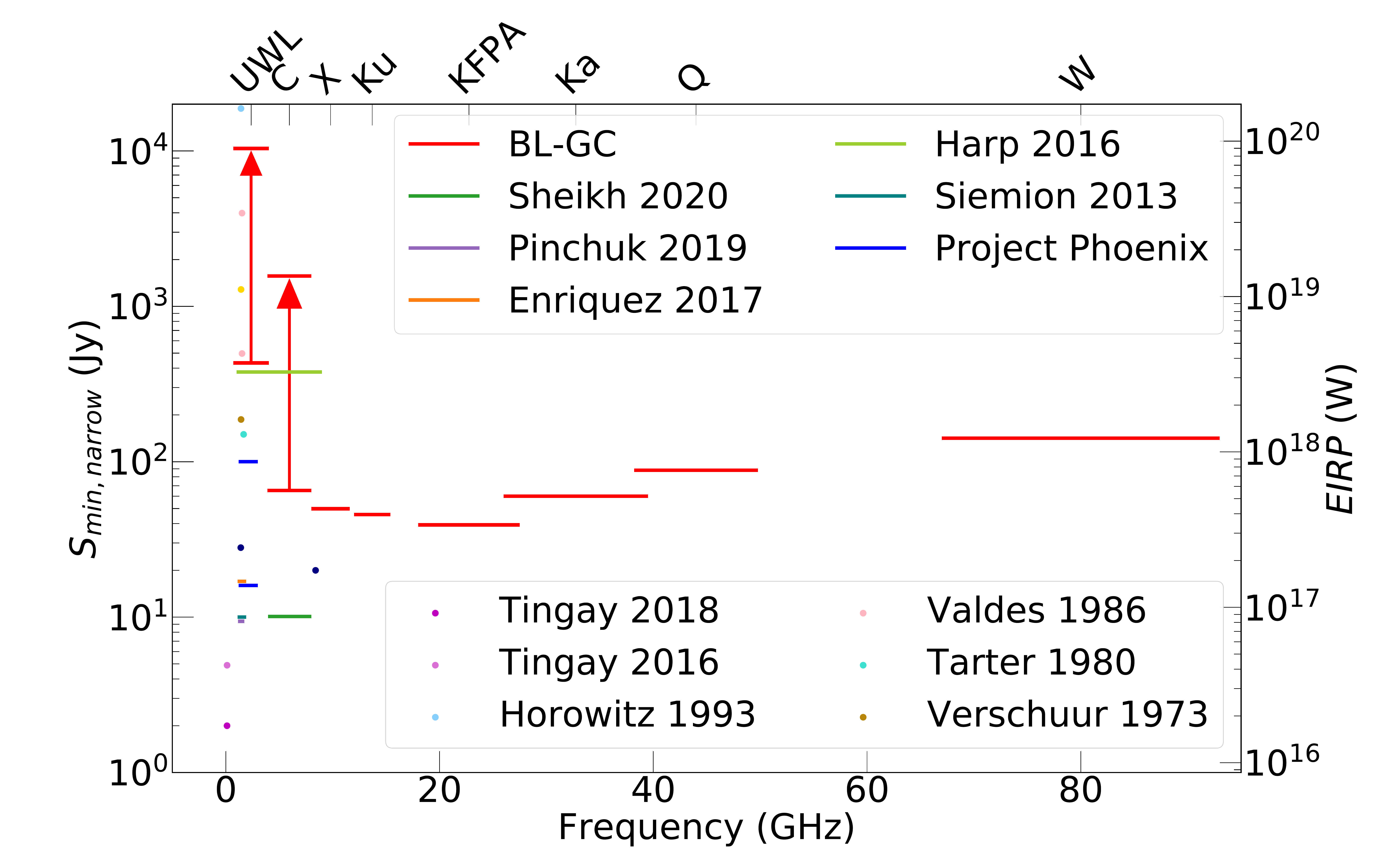}
  \label{fig:narrowband_sensitivity}}
  \hspace*{-0.25cm}
  \subfloat[]{
  \includegraphics[scale=0.13]{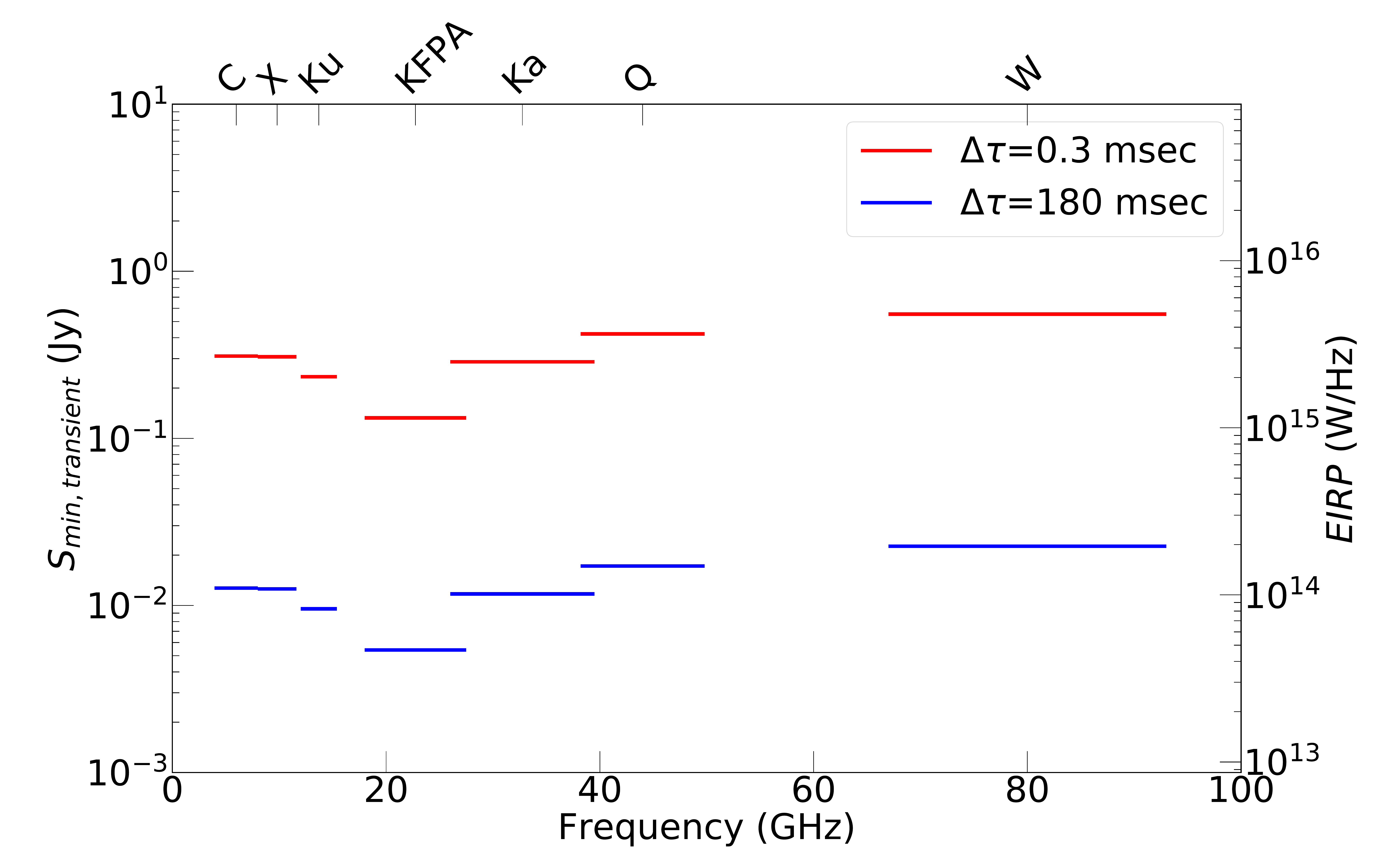}
  \label{fig:transient_sensitivity}}

  \caption{Sensitivity for narrowband and broadband type signals across all observing frequencies for BL GC survey. {\itshape Left:} Minimum detectable flux for narrowband drifting signals as a function of frequency. For our current performed survey, the bounds at UWL and C-band correspond to a changing dechirping efficiency, where the lower bound corresponds to the detection limit for drifting signals with $|\dot{\nu}|\leq$0.16 Hz/sec and the upper bound corresponds to the maximum drift rate ($\pm$4\,Hz/sec) we included in the survey. A clear change in the sensitivity can be seen due to the combined effect from the increased T$_{\rm sys}$ at higher frequencies due to atmospheric contribution and the significant loss of sensitivity at lower frequencies due to excess T$_{\rm GC}$. We are also displaying similar limits from some of the well-known narrowband SETI surveys for comparison \citep{Sheikh_2020, Pinchuk_2019, Enriquez:2017, Tingay_2016, Tingay_2018, Harp_2016, 2013ApJ...767...94S, 1993ApJ...415..218H, 1986Icar...65..152V, 1980Icar...42..136T, 1973Icar...19..329V}. 
  Here, we have assumed a dechirping efficiency of 1 for all the other surveys as calculating such efficiency based on the tools they have used is beyond the scope of this paper. {\itshape Right:} Minimum detectable flux limit for scatter independent minimum and maximum pulse widths for broadband transient signals with both pDM and aDM types at {S/N}$_{\rm min}$ threshold of 10. For both panels, the mirrored coordinate axes show the respective EIRP limits for putative transmitters residing at the distance of the GC. 
  } 
  \label{fig:all_sensitivity}
  
\end{figure}

Here, $\nu_{\rm GHz}$ is the observing frequency. As it can be seen, this will add a huge contribution to the system temperature at lower frequencies. For example, at 0.7 GHz from the \gls{uwl} at Parkes, the expected noise contribution from the \gls{gc} would be of the order of 800 K. Considering this contribution, our survey sensitivity is listed in Figure \ref{fig:all_sensitivity} for both narrow and broadband signals. As it can be seen, Ku and KFPA bands provide maximum sensitivity for both types of signals with contribution from the \gls{gc} dominating at lower frequency, and contribution from the Earth's atmosphere dominating the T$_{sys}$ at higher frequencies. Based on our minimum S$_{\rm min,narrow}$ and S$_{\rm min,broad}$ at various frequencies, we can calculate the corresponding Equivalent Isotropic Radiated Power (EIRP) at all our observing bands as, 

\begin{equation}
    {\rm EIRP}_{\rm GC;(narrow,~broad)} ~ = ~ S_{\rm min;(narrow,~broad)}\times4{\pi}D_{GC}^2. 
\end{equation}
 
For transmitters at the distance of the \gls{gc}, the minimum EIRP$_{\rm GC}$ are shown in Figure \ref{fig:all_sensitivity}. It is evident that the minimum EIRP sensitivities of $3 \times10^{17}$\,W and $4 \times 10^{13}$\,W/Hz are possible to achieve at the KFPA band for the narrowband and broadband transient signals, respectively. For our current survey with the GBT C-band and Parkes-\gls{uwl}, we estimate minimum EIRP detection sensitivity of around $5 \times 10^{17}$\,W ($\geq$ 10$^{19}$\,W for the signals with highest drift rates) and $4 \times 10^{18}$\,W ($\geq$ 10$^{20}$\,W for the signals with highest  drift rates) for the narrowband drifting signals, respectively. Similarly, for the broadband artificially dispersed signals from GBT C-band survey, we estimate a minimum detectable EIRP of the order of $1 \times 10^{14}$\,W/Hz. It is clearly evident that broadband signal allows much better sensitivity to provide better constraints on the weaker signals if they are band limited. However, since such signals are assumed to last for only a fraction of a second, we rely upon the repeatability of such signals which is further discussed in Section \ref{sect:repeat_broadband}. 

\subsection{Survey figure of merit}

\begin{figure}
    \centering
    \includegraphics[scale=0.35]{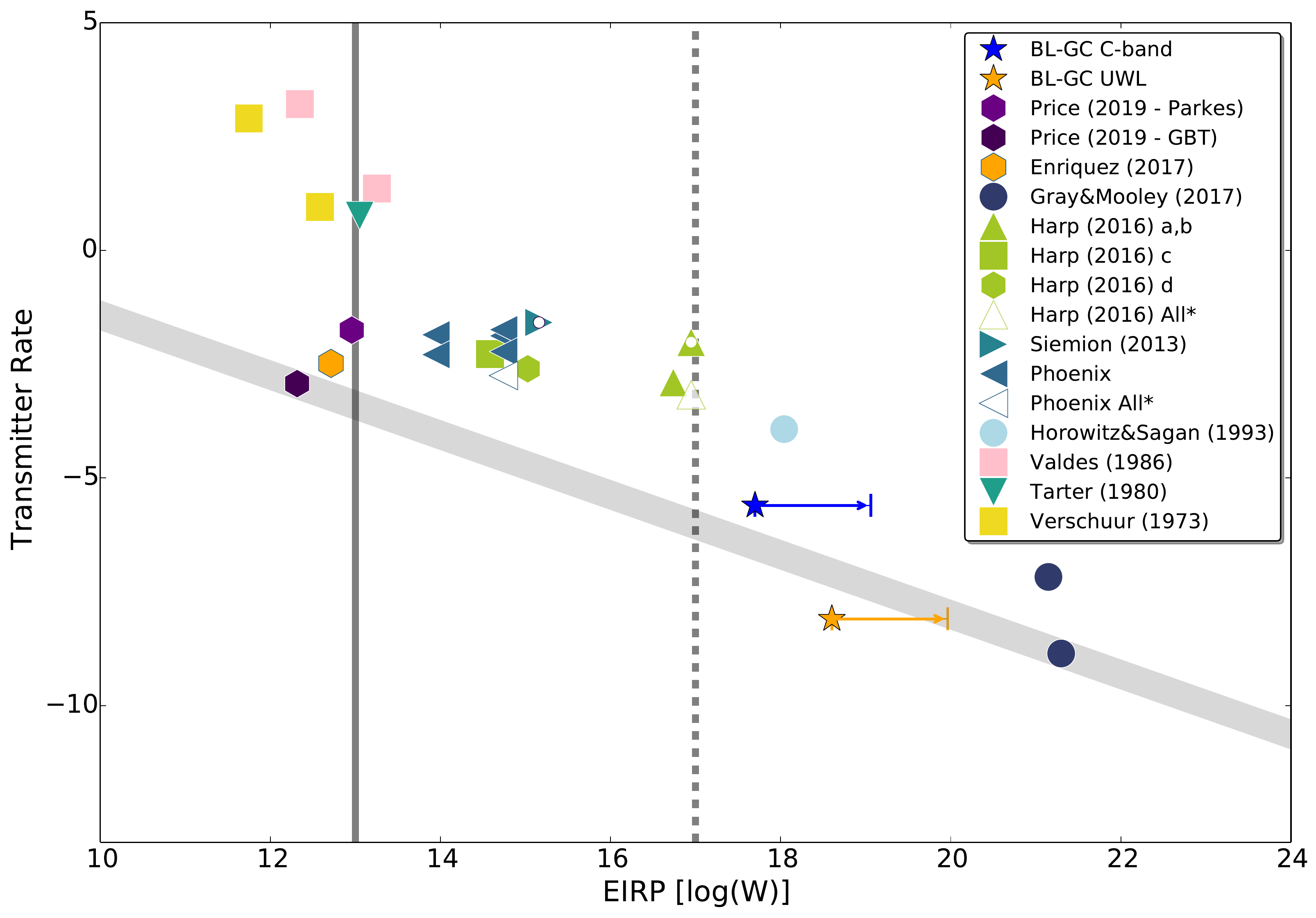}
    \caption{A comparison between the CWTFM-vs-EIRP for some of the prominent SETI surveys compared to our BL-GC survey conducted at C-band (blue star) and \gls{uwl} (orange star) reported in this paper. For our surveys, we have shown the lower bound in EIRP limits which correspond to the dechirping efficiency and extend up to maximum drift rates searched at both bands. We have assumed a dechirping efficiency of 1 for all the other SETI surveys. The gray thick line shows the slope of the transmitter-rate as a function of their EIRP power. The dotted line shows the total energy budget of a Kardashev Type I civilization which is equivalent to the total solar power incident on Earth's surface. The solid vertical line shows the EIRP output of the Arecibo planetary radar.}
    \label{fig:CWTFM_vs_EIRP}
\end{figure}

There have been several SETI surveys that have been conducted in the past six decades \citep[][and references therein]{Enriquez:2017,wkl18}. These surveys have primarily searched for narrowband drifting signals and thus they can be compared with each other based on the total number of stars they surveyed (or total area of the sky covered), total bandwidth utilized, detection sensitivity and other such parameters. For narrowband signals, the Drake figure-of-merit (DFM; \citealt{DFM84}) has been computed by many previous SETI surveys and can be expressed as, 

\begin{equation}
    {\rm DFM}~=~\frac{\Delta{\nu}~\Omega}{S_{min,avg}^\frac{3}{2}}, 
\end{equation}
where $\Omega$ is the area of the sky surveyed, $S_{min,avg}$ is the minimum detectable flux average across all observing frequencies, and $\Delta\nu$ is the total observing bandwidth. For our survey with the GBT across 3.9 -- 8\,GHz, we found the DFM to be $\sim 4 \times 10^{28}$ for the narrowband signals with drift-rates $\leq$0.16 Hz/sec and $\sim 4 \times 10^{26}$ for the highest drift-rate considered for our survey. It should be noted that the DFM has a number of limitations as outlined by \cite{Enriquez:2017} and \cite{mpg+21}. For example, the DFM does not consider a range of drift-rates and dechirping efficiency which heavily impacts S$_{min,avg}$ for a survey under consideration  \citep{mpg+21}. Moreover, the DFM assumes a uniform distribution of ETI transmitters. For surveys like ours, where a large concentration of stars are likely to reside within our beam towards the \gls{gc}, the DFM does not provide an accurate comparison. A better figure-of-merit is the Continuous Waveform Transmitter Rate (CWTFM), as introduced by \cite{Enriquez:2017}, which is also for narrowband signals and can be given as, 

\begin{equation}
    CWTFM~=~\zeta_{\rm AO}\frac{EIRP}{N_{\rm stars}\nu_{\rm frac}}. 
\end{equation}

Here, $\nu_{\rm frac}$ is the fractional bandwidth $\Delta\nu/{\nu}$, $\zeta_{\rm AO}$ is the normalization factor such that CWTFM = 1 when the EIRP is equivalent to the EIRP of the Arecibo radio telescope's planetary radar ($\sim 10^{13}$\,W), $\nu_{\rm frac}$ = 0.5, and N$_{\rm stars}$ = 1000. For our conducted survey, as discussed in Section \ref{sect:number_of_stars}, we can estimated the number of stars using the stellar density model and estimate the corresponding EIRP limits (see Section \ref{sect:survey_sensitivity}). We found that for the UWL and C-band, we surveyed around 60 million and 0.6 million stars, respectively. Figure \ref{fig:CWTFM_vs_EIRP} shows the measured CWTFM for our current C-band and \gls{uwl} surveys compared with some of the prominent SETI surveys reported earlier. It is clearly evident that with our current survey, we are already able to push the limit of transmitter-rate slope lower than what was obtained from previous SETI surveys. By extending this survey to other frequencies from the GBT and covering all the pointings from the Parkes telescopes, we are likely to push the CWFTM slope much lower and able to constrain the presence of technosignatures across a wide range of frequencies. Furthermore, Figure \ref{fig:narrowband_sensitivity} shows a comparison of sensitivity of narrowband signals of our survey with some of the prominent SETI surveys. As it is clearly evident, no other survey has ever explored the frequency coverage we aim to search with our survey. Moreover, these comparisons are only carried out for narrowband drifting signals. Our survey is already the only survey to search for artificially dispersed signals which have never been explored and thus cannot be compared with any previous SETI surveys. 

\subsection{Limits on repeatability of broadband signals from ETIs}
\label{sect:repeat_broadband}

\begin{figure}[h]
    \centering
    \subfloat[]{
    \includegraphics[scale=0.13]{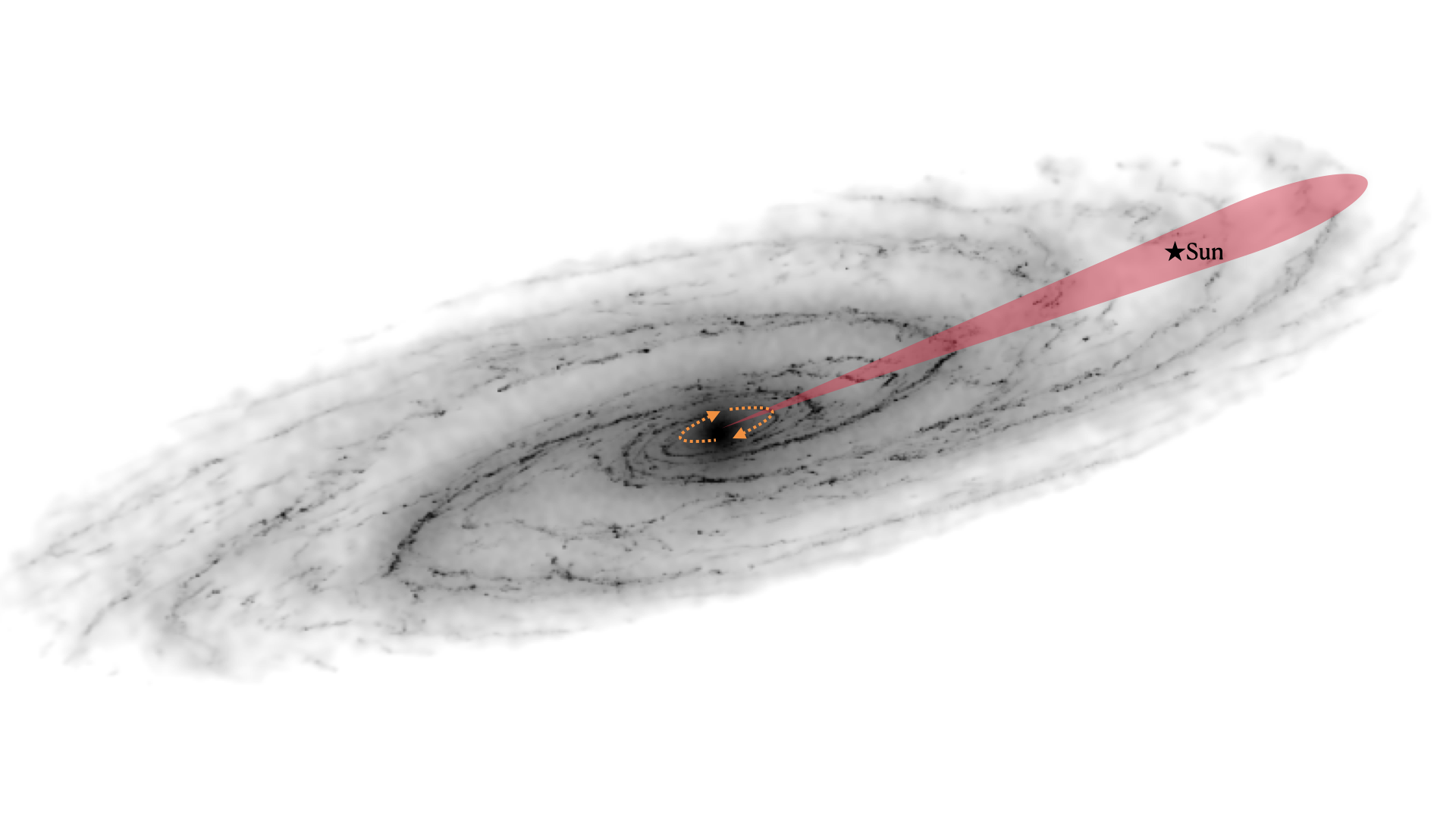}
    }
    \subfloat[]{
    \includegraphics[scale=0.4]{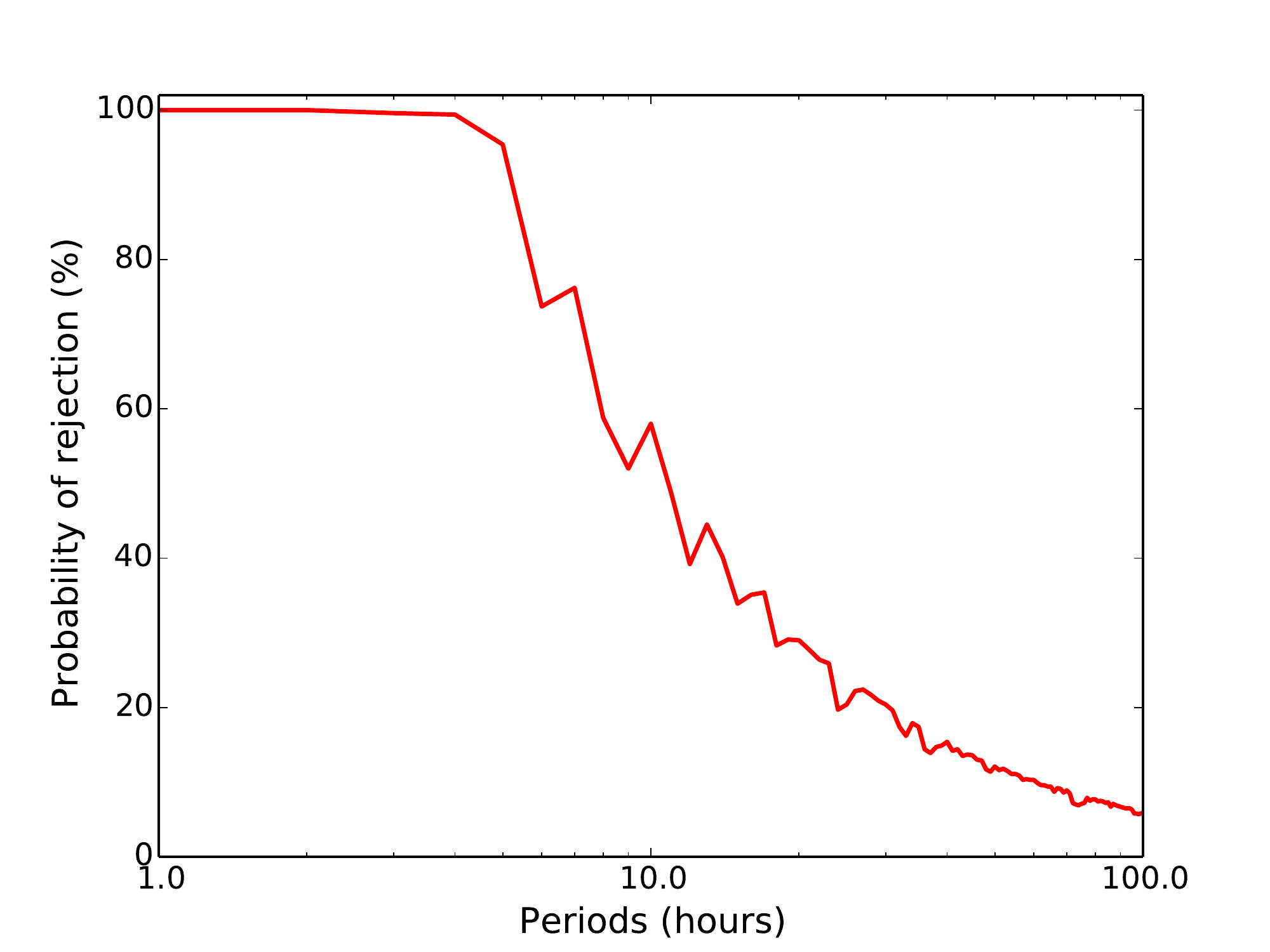}
    }
    \caption{Rejection of repeatability of broadband signals from our BL-GC survey conducted at C-band. {\itshape Left:} 
     Schematic sketch showing a rotating beam in $\rm Z=0$ plane from an artificial source located at the GC illuminating the entire Galaxy (the sketch reproduced and modified by permission from Alex Pettitt and originally from a simulation reported in \citealt{pda+15}). {\itshape Right:} Probability of rejecting repeating bursts of natural or artificial origin for a range of periods. 
     }
    \label{fig:repeat_broadband}
\end{figure}

We searched for three different types of artificially-dispersed broadband transient bursts likely to originate from deliberately transmitted ETI beacons. In future work, we will provide a detailed discussion on harmonic searches for these broadband signals. In this section, we evaluate the repetition rate of these broadband bursts from our GC$(0,0)$ deep observations. Figure \ref{fig:repeat_broadband} shows one of the scenarios of repeating broadband signals originating from a rotating beam from an artificial beacon\footnote{Here, we are only considering rotation of an artificial beam transmitting pulses with aDM dispersion to illuminate the Galaxy. We are not considering sweeping of such an artificial beam to produce aDM bursts such as in the case of pulsars generating pDM bursts. Thus, we are not constrained by the beaming fraction of such a beam.}. We can also infer different mechanisms which do not necessarily require a rotating beam but can still generate repeated bursts of similar nature. Such rotating beams of an artificial nature have been proposed earlier. For example, \cite{bbb10II} suggested such a beacon situated at the GC would be the most energy-efficient way to signal the entire Galaxy. \cite{SHOSTAK1985369} have also considered narrowband signals originating from such a rotating beacon. We did not find any strong aDM candidates; thus, considering length and separation between all our observations, we can reject repetition of all three types of broadband signals. Table \ref{tab:observations} lists all the observing dates and lengths of each of these observations. We arranged them on the time-axis and then simulated pulse trains with a range of periods. We also adjusted the phase (or offset) of these bursts within the corresponding period under consideration. For each period, we calculated the number of instances of offset phases which would have allowed us to see at least one pulse. Here, we assumed that we are likely to detect a burst at every period. Figure \ref{fig:repeat_broadband} shows the probability of rejecting periods for such bursts. We can reject any mechanism which can regularly produce pulses across 3.9 to 8\,GHz with the repetition period of around 4.3 hours with high accuracy (considering imperfect recovery rate of our search pipeline). For larger repetition periods up to 10 hours, we can reject such mechanism with $\geq$50\% probability while for further large periods up to 100 hours, rejection probability is marginal ($<$10\%) from our observations. In the future, we plan to extend this analysis for narrowband signals and will also schedule our observations such as to allow sufficient epoch sampling to reject larger repeating periodicity with better significance. These limits are non-trivial to apply to astrophysical sources such as magnetars which are not known to produce strong transient pulses during every rotation and their beam of emission not necessarily likely to reside in the $\rm Z=0$ plane as we assumed for this analysis. 

\subsection{SGR\,J1745--2900 and constraints on other transients from magnetars near the GC}
\label{sect:sgr1745sp}
\begin{figure}[h]
    \centering
    \includegraphics[scale=7]{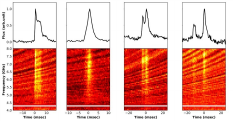}
    \caption{Four examples of different types of single pulses detected across 3.9 to 8.0\,GHz from SGR\,J1745--2900 detected from the SPANDAK pipeline with the BL-GC survey.}
    \label{fig:sp1745}
\end{figure}

We detected 603 single pulses from SGR\,J1745--2900 across 3.9 to 8.0\,GHz with a total of 6.6\,h of observations. These single pulses exhibit a wide variety of pulse shapes with single and multiple components. Figure \ref{fig:sp1745} shows examples of four different types of single pulses seen from SGR\,J1745--2900 from our observations: a narrow single pulse with a scattering tail, a pulse with unresolved multiple-components, and two well resolved dual-component pulses. This is consistent with single pulses seen around similar frequencies by \cite{pealmanSGR1745} and \cite{wharton1745VLA}. A detailed spectro-temporal analysis of these bursts is beyond the scope of this paper and will be reported in future publications. These observations are useful to constrain the presence of other radio magnetars at the GC producing transient bursts. 
 
Scattering due to a relatively dense environment near the GC likely limits the detection of other transients. In order to estimate scatter-broadened pulse widths at our observed frequencies, we utilized 
a narrow single peaked pulse from SGR\,J1745--2900 shown in Figure \ref{fig:sp1745} for further analysis. \cite{Spitler2014} estimated that the scatter broadening ($\tau_{d}$) from SGR\,J1745--2900 is of the order of $1.3 \pm 0.2$\,sec at 1\,GHz with a spectral index of $\alpha = -3.95 \pm 0.2$. Considering a similar spectral index, the expected $\tau_{d}$ will be $\sim 1.4$\,ms at 6\,GHz. Measuring the observed $\tau_{d}$ is challenging, as suggested by \cite{kjm+17}, due to the degeneracy of the unknown intrinsic pulse width and unresolved emission components. To model our underlying observed single pulse, we assumed a simple intrinsic single pulse with a Gaussian profile from the source, convolved with the $s(t)~=~e^{-t/{{\tau}_d}}U(t)$ for transmission through the ISM; where $U(t)$ is a step function with $U(t)=0$ for $t<0$ and $U(t)=1$ for $t\geq0$. For the intrinsic pulse, we considered a pulse width of 1.8\,ms reported from \cite{pealmanSGR1745} at 8.4\,GHz, which is close to the highest end of our observed band. Similar techniques have been used by \cite{kjm+17} and \cite{kmj+19} to estimate $\tau_{d}$. We found the $\tau_{d}\approx4$ ms by fitting this model to the narrow pulse shown in Figure \ref{fig:sp1745}. This is slightly higher and not consistent with the spectral index of $-3.8$ reported earlier at lower frequencies by \cite{Spitler2014}. Similar inconsistencies were also reported by \cite{pealmanSGR1745}. It is also likely that the larger scatter broadening is due to an unresolved emission component at the trailing edge of the pulse. 

The scatter broadening time scale found from the narrow SGR\,J1745--2900 burst can be utilized to place a limit on other FRB-like Galactic magnetars and \gls{msp} located in a similar environment around the GC. We reevaluated our minimum detection limit for the broadband signals calculated in Section \ref{sect:survey_sensitivity}. We find that for our current survey across 3.9 to 8\,GHz for transients, we were sensitive to bursts with peak flux density of around 68\,mJy which is equivalent to the peak luminosity of $\sim 10^{31}$\,erg\,s$^{-1}$ across our band for a source located at the GC. 
Although average flux densities of folded profiles of radio-loud magnetars are relatively lower, single pulses from them do exhibit bright emission. For example, XTE\,J1810--197, a known Galactic radio magnetar, was seen to produce bright transient radio pulses up to 42\,GHz \citep{crh+06}. XTE\,J1810--197 showed many single pulses with a peak flux of $\gtrsim 10$\,Jy at 6.4\,GHz which corresponds to a peak luminosity of around $7 \times 10^{31}$\,erg\,s$^{-1}$ across their 500\,MHz band. Similarly, another radio-loud magnetar, PSR\,J1622--4950, is known to show bursts with a peak flux of around 16\,Jy \citep{lbb+10}, which corresponds to a peak luminosity of $5 \times 10^{32}$\,erg\,s$^{-1}$. This is an order of magnitude larger than our limit. Moreover, the recently detected FRB-like Galactic magnetar, SGR\,1935+2154, also exhibits a bright luminosity on the order of $10^{36}$\,erg\,s$^{-1}$ \citep{chime_sgr1935} which is several orders of magnitude larger than our peak luminosity limit. Since our longest observing run was about 4.3\,h (see Table \ref{tab:observations}), we can reject the presence of any transient signal with a typical burst-rate of $\geq 0.23$\ bursts\,h$^{-1}$ at our radio luminosity limit. A few of the radio magnetars are known to produce transient radio bursts with much larger burst rates. For example, SGR\,J1745--2900 produced $\gtrsim 10$\,bursts\,hr$^{-1}$ as measured from our observations. 

Our scatter-broadened limits are also important as many different theories have been proposed to solve the \textit{missing pulsar} problem. For example, the \gls{gc} has been considered to favor \gls{msp} production \citep{Bartels2016}. As we found a scatter broadening timescale of $\sim 4$\,ms, any \gls{msp} signals with much smaller intrinsic pulse widths than normal pulsars are likely to get scattered and will therefore be undetectable up to our highest frequency of 8\,GHz. Extension of our survey to higher frequencies would be ideal to search for other such \glspl{msp} near the \gls{gc}. 

\section{Conclusion}
\label{sec:conclusion}

In this paper, by extending the habitability model of \cite{gpm11} to the inner 2.5\,kpc region, we demonstrated that a line-of-sight towards the \gls{gc} is likely to render the largest concentration and highest number density of habitable worlds. The \gls{gc} region also provides an ideal Schelling point for advanced civilizations to place a powerful transmitter to efficiently send beacons across the entire Milky Way. Thus, the \gls{gc} is one of the primary targets of the \gls{bl} program \citep{Isaacson:2017ib}. We outlined our survey strategy for the most comprehensive search for evidence of intelligent life ever conducted by utilizing a large fraction of the accessible radio window (0.7 to 93\,GHz) from the GBT and Parkes telescopes. We compared the sensitivity of our survey to some of the prominent SETI surveys and demonstrate that our survey has remarkable sensitivity with a frequency span never before explored for SETI. 

We have reported early results from our survey of the GC region across 1 to 8\,GHz with around 11.2 and 7\,h of observations from the GBT and Parkes telescopes, respectively. We carried out searches for standard narrowband drifting signals, along with, for the first time, three different types of broadband artificially-dispersed transient signals (up to a DM of --5000\,pc\,cm$^{-3}$) using our state-of-the-art GPU accelerated tools. We did not find a strong beacon signal for any of these signal categories searched. However, our survey already placed a constraint on the existence of narrowband transmitters of $\lesssim 1$ in 60 million stars toward the GC across 1 to 4\,GHz with EIRP $> 4 \times 10^{18}$\,W (and $\geq 10^{20}$\, W for the highest drift-rate signals). We were also able to place meaningful constraints for the first time on the existence of narrowband and broadband transmitters in $\lesssim 1$ in half a million stars across 3.9 to 8\,GHz with EIRP $>5 \times 10^{17}$\,W (and $\geq 10^{19}$\, W for the highest drift-rate signals) and $1 \times 10^{14}$\,W/Hz, respectively. With our deep targeted observations of the GC$(0,0)$, we were also able to reject the existence of periodic broadband beacons showing characteristic aDM dispersion pattern with periods $\leq 4.3$\,h with similar EIRPs.  

We also searched for astrophysical transient signals from other radio-loud magnetars near the GC. We searched for broadband pulses up to a DM of 5000\,pc\,cm$^{-3}$ with pulse widths ranging from 0.3 to 90\,ms with {S/N}$_{\rm min}$\,$\geq 6$. We found 603 single pulses from SGR\,J1745--2900 and we determined a scatter broadening on the order of 4\,ms from one of the narrow detected pulses. Based on the scatter broadening timescale, we ruled out the likely existence of other transient sources with a luminosity limit of $\geq 10^{31}$\,erg\,s$^{-1}$ with a burst-rate of $\geq 0.23$\,burst\,hr$^{-1}$. This is around the typical peak luminosity of other Galactic radio-loud magnetars. 

\section{Acknowledgments}
We would like to thank the anonymous referee for a critical review of the paper and for suggesting several improvements to the manuscript. Breakthrough Listen is managed by the Breakthrough Initiatives, sponsored by the Breakthrough Prize Foundation. SC, JMC, and AS acknowledge support from the National Science Foundation (AAG~1815242). VG thanks Alex Pettitt for giving us permission to use a sketch of the milky-way. The Parkes radio telescope is part of the Australia Telescope National Facility which is funded by the Australian Government for operation as a National Facility managed by CSIRO. The Green Bank Observatory is a facility of the National Science Foundation, operated under cooperative agreement by Associated Universities, Inc. We thank the staff at Parkes and Green Bank observatories for their operational support. We thank all the frontline workers during this global pandemic and all the California firefighters for keeping most of us out of harm's way. 

\bibliography{references}

\begin{thebibliography}{}
\expandafter\ifx\csname natexlab\endcsname\relax\def\natexlab#1{#1}\fi
\providecommand{\url}[1]{\href{#1}{#1}}
\providecommand{\dodoi}[1]{doi:~\href{http://doi.org/#1}{\nolinkurl{#1}}}
\providecommand{\doeprint}[1]{\href{http://ascl.net/#1}{\nolinkurl{http://ascl.net/#1}}}
\providecommand{\doarXiv}[1]{\href{https://arxiv.org/abs/#1}{\nolinkurl{https://arxiv.org/abs/#1}}}

\bibitem[{{Backus} \& {Project Phoenix Team}(2004)}]{2004AAS...204.7504B}
{Backus}, P.~R., \& {Project Phoenix Team}. 2004, in Bulletin of the American
  Astronomical Society, Vol.~36, American Astronomical Society Meeting
  Abstracts \#204, 805

\bibitem[{Bartels {et~al.}(2016)Bartels, Krishnamurthy, \&
  Weniger}]{Bartels2016}
Bartels, R., Krishnamurthy, S., \& Weniger, C. 2016, Phys. Rev. Lett., 116,
  051102, \dodoi{10.1103/PhysRevLett.116.051102}

\bibitem[{{Benford} {et~al.}(2010{\natexlab{a}}){Benford}, {Benford}, \&
  {Benford}}]{bbb10II}
{Benford}, G., {Benford}, J., \& {Benford}, D. 2010{\natexlab{a}},
  Astrobiology, 10, 491, \dodoi{10.1089/ast.2009.0394}

\bibitem[{{Benford} {et~al.}(2010{\natexlab{b}}){Benford}, {Benford}, \&
  {Benford}}]{bbb10I}
{Benford}, J., {Benford}, G., \& {Benford}, D. 2010{\natexlab{b}},
  Astrobiology, 10, 475, \dodoi{10.1089/ast.2009.0393}

\bibitem[{{Bochenek} {et~al.}(2020){Bochenek}, {Ravi}, {Belov}, {Hallinan},
  {Kocz}, {Kulkarni}, \& {McKenna}}]{stare2SGR}
{Bochenek}, C.~D., {Ravi}, V., {Belov}, K.~V., {et~al.} 2020, arXiv e-prints,
  arXiv:2005.10828.
\newblock \doarXiv{2005.10828}

\bibitem[{Bower {et~al.}(2014)Bower, Deller, Demorest, Brunthaler, Eatough,
  Falcke, Kramer, Lee, \& Spitler}]{Bower2014}
Bower, G.~C., Deller, A., Demorest, P., {et~al.} 2014, The Astrophysical
  Journal Letters, 780, L2.
\newblock \url{http://stacks.iop.org/2041-8205/780/i=1/a=L2}

\bibitem[{{Bryson} {et~al.}(2021){Bryson}, {Kunimoto}, {Kopparapu}, {Coughlin},
  {Borucki}, {Koch}, {Aguirre}, {Allen}, {Barentsen}, {Batalha}, {Berger},
  {Boss}, {Buchhave}, {Burke}, {Caldwell}, {Campbell}, {Catanzarite},
  {Chandrasekaran}, {Chaplin}, {Christiansen}, {Christensen-Dalsgaard},
  {Ciardi}, {Clarke}, {Cochran}, {Dotson}, {Doyle}, {Duarte}, {Dunham},
  {Dupree}, {Endl}, {Fanson}, {Ford}, {Fujieh}, {Gautier}, {Geary},
  {Gilliland}, {Girouard}, {Gould}, {Haas}, {Henze}, {Holman}, {Howard},
  {Howell}, {Huber}, {Hunter}, {Jenkins}, {Kjeldsen}, {Kolodziejczak},
  {Larson}, {Latham}, {Li}, {Mathur}, {Meibom}, {Middour}, {Morris}, {Morton},
  {Mullally}, {Mullally}, {Pletcher}, {Prsa}, {Quinn}, {Quintana}, {Ragozzine},
  {Ramirez}, {Sanderfer}, {Sasselov}, {Seader}, {Shabram}, {Shporer}, {Smith},
  {Steffen}, {Still}, {Torres}, {Troeltzsch}, {Twicken}, {Uddin}, {Van Cleve},
  {Voss}, {Weiss}, {Welsh}, {Wohler}, \& {Zamudio}}]{Bryson_2021}
{Bryson}, S., {Kunimoto}, M., {Kopparapu}, R.~K., {et~al.} 2021, \aj, 161, 36,
  \dodoi{10.3847/1538-3881/abc418}

\bibitem[{{Brzycki} {et~al.}(2020){Brzycki}, {Siemion}, {Croft}, {Czech},
  {DeBoer}, {DeMarines}, {Drew}, {Gajjar}, {Isaacson}, {Lacki}, {Lebofsky},
  {MacMahon}, {de Pater}, {Price}, \& {Worden}}]{bsc+20}
{Brzycki}, B., {Siemion}, A. P.~V., {Croft}, S., {et~al.} 2020, arXiv e-prints,
  arXiv:2006.04362.
\newblock \doarXiv{2006.04362}

\bibitem[{{Camilo} {et~al.}(2006){Camilo}, {Ransom}, {Halpern}, {Reynolds},
  {Helfand}, {Zimmerman}, \& {Sarkissian}}]{crh+06}
{Camilo}, F., {Ransom}, S.~M., {Halpern}, J.~P., {et~al.} 2006, \nat, 442, 892,
  \dodoi{10.1038/nature04986}

\bibitem[{{Camilo} {et~al.}(2007){Camilo}, {Ransom}, {Pe{\~n}alver},
  {Karastergiou}, {van Kerkwijk}, {Durant}, \& et~al.}]{Camilo2007}
{Camilo}, F., {Ransom}, S.~M., {Pe{\~n}alver}, J., {et~al.} 2007, The
  Astrophysical Journal, 669, 561, \dodoi{10.1086/521548}

\bibitem[{Carroll \& Ostlie(2007)}]{Carroll2007}
Carroll, B.~W., \& Ostlie, D.~A. 2007, {A}n {I}ntroduction to {M}odern
  {A}strophysics, 2nd edn., ed. S.~F.~P. Addison-Wesley

\bibitem[{{Carroll-Nellenback} {et~al.}(2019){Carroll-Nellenback}, {Frank},
  {Wright}, \& {Scharf}}]{cfw+19}
{Carroll-Nellenback}, J., {Frank}, A., {Wright}, J., \& {Scharf}, C. 2019, \aj,
  158, 117, \dodoi{10.3847/1538-3881/ab31a3}

\bibitem[{Cocconi \& Morrison(1959)}]{1959Natur.184..844C}
Cocconi, G., \& Morrison, P. 1959, Nature, 184, 844, \dodoi{10.1038/184844a0}

\bibitem[{{Cole} \& {Ekers}(1979)}]{CoEk79}
{Cole}, T.~W., \& {Ekers}, R.~D. 1979, Proceedings of the Astronomical Society
  of Australia, 3, 328

\bibitem[{Comisso \& Asenjo(2021)}]{PhysRevD.103.023014}
Comisso, L., \& Asenjo, F.~A. 2021, Phys. Rev. D, 103, 023014,
  \dodoi{10.1103/PhysRevD.103.023014}

\bibitem[{{Cordes} {et~al.}(1997){Cordes}, {Lazio}, \&
  {Sagan}}]{1997ApJ...487..782C}
{Cordes}, J.~M., {Lazio}, J.~W., \& {Sagan}, C. 1997, \apj, 487, 782,
  \dodoi{10.1086/304620}

\bibitem[{{Cordes} \& {Lazio}(1991)}]{1991ApJ...376..123C}
{Cordes}, J.~M., \& {Lazio}, T.~J. 1991, The Astrophysical Journal, 376, 123,
  \dodoi{10.1086/170261}

\bibitem[{{Cordes} \& {Lazio}(1997)}]{1997ApJ...475..557C}
{Cordes}, J.~M., \& {Lazio}, T. J.~W. 1997, \apj, 475, 557,
  \dodoi{10.1086/303569}

\bibitem[{{Cordes} \& {Lazio}(2002)}]{Cordes2002}
{Cordes}, J.~M., \& {Lazio}, T.~J.~W. 2002, arXiv: astro-ph/0207156

\bibitem[{{Degenaar} {et~al.}(2013){Degenaar}, {Reynolds}, {Miller}, {Kennea},
  \& {Wijnands}}]{Degenaar2013}
{Degenaar}, N., {Reynolds}, M.~T., {Miller}, J.~M., {Kennea}, J.~A., \&
  {Wijnands}, R. 2013, The Astronomer's Telegram, 5006, 1

\bibitem[{{Deneva} {et~al.}(2009){Deneva}, {Cordes}, \&
  {Lazio}}]{2009ApJ...702L.177D}
{Deneva}, J.~S., {Cordes}, J.~M., \& {Lazio}, T.~J.~W. 2009, The Astrophysical
  Journal Letters, 702, L177, \dodoi{10.1088/0004-637X/702/2/L177}

\bibitem[{{Dexter} \& {O'Leary}(2014)}]{Dexter2014}
{Dexter}, J., \& {O'Leary}, R.~M. 2014, The Astrophysical Journal, 783, L7,
  \dodoi{10.1088/2041-8205/783/1/L7}

\bibitem[{Di~Stefano \& Ray(2016)}]{DiStefano:2016}
Di~Stefano, R., \& Ray, A. 2016, The Astrophysical Journal, 827, 54,
  \dodoi{10.3847/0004-637X/827/1/54}

\bibitem[{{Drake} {et~al.}(1984){Drake}, {Wolfe}, \& {Seeger}}]{DFM84}
{Drake}, F., {Wolfe}, J.~H., \& {Seeger}, C.~L. 1984, SETI science working
  group report, 2244

\bibitem[{{Dressing} \& {Charbonneau}(2015)}]{dc15}
{Dressing}, C.~D., \& {Charbonneau}, D. 2015, \apj, 807, 45,
  \dodoi{10.1088/0004-637X/807/1/45}

\bibitem[{{Eatough} {et~al.}(2013{\natexlab{a}}){Eatough}, {Karuppusamy},
  {Kramer}, {Klein}, {Champion}, \& {Kraus}}]{Eatough2013ATel}
{Eatough}, R., {Karuppusamy}, R., {Kramer}, M., {et~al.} 2013{\natexlab{a}},
  The Astronomer's Telegram, 5040

\bibitem[{{Eatough} {et~al.}(2013{\natexlab{b}}){Eatough}, {Falcke},
  {Karuppusamy}, {Lee}, {Champion}, {Keane}, \& et~al.}]{Eatough2013}
{Eatough}, R.~P., {Falcke}, H., {Karuppusamy}, R., {et~al.} 2013{\natexlab{b}},
  Nature, 501, 391, \dodoi{10.1038/nature12499}

\bibitem[{{Enriquez} {et~al.}(2017){Enriquez}, {Siemion}, {Foster}, {Gajjar},
  {Hellbourg}, {Hickish}, {Isaacson}, {Price}, {Croft}, {DeBoer}, {Lebofsky},
  {MacMahon}, \& {Werthimer}}]{Enriquez:2017}
{Enriquez}, J.~E., {Siemion}, A., {Foster}, G., {et~al.} 2017, \apj, 849, 104,
  \dodoi{10.3847/1538-4357/aa8d1b}

\bibitem[{{Gajjar} {et~al.}(2020){Gajjar}, {Perez}, {Siemion}, {MacMahon},
  {Lebofsky}, {Croft}, \& {Price}}]{gps+20}
{Gajjar}, V., {Perez}, K., {Siemion}, A., {et~al.} 2020, The Astronomer's
  Telegram, 13575, 1

\bibitem[{{Gajjar} {et~al.}(2018){Gajjar}, {Siemion}, {Price}, {Law},
  {Michilli}, {Hessels}, {Chatterjee}, {Archibald}, {Bower}, {Brinkman},
  {Burke-Spolaor}, {Cordes}, {Croft}, {Enriquez}, {Foster}, {Gizani},
  {Hellbourg}, {Isaacson}, {Kaspi}, {Lazio}, {Lebofsky}, {Lynch}, {MacMahon},
  {McLaughlin}, {Ransom}, {Scholz}, {Seymour}, {Spitler}, {Tendulkar},
  {Werthimer}, \& {Zhang}}]{gaj18apj}
{Gajjar}, V., {Siemion}, A.~P.~V., {Price}, D.~C., {et~al.} 2018, \apj, 863, 2,
  \dodoi{10.3847/1538-4357/aad005}

\bibitem[{{Gajjar} {et~al.}(2019){Gajjar}, {Siemion}, {Croft}, {Brzycki},
  {Burgay}, {Carozzi}, {Concu}, {Czech}, {DeBoer}, {DeMarines}, {Drew},
  {Enriquez}, {Fawcett}, {Gallagher}, {Gerret}, {Gizani}, {Hellbourg},
  {Holder}, {Isaacson}, {Kudale}, {Lacki}, {Lebofsky}, {Li}, {MacMahon},
  {McCauley}, {Melis}, {Molinari}, {Murphy}, {Perrodin}, {Pilia}, {Price},
  {Webb}, {Werthimer}, {Williams}, {Worden}, {Zarka}, \& {Zhang}}]{gsc+19}
{Gajjar}, V., {Siemion}, A., {Croft}, S., {et~al.} 2019, in \baas, Vol.~51,
  223.
\newblock \doarXiv{1907.05519}

\bibitem[{{Genzel} {et~al.}(2010){Genzel}, {Eisenhauer}, \&
  {Gillessen}}]{Genzel_GC_BH}
{Genzel}, R., {Eisenhauer}, F., \& {Gillessen}, S. 2010, Reviews of Modern
  Physics, 82, 3121, \dodoi{10.1103/RevModPhys.82.3121}

\bibitem[{{Ghez} {et~al.}(2008){Ghez}, {Salim}, {Weinberg}, {Lu}, {Do}, {Dunn},
  {Matthews}, {Morris}, {Yelda}, {Becklin}, {Kremenek}, {Milosavljevic}, \&
  {Naiman}}]{Ghez_BH_GC}
{Ghez}, A.~M., {Salim}, S., {Weinberg}, N.~N., {et~al.} 2008, \apj, 689, 1044,
  \dodoi{10.1086/592738}

\bibitem[{{Gowanlock} {et~al.}(2011){Gowanlock}, {Patton}, \&
  {McConnell}}]{gpm11}
{Gowanlock}, M.~G., {Patton}, D.~R., \& {McConnell}, S.~M. 2011, Astrobiology,
  11, 855, \dodoi{10.1089/ast.2010.0555}

\bibitem[{{Gravity Collaboration} {et~al.}(2019){Gravity Collaboration},
  {Abuter}, {Amorim}, {Baub{\"o}ck}, {Berger}, {Bonnet}, {Brand ner},
  {Cl{\'e}net}, {Coud{\'e} Du Foresto}, {de Zeeuw}, {Dexter}, {Duvert},
  {Eckart}, {Eisenhauer}, {F{\"o}rster Schreiber}, {Garcia}, {Gao}, {Gendron},
  {Genzel}, {Gerhard}, {Gillessen}, {Habibi}, {Haubois}, {Henning}, {Hippler},
  {Horrobin}, {Jim{\'e}nez-Rosales}, {Jocou}, {Kervella}, {Lacour},
  {Lapeyr{\`e}re}, {Le Bouquin}, {L{\'e}na}, {Ott}, {Paumard}, {Perraut},
  {Perrin}, {Pfuhl}, {Rabien}, {Rodriguez Coira}, {Rousset}, {Scheithauer},
  {Sternberg}, {Straub}, {Straubmeier}, {Sturm}, {Tacconi}, {Vincent}, {von
  Fellenberg}, {Waisberg}, {Widmann}, {Wieprecht}, {Wiezorrek}, {Woillez}, \&
  {Yazici}}]{Distance_to_gc}
{Gravity Collaboration}, {Abuter}, R., {Amorim}, A., {et~al.} 2019, \aap, 625,
  L10, \dodoi{10.1051/0004-6361/201935656}

\bibitem[{Hankins \& Eilek(2007)}]{Hankins_2007}
Hankins, T.~H., \& Eilek, J.~A. 2007, The Astrophysical Journal, 670, 693,
  \dodoi{10.1086/522362}

\bibitem[{{Harp} {et~al.}(2012){Harp}, {Ackermann}, {Blair}, {Arbunich},
  {Backus}, {Tarter}, \& {the ATA Team}}]{2012arXiv1211.6470H}
{Harp}, G.~R., {Ackermann}, R.~F., {Blair}, S.~K., {et~al.} 2012, ArXiv
  e-prints.
\newblock \doarXiv{1211.6470}

\bibitem[{Harp {et~al.}(2016)Harp, Richards, Tarter, Dreher, Jordan, Shostak,
  Smolek, Kilsdonk, Wilcox, Wimberly, Ross, Barott, Ackermann, \&
  Blair}]{Harp_2016}
Harp, G.~R., Richards, J., Tarter, J.~C., {et~al.} 2016, The Astronomical
  Journal, 152, 181, \dodoi{10.3847/0004-6256/152/6/181}

\bibitem[{{Harp} {et~al.}(2018){Harp}, {Ackermann}, {Astorga}, {Arbunich},
  {Barrios}, {Hightower}, {Meitzner}, {Barott}, {Nolan}, {Messerschmitt},
  {Vakoch}, {Shostak}, \& {Tarter}}]{Harp:2018apj}
{Harp}, G.~R., {Ackermann}, R.~F., {Astorga}, A., {et~al.} 2018, \apj, 869, 66,
  \dodoi{10.3847/1538-4357/aaeb98}

\bibitem[{{Hobbs} {et~al.}(2020){Hobbs}, {Manchester}, {Dunning}, {Jameson},
  {Roberts}, {George}, {Green}, {Tuthill}, {Toomey}, {Kaczmarek}, {Mader},
  {Marquarding}, {Ahmed}, {Amy}, {Bailes}, {Beresford}, {Bhat}, {Bock},
  {Bourne}, {Bowen}, {Brothers}, {Cameron}, {Carretti}, {Carter}, {Castillo},
  {Chekkala}, {Cheng}, {Chung}, {Craig}, {Dai}, {Dawson}, {Dempsey}, {Doherty},
  {Dong}, {Edwards}, {Ergesh}, {Gao}, {Han}, {Hayman}, {Indermuehle},
  {Jeganathan}, {Johnston}, {Kanoniuk}, {Kesteven}, {Kramer}, {Leach},
  {Mcintyre}, {Moss}, {Os{\l}owski}, {Phillips}, {Pope}, {Preisig}, {Price},
  {Reeves}, {Reilly}, {Reynolds}, {Robishaw}, {Roush}, {Ruckley}, {Sadler},
  {Sarkissian}, {Severs}, {Shannon}, {Smart}, {Smith}, {Smith}, {Sobey},
  {Staveley-Smith}, {Tzioumis}, {van Straten}, {Wang}, {Wen}, \&
  {Whiting}}]{hobbs_UWL2020}
{Hobbs}, G., {Manchester}, R.~N., {Dunning}, A., {et~al.} 2020, \pasa, 37,
  e012, \dodoi{10.1017/pasa.2020.2}

\bibitem[{{Horowitz} \& {Sagan}(1993)}]{1993ApJ...415..218H}
{Horowitz}, P., \& {Sagan}, C. 1993, \apj, 415, 218, \dodoi{10.1086/173157}

\bibitem[{{Howell} {et~al.}(2014){Howell}, {Sobeck}, {Haas}, {Still},
  {Barclay}, {Mullally}, {Troeltzsch}, {Aigrain}, {Bryson}, {Caldwell},
  {Chaplin}, {Cochran}, {Huber}, {Marcy}, {Miglio}, {Najita}, {Smith},
  {Twicken}, \& {Fortney}}]{hsh+14}
{Howell}, S.~B., {Sobeck}, C., {Haas}, M., {et~al.} 2014, \pasp, 126, 398,
  \dodoi{10.1086/676406}

\bibitem[{Isaacson {et~al.}(2017)Isaacson, Siemion, Marcy, Lebofsky, Price,
  MacMahon, Croft, DeBoer, Hickish, Werthimer, Sheikh, Hellbourg, \&
  Enriquez}]{Isaacson:2017ib}
Isaacson, H., Siemion, A. P.~V., Marcy, G.~W., {et~al.} 2017, Publications of
  the Astronomical Society of the Pacific, 129, 054501

\bibitem[{{Isaacson} {et~al.}(2017){Isaacson}, {Siemion}, {Marcy}, {Lebofsky},
  {Price}, {MacMahon}, {Croft}, {DeBoer}, {Hickish}, {Werthimer}, {Sheikh},
  {Hellbourg}, \& {Enriquez}}]{2017PASP..129e4501I}
{Isaacson}, H., {Siemion}, A.~P.~V., {Marcy}, G.~W., {et~al.} 2017, PASP, 129,
  054501, \dodoi{10.1088/1538-3873/aa5800}

\bibitem[{{Jim{\'e}nez-Torres} {et~al.}(2013){Jim{\'e}nez-Torres}, {Pichardo},
  {Lake}, \& {Segura}}]{Torres_habitability}
{Jim{\'e}nez-Torres}, J.~J., {Pichardo}, B., {Lake}, G., \& {Segura}, A. 2013,
  Astrobiology, 13, 491, \dodoi{10.1089/ast.2012.0842}

\bibitem[{{Johnston} {et~al.}(2006){Johnston}, {Kramer}, {Lorimer}, {Lyne},
  {McLaughlin}, {Klein}, \& {Manchester}}]{2006MNRAS.373L...6J}
{Johnston}, S., {Kramer}, M., {Lorimer}, D.~R., {et~al.} 2006, MNRAS, 373, L6,
  \dodoi{10.1111/j.1745-3933.2006.00232.x}

\bibitem[{{Kennea} {et~al.}(2013){Kennea}, {Krimm}, {Barthelmy}, {Gehrels},
  {Markwardt}, {Cummings}, \& et~al.}]{Kennea2013}
{Kennea}, J.~A., {Krimm}, H., {Barthelmy}, S., {et~al.} 2013, The Astronomer's
  Telegram, 5009, 1

\bibitem[{{Krishnakumar} {et~al.}(2017){Krishnakumar}, {Joshi}, \&
  {Manoharan}}]{kjm+17}
{Krishnakumar}, M.~A., {Joshi}, B.~C., \& {Manoharan}, P.~K. 2017, \apj, 846,
  104, \dodoi{10.3847/1538-4357/aa7af2}

\bibitem[{{Krishnakumar} {et~al.}(2019){Krishnakumar}, {Maan}, {Joshi}, \&
  {Manoharan}}]{kmj+19}
{Krishnakumar}, M.~A., {Maan}, Y., {Joshi}, B.~C., \& {Manoharan}, P.~K. 2019,
  \apj, 878, 130, \dodoi{10.3847/1538-4357/ab20c5}

\bibitem[{Law {et~al.}(2008)Law, Yusef‐Zadeh, Cotton, \&
  Maddalena}]{Law_2008}
Law, C.~J., Yusef‐Zadeh, F., Cotton, W.~D., \& Maddalena, R.~J. 2008, The
  Astrophysical Journal Supplement Series, 177, 255–274,
  \dodoi{10.1086/533587}

\bibitem[{{Lebofsky} {et~al.}(2019){Lebofsky}, {Croft}, {Siemion}, {Price},
  {Enriquez}, {Isaacson}, {MacMahon}, {Anderson}, {Brzycki}, {Cobb}, {Czech},
  {DeBoer}, {DeMarines}, {Drew}, {Foster}, {Gajjar}, {Gizani}, {Hellbourg},
  {Korpela}, {Lacki}, {Sheikh}, {Werthimer}, {Worden}, {Yu}, \&
  {Zhang}}]{leb19_bl_data_format}
{Lebofsky}, M., {Croft}, S., {Siemion}, A. P.~V., {et~al.} 2019, \pasp, 131,
  124505, \dodoi{10.1088/1538-3873/ab3e82}

\bibitem[{{Levin} {et~al.}(2010{\natexlab{a}}){Levin}, {Bailes}, {Bates},
  {Bhat}, {Burgay}, {Burke-Spolaor}, \& et~al.}]{Levin2010}
{Levin}, L., {Bailes}, M., {Bates}, S., {et~al.} 2010{\natexlab{a}}, The
  Astrophysical Journal Letters, 721, L33, \dodoi{10.1088/2041-8205/721/1/L33}

\bibitem[{{Levin} {et~al.}(2010{\natexlab{b}}){Levin}, {Bailes}, {Bates},
  {Bhat}, {Burgay}, {Burke-Spolaor}, {D'Amico}, {Johnston}, {Keith}, {Kramer},
  {Milia}, {Possenti}, {Rea}, {Stappers}, \& {van Straten}}]{lbb+10}
---. 2010{\natexlab{b}}, \apjl, 721, L33, \dodoi{10.1088/2041-8205/721/1/L33}

\bibitem[{{Li} {et~al.}(2020){Li}, {Gajjar}, {Wang}, {Siemion}, {Zhang},
  {Zhang}, {Yue}, {Zhu}, {Jin}, {Li}, {Berger}, {Brzycki}, {Cobb}, {Croft},
  {Czech}, {DeBoer}, {DeMarines}, {Drew}, {Emilio Enriquez}, {Gizani},
  {Korpela}, {Isaacson}, {Lebofsky}, {Lacki}, {MacMahon}, {Nanez}, {Niu},
  {Pei}, {Price}, {Werthimer}, {Worden}, {Gerry Zhang}, {Zhang}, \& {FAST
  Collaboration}}]{li_gajjar_2020}
{Li}, D., {Gajjar}, V., {Wang}, P., {et~al.} 2020, Research in Astronomy and
  Astrophysics, 20, 078, \dodoi{10.1088/1674-4527/20/5/78}

\bibitem[{{Lineweaver} {et~al.}(2004){Lineweaver}, {Fenner}, \&
  {Gibson}}]{lineweaver2004GHZ}
{Lineweaver}, C.~H., {Fenner}, Y., \& {Gibson}, B.~K. 2004, Science, 303, 59,
  \dodoi{10.1126/science.1092322}

\bibitem[{Lipman {et~al.}(2018)Lipman, Isaacson, Siemion, Lebofsky, Price,
  MacMahon, Croft, DeBoer, Hickish, Werthimer, Hellbourg, Enriquez, \&
  Gizani}]{Lipman:2018wv}
Lipman, D., Isaacson, H., Siemion, A. P.~V., {et~al.} 2018, arXiv.org

\bibitem[{{Liu} {et~al.}(2012){Liu}, {Wex}, {Kramer}, {Cordes}, \&
  {Lazio}}]{Liu2012}
{Liu}, K., {Wex}, N., {Kramer}, M., {Cordes}, J.~M., \& {Lazio}, T.~J.~W. 2012,
  ApJ, 747, 1, \dodoi{10.1088/0004-637X/747/1/1}

\bibitem[{{MacMahon} {et~al.}(2018){MacMahon}, {Price}, {Lebofsky}, {Siemion},
  {Croft}, {DeBoer}, {Enriquez}, {Gajjar}, {Hellbourg}, {Isaacson},
  {Werthimer}, {Abdurashidova}, {Bloss}, {Brandt}, {Creager}, {Ford}, {Lynch},
  {Maddalena}, {McCullough}, {Ray}, {Whitehead}, \&
  {Woody}}]{2018PASP..130d4502M}
{MacMahon}, D.~H.~E., {Price}, D.~C., {Lebofsky}, M., {et~al.} 2018, PASP, 130,
  044502, \dodoi{10.1088/1538-3873/aa80d2}

\bibitem[{{Macquart} {et~al.}(2010){Macquart}, {Kanekar}, {Frail}, \&
  {Ransom}}]{Macquart2010}
{Macquart}, J.-P., {Kanekar}, N., {Frail}, D.~A., \& {Ransom}, S.~M. 2010, The
  Astrophysical Journal, 715, 939, \dodoi{10.1088/0004-637X/715/2/939}

\bibitem[{{Margot} {et~al.}(2021){Margot}, {Pinchuk}, {Geil}, {Alexander},
  {Arora}, {Biswas}, {Cebreros}, {Desai}, {Duclos}, {Dunne}, {Lin Fu}, {Goel},
  {Gonzales}, {Gonzalez}, {Jain}, {Lam}, {Lewis}, {Lewis}, {Li}, {MacDougall},
  {Makarem}, {Manan}, {Molina}, {Nagib}, {Neville}, {O'Toole}, {Rockwell},
  {Rokushima}, {Romanek}, {Schmidgall}, {Seth}, {Shah}, {Shimane}, {Singhal},
  {Tokadjian}, {Villafana}, {Wang}, {Yun}, {Zhu}, \& {Lynch}}]{mpg+21}
{Margot}, J.-L., {Pinchuk}, P., {Geil}, R., {et~al.} 2021, \aj, 161, 55,
  \dodoi{10.3847/1538-3881/abcc77}

\bibitem[{{Mori} {et~al.}(2013){Mori}, {Gotthelf}, {Zhang}, {An}, {Baganoff},
  {Barri{\`e}re}, \& et~al.}]{Mori2013}
{Mori}, K., {Gotthelf}, E.~V., {Zhang}, S., {et~al.} 2013, The Astrophysical
  Journal Letters, 770, L23, \dodoi{10.1088/2041-8205/770/2/L23}

\bibitem[{Morrison \& Gowanlock(2015)}]{Morrison:2015}
Morrison, I.~S., \& Gowanlock, M.~G. 2015, Astrobiology, 15, 683,
  \dodoi{10.1089/ast.2014.1192}

\bibitem[{Newman \& Sagan(1981)}]{NEWMAN_SAGAN}
Newman, W.~I., \& Sagan, C. 1981, Icarus, 46, 293,
  \dodoi{https://doi.org/10.1016/0019-1035(81)90135-4}

\bibitem[{Oliver \& Billingham(1971)}]{NASA:2003p185}
Oliver, B.~M., \& Billingham, J., eds. 1971, {Project Cyclops: A Design Study
  of a System for Detecting Extraterrestrial Intelligent Life} (NASA)

\bibitem[{{Pearlman} {et~al.}(2018){Pearlman}, {Majid}, {Prince}, {Kocz}, \&
  {Horiuchi}}]{pealmanSGR1745}
{Pearlman}, A.~B., {Majid}, W.~A., {Prince}, T.~A., {Kocz}, J., \& {Horiuchi},
  S. 2018, \apj, 866, 160, \dodoi{10.3847/1538-4357/aade4d}

\bibitem[{{Pettitt} {et~al.}(2015){Pettitt}, {Dobbs}, {Acreman}, \&
  {Bate}}]{pda+15}
{Pettitt}, A.~R., {Dobbs}, C.~L., {Acreman}, D.~M., \& {Bate}, M.~R. 2015,
  \mnras, 449, 3911, \dodoi{10.1093/mnras/stv600}

\bibitem[{{Pfahl} \& {Loeb}(2004)}]{PfahlLoeb2004}
{Pfahl}, E., \& {Loeb}, A. 2004, The Astrophysical Journal, 615, 253,
  \dodoi{10.1086/423975}

\bibitem[{Pinchuk {et~al.}(2019)Pinchuk, Margot, Greenberg, Ayalde, Bloxham,
  Boddu, Chinchilla-Garcia, Cliffe, Gallagher, Hart, \& et~al.}]{Pinchuk_2019}
Pinchuk, P., Margot, J.-L., Greenberg, A.~H., {et~al.} 2019, The Astronomical
  Journal, 157, 122, \dodoi{10.3847/1538-3881/ab0105}

\bibitem[{{Prestage} {et~al.}(2015){Prestage}, {Bloss}, {Brandt}, {Chen},
  {Creager}, {Demorest}, {Ford}, {Jones}, {Kepley}, {Kobelski}, {Marganian},
  {Mello}, {McMahon}, {McCullough}, {Ray}, {Roshi}, {Werthimer}, \&
  {Whitehead}}]{vegas_2015}
{Prestage}, R.~M., {Bloss}, M., {Brandt}, J., {et~al.} 2015, in 2015 URSI-USNC
  Radio Science Meeting, 4, \dodoi{10.1109/USNC-URSI.2015.7303578}

\bibitem[{Price {et~al.}(2018)Price, MacMahon, Lebofsky, Croft, DeBoer,
  Enriquez, Foster, Gajjar, Gizani, Hellbourg, Isaacson, Siemion, Werthimer,
  Green, Amy, Ball, Bock, Craig, Edwards, Jameson, Mader, Preisig, Smith,
  Reynolds, \& Sarkissian}]{Price:2018bv}
Price, D.~C., MacMahon, D. H.~E., Lebofsky, M., {et~al.} 2018, Publications of
  the Astronomical Society of Australia, 35, 213

\bibitem[{{Price} {et~al.}(2020){Price}, {Enriquez}, {Brzycki}, {Croft},
  {Czech}, {DeBoer}, {DeMarines}, {Foster}, {Gajjar}, {Gizani}, {Hellbourg},
  {Isaacson}, {Lacki}, {Lebofsky}, {MacMahon}, {Pater}, {Siemion}, {Werthimer},
  {Green}, {Kaczmarek}, {Maddalena}, {Mader}, {Drew}, \& {Worden}}]{price2020}
{Price}, D.~C., {Enriquez}, J.~E., {Brzycki}, B., {et~al.} 2020, \aj, 159, 86,
  \dodoi{10.3847/1538-3881/ab65f1}

\bibitem[{Rajwade {et~al.}(2017)Rajwade, Lorimer, \& Anderson}]{Rajwade_2017}
Rajwade, K.~M., Lorimer, D.~R., \& Anderson, L.~D. 2017, Monthly Notices of the
  Royal Astronomical Society, 471, 730–739, \dodoi{10.1093/mnras/stx1661}

\bibitem[{{Schelling}(1960)}]{schelling_1960}
{Schelling}, C.~T. 1960, {The strategy of conflict} (Oxford University Press)

\bibitem[{Sheikh {et~al.}(2020)Sheikh, Siemion, Enriquez, Price, Isaacson,
  Lebofsky, Gajjar, \& Kalas}]{Sheikh_2020}
Sheikh, S.~Z., Siemion, A., Enriquez, J.~E., {et~al.} 2020, The Astronomical
  Journal, 160, 29, \dodoi{10.3847/1538-3881/ab9361}

\bibitem[{Sheikh {et~al.}(2019)Sheikh, Wright, Siemion, \&
  Enriquez}]{Sheikh_2019}
Sheikh, S.~Z., Wright, J.~T., Siemion, A., \& Enriquez, J.~E. 2019, The
  Astrophysical Journal, 884, 14, \dodoi{10.3847/1538-4357/ab3fa8}

\bibitem[{Shostak \& Tarter(1985)}]{SHOSTAK1985369}
Shostak, G., \& Tarter, J. 1985, Acta Astronautica, 12, 369 ,
  \dodoi{https://doi.org/10.1016/0094-5765(85)90071-2}

\bibitem[{Siemion {et~al.}(2010)Siemion, Korff, McMahon, Korpela, Werthimer,
  Anderson, Bower, Cobb, Foster, Lebofsky, Leeuwen, \&
  Wagner}]{Siemion:2010p6845}
Siemion, A., Korff, J.~V., McMahon, P., {et~al.} 2010, Acta Astronautica, 67,
  1342.
\newblock \url{http://linkinghub.elsevier.com/retrieve/pii/S0094576510000299}

\bibitem[{{Siemion} {et~al.}(2013){Siemion}, {Demorest}, {Korpela},
  {Maddalena}, {Werthimer}, {Cobb}, {Howard}, {Langston}, {Lebofsky}, {Marcy},
  \& {Tarter}}]{2013ApJ...767...94S}
{Siemion}, A.~P.~V., {Demorest}, P., {Korpela}, E., {et~al.} 2013, The
  Astrophysical Journal, 767, 94, \dodoi{10.1088/0004-637X/767/1/94}

\bibitem[{{Spitler} {et~al.}(2014){Spitler}, {Lee}, {Eatough}, {Kramer},
  {Karuppusamy}, {Bassa}, \& et~al.}]{Spitler2014}
{Spitler}, L.~G., {Lee}, K.~J., {Eatough}, R.~P., {et~al.} 2014, The
  Astrophysical Journal Letters, 780, L3, \dodoi{10.1088/2041-8205/780/1/L3}

\bibitem[{{Surnis} {et~al.}(2020{\natexlab{a}}){Surnis}, {Joshi}, {Bagchi},
  {McLaughlin}, {Lorimer}, {Blumer}, {Gajjar}, {Perera}, {Manoharan}, \&
  {Stappers}}]{surnisSGR1}
{Surnis}, M., {Joshi}, B.~C., {Bagchi}, M., {et~al.} 2020{\natexlab{a}}, The
  Astronomer's Telegram, 13777, 1

\bibitem[{{Surnis} {et~al.}(2020{\natexlab{b}}){Surnis}, {Joshi}, {Bagchi},
  {Gajjar}, {Sand}, {McLaughlin}, {Lorimer}, {Blumer}, {Perera}, {Manoharan},
  \& {Stappers}}]{surnisSGR2}
---. 2020{\natexlab{b}}, The Astronomer's Telegram, 13799, 1

\bibitem[{Tarter(2003)}]{Tarter:2003p266}
Tarter, J. 2003, Annual Reviews of Astronomy and Astrophysics, 39, 511.
\newblock
  \url{http://arjournals.annualreviews.org/doi/abs/10.1146/annurev.astro.39.1.511?prevSearch=%255Bauthor%253A%2Btarter%2Bjill%255D&searchHistoryKey=}

\bibitem[{{Tarter} {et~al.}(1980){Tarter}, {Cuzzi}, {Black}, \&
  {Clark}}]{1980Icar...42..136T}
{Tarter}, J., {Cuzzi}, J., {Black}, D., \& {Clark}, T. 1980, \icarus, 42, 136,
  \dodoi{10.1016/0019-1035(80)90251-1}

\bibitem[{{The CHIME/FRB Collaboration} {et~al.}(2020){The CHIME/FRB
  Collaboration}, {:}, {Andersen}, {Band ura}, {Bhardwaj}, {Bij}, {Boyce},
  {Boyle}, {Brar}, {Cassanelli}, {Chawla}, {Chen}, {Cliche}, {Cook},
  {Cubranic}, {Curtin}, {Denman}, {Dobbs}, {Dong}, {Fandino}, {Fonseca},
  {Gaensler}, {Giri}, {Good}, {Halpern}, {Hill}, {Hinshaw}, {H{\"o}fer},
  {Josephy}, {Kania}, {Kaspi}, {Landecker}, {Leung}, {Li}, {Lin}, {Masui},
  {Mckinven}, {Mena-Parra}, {Merryfield}, {Meyers}, {Michilli}, {Milutinovic},
  {Mirhosseini}, {M{\"u}nchmeyer}, {Naidu}, {Newburgh}, {Ng}, {Patel}, {Pen},
  {Pinsonneault-Marotte}, {Pleunis}, {Quine}, {Rafiei-Ravandi}, {Rahman},
  {Ransom}, {Renard}, {Sanghavi}, {Scholz}, {Shaw}, {Shin}, {Siegel}, {Singh},
  {Smegal}, {Smith}, {Stairs}, {Tan}, {Tendulkar}, {Tretyakov}, {Vanderlinde},
  {Wang}, {Wulf}, \& {Zwaniga}}]{chime_sgr1935}
{The CHIME/FRB Collaboration}, {:}, {Andersen}, B.~C., {et~al.} 2020, arXiv
  e-prints, arXiv:2005.10324.
\newblock \doarXiv{2005.10324}

\bibitem[{{Tingay} {et~al.}(2016){Tingay}, {Tremblay}, {Walsh}, \&
  {Urquhart}}]{Tingay_2016}
{Tingay}, S.~J., {Tremblay}, C., {Walsh}, A., \& {Urquhart}, R. 2016, \apjl,
  827, L22, \dodoi{10.3847/2041-8205/827/2/L22}

\bibitem[{{Tingay} {et~al.}(2018){Tingay}, {Tremblay}, \&
  {Croft}}]{Tingay_2018}
{Tingay}, S.~J., {Tremblay}, C.~D., \& {Croft}, S. 2018, \apj, 856, 31,
  \dodoi{10.3847/1538-4357/aab363}

\bibitem[{Torne {et~al.}(2015)Torne, Eatough, Karuppusamy, Kramer, Paubert,
  Klein, Desvignes, Champion, Wiesemeyer, Kramer, Spitler, Thum, Güsten,
  Schuster, \& Cognard}]{tek+15}
Torne, P., Eatough, R.~P., Karuppusamy, R., {et~al.} 2015, Monthly Notices of
  the Royal Astronomical Society: Letters, 451, L50,
  \dodoi{10.1093/mnrasl/slv063}

\bibitem[{{Valdes} \& {Freitas}(1986)}]{1986Icar...65..152V}
{Valdes}, F., \& {Freitas}, R.~A., J. 1986, \icarus, 65, 152,
  \dodoi{10.1016/0019-1035(86)90069-2}

\bibitem[{{Verschuur}(1973)}]{1973Icar...19..329V}
{Verschuur}, G.~L. 1973, \icarus, 19, 329, \dodoi{10.1016/0019-1035(73)90109-7}

\bibitem[{von Korff(2010)}]{vonKorff:2010p3275}
von Korff, J. 2010, UC Berkeley PhD Thesis in Physics.
\newblock
  \url{http://seti.berkeley.edu/sites/default/files/vonkorff_thesis_full_051010.pdf}

\bibitem[{{Wharton} {et~al.}(2019){Wharton}, {Chatterjee}, {Cordes}, {Bower},
  {Butler}, {Deller}, {Demorest}, {Lazio}, \& {Ransom}}]{wharton1745VLA}
{Wharton}, R.~S., {Chatterjee}, S., {Cordes}, J.~M., {et~al.} 2019, \apj, 875,
  143, \dodoi{10.3847/1538-4357/ab100a}

\bibitem[{Worden {et~al.}(2017)Worden, Drew, Siemion, Werthimer, DeBoer, Croft,
  MacMahon, Lebofsky, Isaacson, Hickish, Price, Gajjar, \&
  Wright}]{2017AcAau.139...98W}
Worden, S.~P., Drew, J., Siemion, A., {et~al.} 2017, Acta Astronautica, 139, 98

\bibitem[{{Wordsworth} \& {Pierrehumbert}(2014)}]{wp14}
{Wordsworth}, R., \& {Pierrehumbert}, R. 2014, \apjl, 785, L20,
  \dodoi{10.1088/2041-8205/785/2/L20}

\bibitem[{{Wright}(2018)}]{wright18}
{Wright}, J.~T. 2018, {Exoplanets and SETI}, 186,
  \dodoi{10.1007/978-3-319-55333-7_186}

\bibitem[{{Wright} {et~al.}(2018){Wright}, {Kanodia}, \& {Lubar}}]{wkl18}
{Wright}, J.~T., {Kanodia}, S., \& {Lubar}, E. 2018, \aj, 156, 260,
  \dodoi{10.3847/1538-3881/aae099}

\bibitem[{Yao {et~al.}(2017)Yao, Manchester, \& Wang}]{YMW16}
Yao, J.~M., Manchester, R.~N., \& Wang, N. 2017, \apj, 835, 29,
  \dodoi{10.3847/1538-4357/835/1/29}

\bibitem[{{Zhang} {et~al.}(2020){Zhang}, {Jiang}, {Men}, {Wang}, {Xu}, {Xu},
  {Niu}, {Zhou}, {Guan}, {Han}, {Jiang}, {Lee}, {Li}, {Lin}, {Niu}, {Wang},
  {Wang}, {Xu}, {Yu}, {Zhang}, \& {Zhu}}]{fastSGR1935}
{Zhang}, C.~F., {Jiang}, J.~C., {Men}, Y.~P., {et~al.} 2020, The Astronomer's
  Telegram, 13699, 1

\end{thebibliography}
\bibliographystyle{aasjournal}



\end{document}